\documentclass[3p,times]{elsarticle}
\usepackage{hyperref}
\hypersetup{
    colorlinks,
    citecolor=blue,
    filecolor=blue,
    linkcolor=blue,
    urlcolor=blue
}

\journal{Progress in Particle and Nuclear Physics}

\usepackage{natbib}
\biboptions{sort&compress}

\usepackage{wrapfig}
\usepackage{epsfig}
\usepackage{amssymb}
\usepackage{amsfonts}
\usepackage{amsmath}
\usepackage{fixmath}
\usepackage{lineno}
\usepackage{color}
\usepackage{xcolor}
\usepackage{multirow}
\usepackage{longtable}
\usepackage{lscape}
\usepackage{tabularx}
\usepackage[utf8x]{inputenc}
\usepackage{bigstrut}
\usepackage{soul}
\usepackage{bm}
\usepackage{comment}
\usepackage{ulem}

\DeclareGraphicsExtensions{.png,.pdf,.jpg,.mps,.eps}
\graphicspath{{gluon_paper/figs_eps/}}
\definecolor{patriarch}{rgb}{0.5, 0.0, 0.5}

\newcommand{\ktsq}{\bm{{k}}_{T}^2}

\newcommand{\ktv}{\bm{{k}}_{T}}
\newcommand{\qtsq}{\bm{{q}}_{T}^2}

\begin{document}
%\linenumbers

% declarations for front matter
\begin{frontmatter}
\title{On the physics potential to study the gluon content of proton and deuteron at NICA SPD}
%\author{list of authors and contributors }
\author[1]{A.~Arbuzov}
\author[2,3]{A. Bacchetta}
\author[4]{M.~Butenschoen}
\author[2,3,5,6]{F.G. Celiberto}
\author[7,8]{U. D'Alesio}
\author[1]{M.~Deka}
\author[1]{I.~Denisenko}
\author[9]{M.~G.~Echevarria}
\author[1]{A.~Efremov}
\author[1,10]{N.Ya.~Ivanov}
\author[1,11]{A.~Guskov}
\author[12,1]{A.~Karpishkov}
\author[1,13]{Ya.~Klopot}
\author[4]{B.~A.~Kniehl}
\author[10,15]{A.~Kotzinian}
\author[16]{S.~Kumano}
\author[17]{J.P.~Lansberg}
\author[18]{Keh-Fei~Liu}
\author[8]{F.~Murgia}
\author[12]{M.~Nefedov}
\author[1,14,15]{B.~Parsamyan}
\author[7,8]{C.~Pisano}
\author[3]{M. Radici}
\author[1]{A.~Rymbekova}
\author[12,1]{V.~Saleev}
\author[12,1]{A.~Shipilova}
\author[19]{Qin-Tao~Song}
\author[1]{O.~Teryaev}

\address[1]{\small Joint Institute for Nuclear Research, 141980 Dubna, Moscow region, Russia}
\address[2]{\small Dipartimento di Fisica, Universit\`a di Pavia, via Bassi 6, I-27100 Pavia, Italy}
\address[3]{\small INFN Sezione di Pavia, via Bassi 6, I-27100 Pavia, Italy}
\address[4]{\small II. Institut f\"ur Theoretische Physik, Universit\"at Hamburg, Luruper Chaussee 149, 22761 Hamburg, Germany}
\address[5]{\small European Centre for Theoretical Studies in Nuclear Physics and Related Areas (ECT*), I-38123 Villazzano, Trento, Italy}
\address[6]{\small Fondazione Bruno Kessler (FBK), I-38123 Povo, Trento, Italy} 
\address[7]{\small Dipartimento di Fisica, Università di Cagliari, I-09042 Monserrato, Italy}
\address[8]{\small INFN Sezione di Cagliari, I-09042 Monserrato, Italy}
\address[9]{\small Dpto. de F\'{i}sica y Matem\'aticas, Universidad de Alcal\'{a}, 28805 Alcal\'{a} de Henares (Madrid), Spain}
\address[10]{\small Yerevan Physics Institute, 0036 Yerevan, Armenia}
\address[11]{\small Moscow Institute of Physics and Technology, Moscow Region, 141700, Russia}
\address[12]{\small Samara National Research University, 443000  Samara, Russia}
\address[13]{\small Bogolyubov Institute for Theoretical Physics, 03143 Kiev, Ukraine}
\address[14]{\small Dipartimento di Fisica, Universit\`a di Torino, Via Peitro Giuria 1, 10125 Torino, Italy}
\address[15]{\small INFN Sezione di Torino, Via Peitro Giuria 1, 10125 Torino, Italy}
\address[16]{\small Institute of Particle and Nuclear Studies, High Energy Accelerator Research Organization (KEK), Oho 1-1, Tsukuba, Ibaraki, 305-0801, Japan}
\address[17]{\small Universit\'e Paris-Saclay, CNRS, IJCLab, 91405 Orsay, France}
\address[18]{\small Department of Physics and Astronomy, University of Kentucky, Lexington, KY 40506, USA}
\address[19]{\small School of Physics and Microelectronics, Zhengzhou University, Zhengzhou, Henan 450001, China}

\date{\today}
%\maketitle

%\abstract
\begin{abstract}
{
The Spin Physics Detector (SPD) is a future multipurpose experiment foreseen to run at the NICA collider, which is currently under construction at the Joint Institute for Nuclear Research (JINR, Dubna, Russia). 
%
%The main purpose of the experiment is the study of the nucleon spin structure in collisions of
%
The physics program of the experiment is based on collisions of longitudinally and transversely polarized protons and deuterons at $\sqrt{s}$ up to 27 GeV and luminosity up to 10$^{32}$ cm$^{-2}$ s$^{-1}$.
%
%It will operate as a universal facility for comprehensive study of unpolarized and polarized gluon content in the proton and deuteron. Such complementing probes as charmonia, open charm and prompt-photon production processes will be used for that.
The SPD will operate as a universal facility for the comprehensive study of the unpolarized and polarized gluon content of the nucleon, using complementary probes such as: charmonia, open charm, and prompt photon production processes.
%Possible physics tasks such as the access to the gluon helicity, gluon Sivers and Boer-Mulders function and gluon transversity in the deuteron via the measurement of single and double spin asymmetries and other gluon-related tasks will be discussed.

The aim of this work is to provide a thorough review of the physics objectives that can potentially be addressed at the SPD, underlining related theoretical aspects and discussing relevant experimental results when available.
%
%\textcolor{red}{, as well as of theoretical and experimental peculiarities of processes which will be used to probe the proton structure: production of charmonia, open charm and prompt photons. M.N.} 
%
Among different pertinent phenomena particular attention is drawn to the study of the gluon helicity, gluon Sivers and Boer-Mulders functions in the nucleon, as well as the gluon transversity distribution in the deuteron, via the measurement of single and double spin asymmetries.

%
%proton and deuteron in the high-$x$ region using the set of complementing probes: charmonia, open charm and prompt photons.
%
%The SPD project at NICA is expected to cover the period of about 10 years after 2025. To be attractive for physicists world-wide, the physics scope of the facility is open for future exciting ideas.
%
%
}
\end{abstract}
\end{frontmatter}
\tableofcontents

\section{Introduction}
Gluons, along with quarks, are the fundamental constituents of the nucleon. 
%They are responsible for generation of almost all its mass and carry about half of its momentum. 
 They play a key role in generation of the mass of the  nucleon and carry about half of its momentum in hard (semi)inclusive processes. 
The spin of the nucleon is also defined by its constituents and is built up from the intrinsic spin of the valence and sea quarks (spin-1/2) and gluons (spin-1), and their orbital angular momenta. Notwithstanding the progress achieved during the last decades in the understanding of the quark contribution to the nucleon spin, the gluon sector is much less developed. One of the difficulties is the lack of the direct probes to access the gluon content in high-energy processes. 
While the quark contribution to the nucleon spin was determined quite precisely in semi-inclusive deep-inelastic scattering (SIDIS) experiments like EMC, CLAS, HERMES, and COMPASS, the gluon contribution, determined through the gluon helicity Parton Distribution Function is still not well-constrained experimentally and is expected to be significant.

In recent years, the three-dimensional partonic structure of the nucleon, in particular the spatial and momentum distributions of its constituents, became a subject of a careful study. Precise mapping of the three-dimensional structure of the nucleon is crucial for our understanding of Quantum Chromodynamics (QCD). One of the ways to go beyond the usual collinear approximation in the momentum space, is to take into consideration intrinsic transverse-motion of partons in the nucleon \textit{i.e.} assuming non-zero transverse momentum vector $\ktv$ for partons. Then the spin-structure of the nucleon in semi-inclusive hard processes is described by the so-called Transverse-Momentum-Dependent Parton Distribution Functions (TMD PDFs)~\cite{Kotzinian:1994dv,Mulders:1995dh,Boer:1997nt,Goeke:2005hb,Bacchetta:2006tn,Angeles-Martinez:2015sea}.
%While some experimental results exist for such LO TMD PDFs like the unpolarized, transversity, Sivers and Boer-Mulders functions of quarks, only a few  tentative attempts to measure TMD PDFs of gluons have been performed.

One of the most powerful tools to study quark TMD PDFs are the measurements of the nucleon spin (in)dependent azimuthal asymmetries in SIDIS~\cite{Kotzinian:1994dv,Goeke:2005hb,Bacchetta:2006tn,Anselmino:2011ch, Bastami:2018xqd} and Drell--Yan processes \cite{Arnold:2008kf,Bastami:2020asv}. Complementary information on TMD fragmentation process, necessary for the interpretation of SIDIS data, is obtained from $e^+e^-$ measurements~\cite{Metz:2016swz}.
Being an actively developing field, TMD physics triggers a lot of experimental and theoretical interest all over the world, stimulating new measurements and developments in TMD extraction techniques oriented on existing and future data from lepton-nucleon,  hadron-hadron and electron-positron facilities at CERN, DESY, JLab, FNAL, BNL, and KEK. For recent reviews on experimental and theoretical advances on TMDs see Refs.~\cite{Anselmino:2020vlp,Avakian:2019drf,Perdekamp:2015vwa,Boglione:2015zyc,Aidala:2012mv,Lansberg:2017tlc, Scarpa:2019fol, Abdulov:2019txc}.
These efforts improved significantly the phenomenological modeling of quark Sivers, transversity and Boer-Mulders TMD PDFs.

Recently a remarkable progress has been achieved in modeling of gluon TMD PDFs and phenomenological calculations for azimuthal asymmetries in gluon-sensitive channels including \textit{e.g.} open heavy-flavor and charmonia % JPL: charmonium is used as an adjective is thus invariable
production in hard processes, see Refs.~\cite{DAlesio:2019qpk,DAlesio:2020eqo,DAlesio:2017rzj,Bacchetta:2020vty} and the references therein. However experimental data relevant for the study of gluon TMD PDFs are still scarce~\cite{Adare:2013ekj,Adare:2010bd,Aidala:2018gmp,Aidala:2017pum,Adolph:2017pgv,Szabelski:2016wym}.

Whereas the experimental efforts are mostly focused on the study of the partonic content of the proton, the gluon structure of the deuteron hides interesting peculiarities.
The simplest model of the deuteron describes it as a weakly-bound state of a proton and a neutron mainly in the S-wave with a small admixture of the D-wave state \cite{Garcon:2001sz}. This approach is not fully reliable in the description of the deuteron structure at large $Q^2$ \footnote{We use $Q^2$ (or $\mu^2$) as a generic notation for the hard scale of a reaction: the invariant mass square of lepton pairs in Drell-Yan processes, $Q^2$, transverse momentum square $p_T^2$ of produced hadron or its mass square $M^2$.}. In particular, possible non-nucleonic degrees of freedom in deuteron could play an important role in the understanding of the nuclear modification of PDFs (the EMC effect~\cite{Aubert:1983xm}).
Since the gluon transversity operator requires two-unit helicity-flip, it cannot be defined for spin-1/2 nucleons~\cite{Barone:2001sp}. Hence, proton and neutron gluon transversity functions can not contribute directly to the gluon transversity of the deuteron. A non-zero deuteron transversity would then be an indication of a non-nucleonic component, or some other exotic hadronic mechanisms within the deuteron.

Most of the existing experimental results on spin-dependent gluon distributions in nucleon are obtained in the experiments at DESY (HERMES), CERN (COMPASS), and BNL (STAR and PHENIX). Study of polarized gluon content of the proton and nuclei is an important part of future projects in Europe 
%Track changes is on 65
 and the United States such as AFTER@LHC and LHCSpin at CERN, and EIC at BNL~\cite{Brodsky:2012vg,Hadjidakis:2018ifr,Aidala:2019pit,Accardi:2012qut}.
 
\begin{figure}[!h]
%  \center{\includegraphics[width=0.8\linewidth]{SPD_Cryo.eps}}
  \center{\epsfig{file=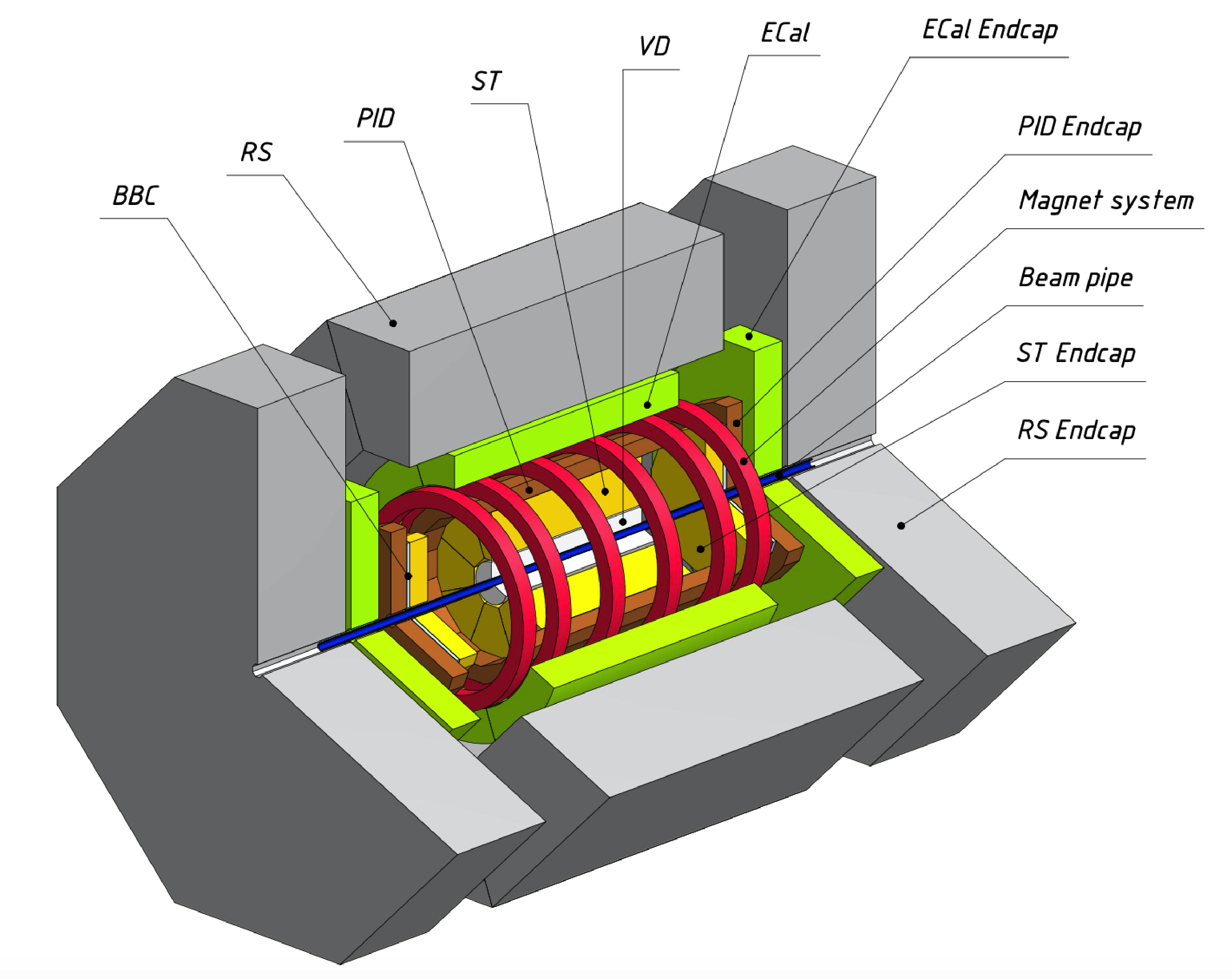,scale=0.3}}
    \caption{General layout of the SPD setup \cite{Abazov:2021hku}.}
  \label{fig:SPD}  
\end{figure}

Experiments with hadronic collisions are used to access gluons at the Born level without involvement of electromagnetic couplings, which is an important advantage over SIDIS measurements. The Spin Physics Detector (SPD) project~\cite{Savin:2015paa,Tsenov:2019pok,Guskov:2019qqt,SPDsite} at the NICA collider that is under construction at JINR (Dubna, Russia) aims to investigate the nucleon spin structure and polarization phenomena in polarized $p$-$p$ and $d$-$d$ collisions. The planned center-of-mass energy can reach up to 27~GeV while the luminosity $L$ is expected to be of order of 10$^{32}$ cm$^{-2}$ s$^{-1}$ in $p$-$p$ collisions at maximal energy~\cite{Meshkov:2019utc}. The $d$-$d$ collisions can be performed at $\sqrt{s_{NN}}\leq 13.5$ GeV  with about one order of magnitude lower luminosity. Asymmetric $p$-$d$ collisions at $\sqrt{s_{NN}}\leq 19$ GeV %with only one beam being polarized 
are also under discussion.
It is planned to achieve beam polarization of up to 70\%. 
%The SPD setup is planned as a multipurpose universal  4$\pi$ detector with advanced tracking and particle identification capabilities like a TOF system, electromagnetic calorimeter and muon (range) system.
The SPD experimental setup (see Fig. \ref{fig:SPD}) is being designed as a universal $4\pi$-acceptance detector equipped with advanced tracking and particle identification systems. The silicon vertex detector (VD) will provide good resolution for the vertex position ($\lesssim$100 $\mu$m) ensuring reliable identification of secondary vertices of $D$-meson decays. The straw-tube based tracking system (ST) to be placed in the solenoidal magnetic field (up to 1 T at the detector axis) will provide the transverse momentum resolution $\sigma_{p_T}/p_T\approx 2\%$ for particles with momentum about 1 GeV. The time-of-flight system (PID in the Fig.~\ref{fig:SPD}) with a time resolution of about 60 ps will provide $3\sigma$ $\pi/K$ and $K/p$ separation for hadrons with momenta up to about 1.2 GeV and 2.2 GeV, respectively. The usage of an aerogel-based Cherenkov detector could extend the momentum range of the PID-system. The detection of photons will be performed by the sampling electromagnetic calorimeter (ECal) with an energy resolution of $5\%/\sqrt{E/\rm{GeV}}$. To reduce multiple scattering and photon conversion effects, the detector material will be minimized throughout the inner part of the spectrometer. The muon (range) system (RS) will be set up for muon identification. It can also act as a rough hadron calorimeter. A pair of beam-beam counters (BBC) and zero-degree calorimeters will be responsible for the local control of polarimetry and luminosity.  To minimize possible systematic effects, SPD will be equipped with a triggerless data acquisition system \cite{Abazov:2021hku}.
Spin physics program at the SPD is expected to start after year 2025 and to extend for about 10 years. 

This Review has the following structure. Section \ref{probes} describes peculiarities of gluon probes such as charmonium (\ref{probes_ch}), open charm (\ref{probes_och}), and prompt photon production (\ref{probes_pp}) in the SPD kinematic domain. Section~\ref{sec:ThIntro}: subsection~\ref{sec:gluon_TMD_intro}  serves as a theoretical and phenomenological introduction into physics of gluon TMD PDFs, in~\ref{quarkonia_th} heavy quarkonium hadroproduction is discussed, in~\ref{un_tmd} peculiarities of TMD-factorization for heavy-quarkonium production are addressed, while the proton spin problem from the point of view of contemporary lattice QCD is reviewed in~\ref{lqcd}. Beam-polarization-independent measurements are discussed in section~\ref{unpol_measurements}. In subsections~\ref{un_pdf},~\ref{un_BM} and~\ref{un_non} we address the determination of the gluon PDFs at high $x$, linearly polarized gluons in unpolarized nucleon and possible non-nucleonic content in deuteron, respectively.
%Relationship between the hadron structure and charmonia production mechanisms is given in \ref{quarkonia_th}.
Section \ref{pol} is dedicated to beam-polarization-dependent measurements that are feasible  at SPD. Gluon helicity function $\Delta g(x)$ and related measurements are discussed in~\ref{hel}. The study of TMD and twist-3 effects is the main topic of Sec.~\ref{tmd_t3}. Section~\ref{trans} is dedicated to the gluon transversity in deuteron. Gluon contribution to the tensor structure of deuteron is reviewed in~\ref{TensorD}.
%Current lattice QCD results on gluon content of nucleon are shortly presented in Sec. \ref{lqcd}.
Section~\ref{sum} summarizes the discussion on the physics program with gluons at the SPD.

\section{Gluon probes at NICA SPD \label{probes}}

Within perturbative QCD, the partonic content of a nucleon is described by (TMD)PDFs. To ensure the validity of this description, one needs to restrict to hard inclusive reactions on nucleons. The hard scale ($Q^2$ or $\mu^2$) in these reactions is provided for (SI)DIS by a) the virtuality of exchanged photon, $Q^2$, or b) by the high transverse momentum of selected final particle or jet, $p_T$ , or c) by the heavy mass of the observed hadron, for example $m_{J/\psi}$. For the hadronic reactions, the selections b) and c) provide access to the partonic content of a nucleon.

\begin{figure}[!t]
%  \center{\includegraphics[width=0.8\linewidth]{gluon_paper/figs/SPD_kinematic_after_2.eps}}
  \center{\epsfig{file=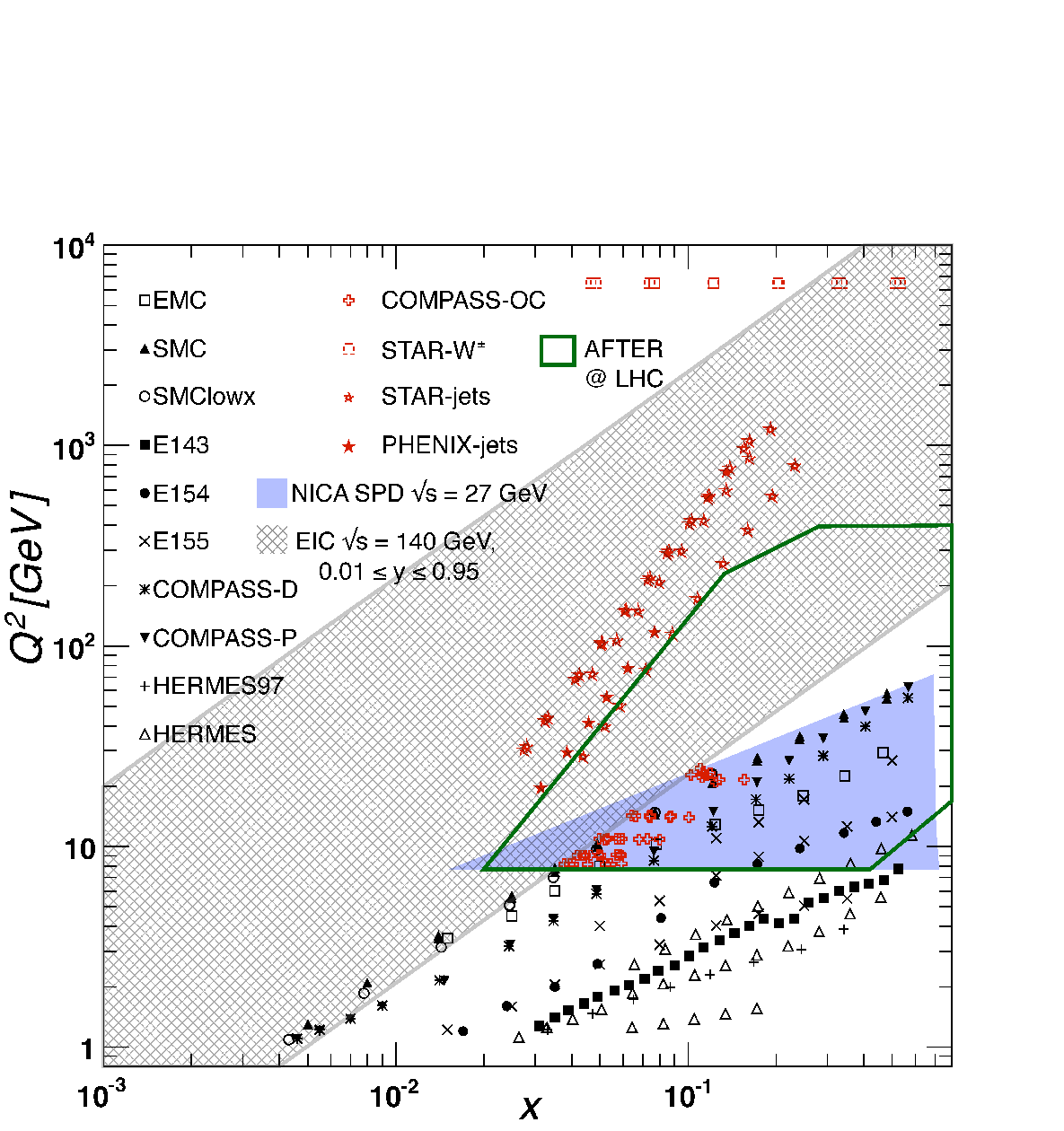,scale=0.7}}
    \caption{The kinematic coverage, in the ($x$, $Q^2$) plane, of the hadronic cross-section data used for determination of the NNPDFpol1.1 set of the polarized PDFs~\cite{Nocera:2014gqa}. The kinematic domains expected to be covered by the NICA SPD at $\sqrt{s}=27$ GeV and AFTER \cite{Hadjidakis:2018ifr} by charmonium, open-charm and prompt-photon production are shown. The EIC kinematic domain for $\sqrt{s}=140$ GeV and $0.01\leq y\leq 0.95$ is also presented according to Ref. \cite{Accardi:2012qut}. The figure is adapted from~\cite{Nocera:2014gqa}~\textcopyright~(2014) by Elsevier.}
  \label{fig:kinematic}  
\end{figure}

\begin{figure}[!h]

 %\center{ \includegraphics[width=0.5\linewidth]{cross_sections}}
 \center{ \epsfig{file=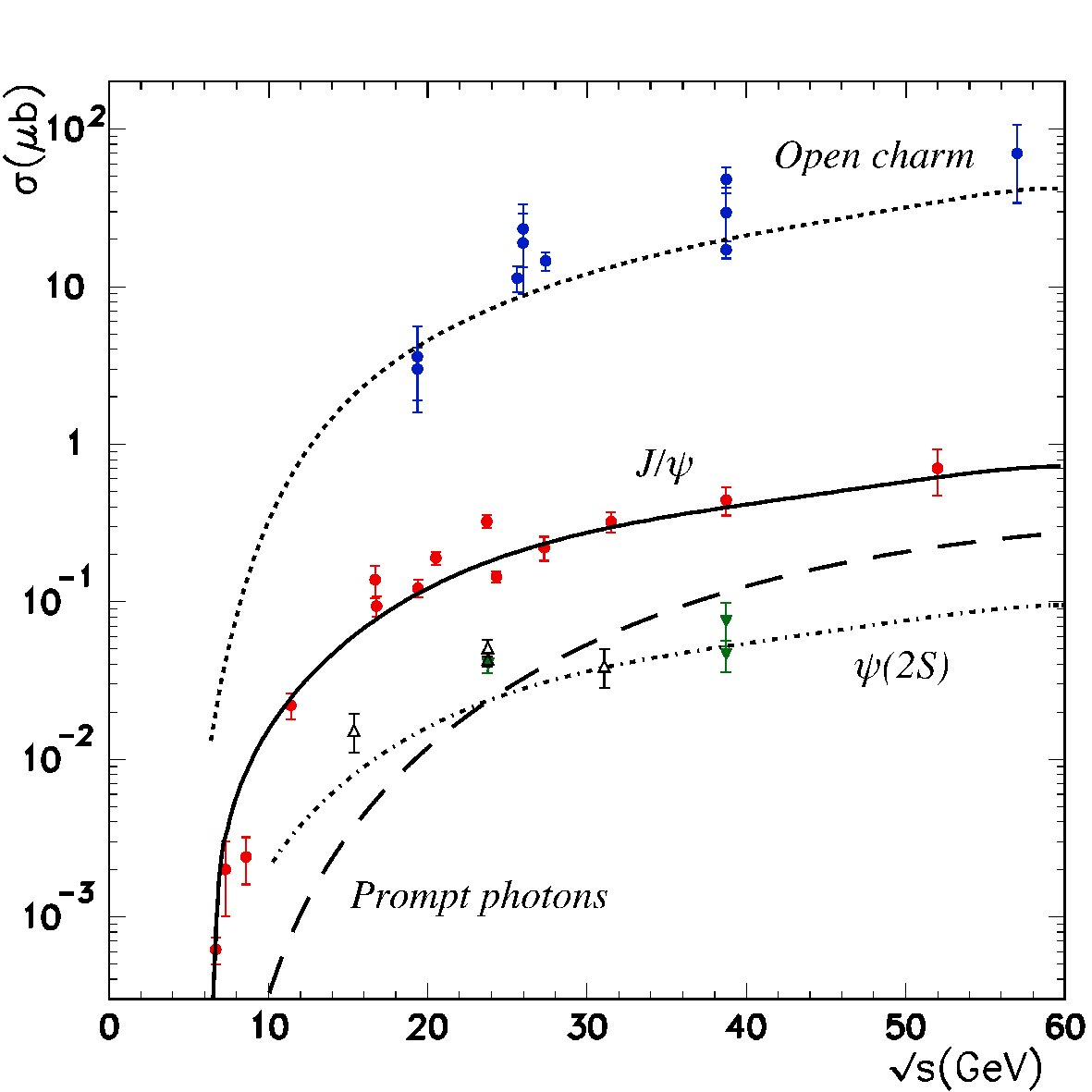,scale=0.5}}
    \caption{Cross-section for open charm, $J/\psi$ and $\psi(2S)$ production from CEM-NLO model (colour evaporation model combined with NLO pQCD matrix elements) and prompt photon production cross-section for $p_T>3$~GeV as a function of center-of-mass energy. Model-calculations are compared with available experimental data sets. The figure is adapted from Ref.~\cite{BrennerMariotto:2001sv}~\textcopyright (2002) by The European Physical Journal.}  %for details see~\cite{BrennerMariotto:2001sv}
  \label{fig:csec}  
\end{figure}

The (un)polarized gluon content of the proton and deuteron at intermediate and high values of Bjorken $x$ will be investigated at the SPD using three main probes: the inclusive production of charmonia, open charm, and prompt photons. 
The study of these processes is complementary to the usual approaches to access the partonic structure of the nucleon in hadronic collisions such as the inclusive production of hadrons at high transverse momentum and the Drell-Yan process. Unfortunately, the latter channel is unlikely to be accessible at SPD due to the small cross-section and unfavourable background conditions.
For an efficient detection of the aforementioned gluon probes, the SPD setup is planned to be equipped with muon-identification system, an electromagnetic calorimeter, a time-of-flight system, and a silicon vertex detector. Nearly a $4\pi$ coverage of the setup and a low material budget in the inner part of the setup should provide a large acceptance for the detection of the desired final states.  
%{The kinematic range in $x$ and $Q^2$ that could be covered by SPD with the gluon probes in comparison with the corresponding ranges of previous, present and future experiments is shown in Fig.}\ref {fig:kinematic}(a). 
In Fig.~\ref{fig:kinematic}, the kinematic phase-space in $x$ and $Q^2$ to be accessed by the SPD is compared to the corresponding ranges of previous, present and future experiments. The parameters of the experimental facilities planning to contribute to gluon physics with polarized beams are listed in Tab.~\ref{tab:facilities}. 
 Figure~\ref{fig:csec} illustrates the behavior of the cross sections for the inclusive production of $J/\psi$, $\psi'$, $D$-mesons and high-$p_T$ prompt photons in $p$-$p$ collisions as a function of $\sqrt{s}$.

%{If we accept the footnote 1, in the Fig.1 (a) we can leave only $Q^2$ on y-axis and increase some bit the distance axis labels from corresponding axes.}

\begin{table}[t!]
\caption{Main present and future gluon-spin-physics experiments.}
\begin{center}
\begin{tabular}{|l|c|c|c|c|c|}
\hline \bigstrut 
 Experimental  & SPD         & RHIC~\cite{RHIC} & EIC~\cite{Accardi:2012qut}& AFTER  & LHCspin \bigstrut \\  
 facility            &  @NICA ~\cite{Meshkov:2019utc}    &       &         &   @LHC~\cite{Hadjidakis:2018ifr}  &    \cite{Aidala:2019pit}    \bigstrut \\
\hline  
Scientific center & JINR     & BNL               & BNL & CERN & CERN   \bigstrut \\
\hline
Operation mode & collider & collider        & collider & fixed  & fixed \bigstrut \\
                           &             &                   &             & target & target  \bigstrut \\
\hline                           
Colliding particles& $p^{\uparrow}$-$p^{\uparrow}$ & $p^{\uparrow}$-$p^{\uparrow}$  & $e^{\uparrow}$-$p^{\uparrow},d^{\uparrow},^3$He$^{\uparrow}$ & $p$-$p^{\uparrow}$,$d^{\uparrow}$  & $p$-$p^{\uparrow}$ \bigstrut  \\                       
\& polarization      & $d^{\uparrow}$-$d^{\uparrow}$&                                 &                &                 &        \bigstrut \\
     & $p^{\uparrow}$-$d$, $p$-$d^{\uparrow}$&                                 &                &                      &   \bigstrut \\
\hline
Center-of-mass            &  $\leq$27 ($p$-$p$) &  63, 200,      & 20-140 ($ep$)&  115 & 115 \bigstrut \\
energy $\sqrt{s_{NN}}$, GeV&  $\leq$13.5 ($d$-$d$)     &   500      &       &   &  \bigstrut \\
                           &  $\leq$19 ($p$-$d$)     &         &       &   &  \bigstrut \\
\hline
Max. luminosity,                  & $\sim$1 ($p$-$p$)  &  2  & 1000 &up to & 4.7  \bigstrut \\   
10$^{32}$ cm$^{-2}$ s$^{-1}$  & $\sim$0.1 ($d$-$d$) &        &  &   $\sim$10 ($p$-$p$) &  \bigstrut \\
\hline
Physics run            & $>$2025  & running & $>$2030  &  $>$2025 & $>$2025 \bigstrut \\
\hline
\end{tabular}
\end{center}
\label{tab:facilities}
\end{table}%
\begin{figure}[!t]
  \begin{minipage}[ht]{0.325\linewidth}
    \center{\epsfig{file=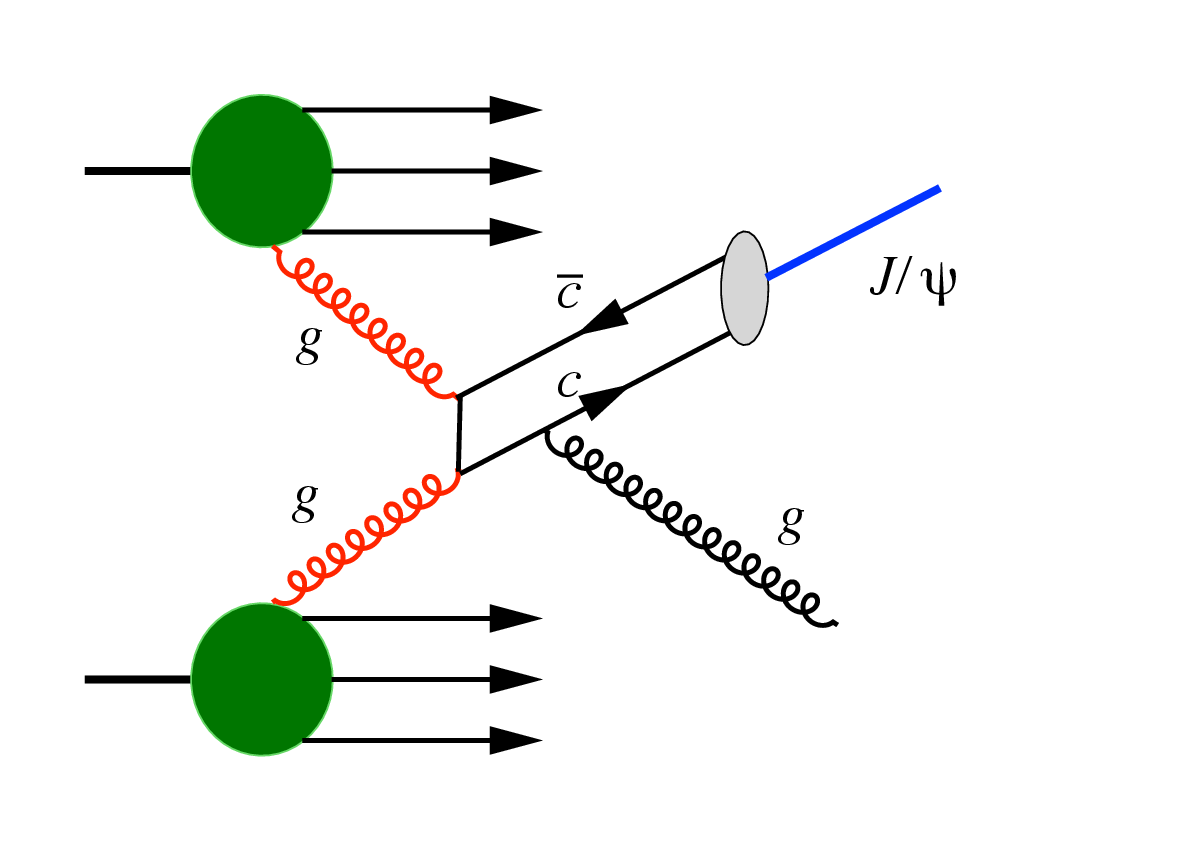,scale=0.3} \\ (a)}
  \end{minipage}
  \begin{minipage}[ht]{0.325\linewidth}
    \center{\epsfig{file=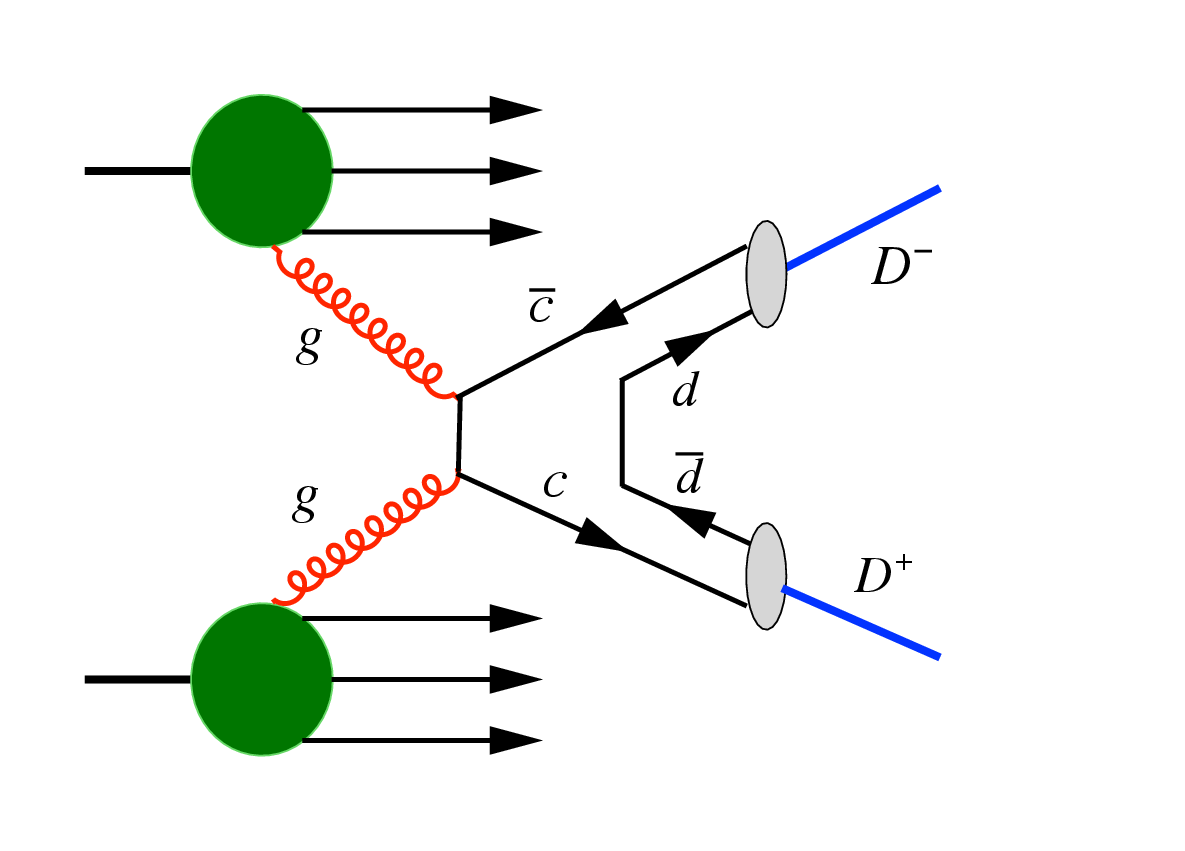,scale=0.3}\\ (b)}
  \end{minipage}
  \begin{minipage}[ht]{0.335\linewidth}
    \center{\epsfig{file=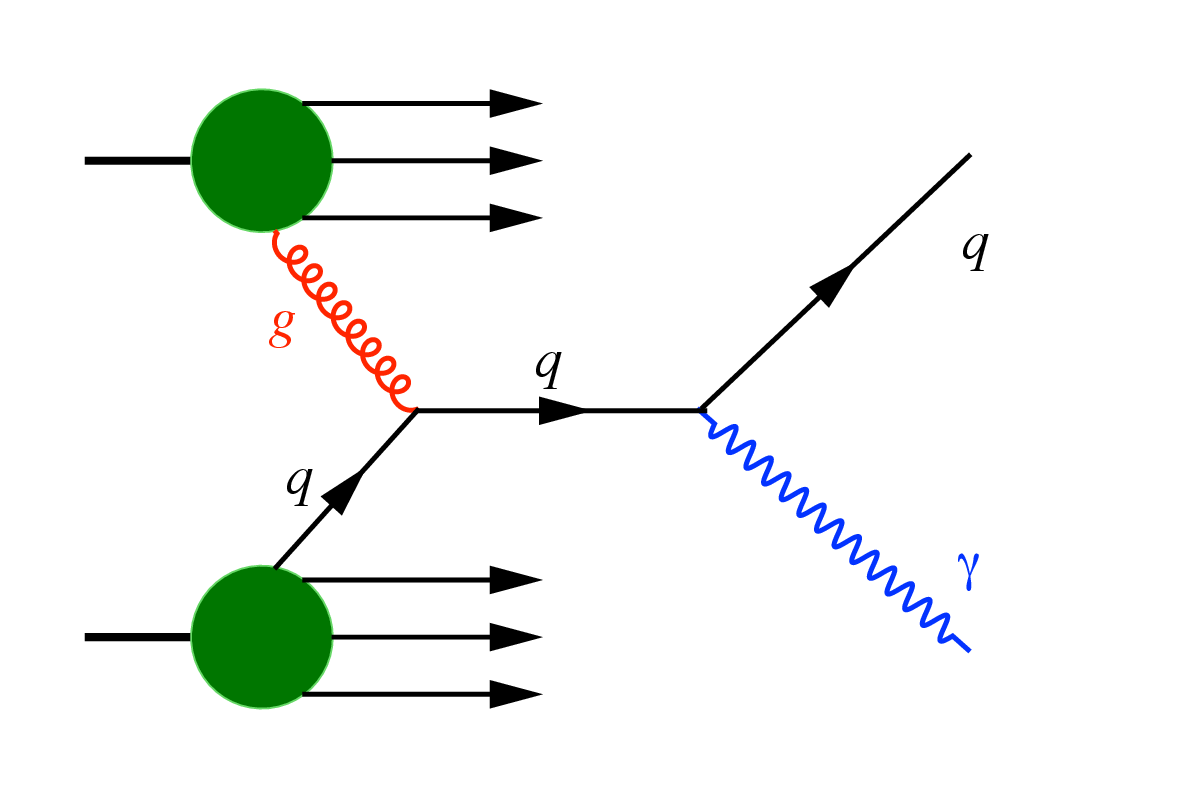,scale=0.3} \\ (c)}
  \end{minipage}
  \caption{Diagrams illustrating three probes to access the gluon content of proton and deuteron in polarized collisions at NICA SPD: production of (a) charmonium (color-singlet model for $J/\psi$, $\psi(2S)$), (b) open charm, (c) prompt photons. }
  \label{fig:diagrams-gg}  
\end{figure}
\subsection{Charmonium  production\label{probes_ch}}

 From the experimental point of view, for the SPD energy range, the hadronic production of charmonia seems to be particularly suited to access gluon content in hadrons due to the clean signal from $J/\psi\to\mu^+\mu^-$ decay ($BF=0.06$). The production of prompt $J/\psi$-mesons looks most attractive, since the corresponding cross section is experimentally known to be significant. A large data set of $J/\psi\to\mu^+ \mu^-$ events is accumulated in beam-dump experiments (where only a muon pair is detected) with proton and pion beams at $\sqrt{s}$ close to 20 GeV. However $J/\psi$-meson is not the cleanest probe of the proton structure, since a significant fraction  of $J/\psi$-mesons observed in hadronic collisions is produced indirectly through decays of $\chi_{cJ}$ and $\psi(2S)$ (the so-called feed-down contribution). The feed-down fraction is $p_T$ and collision-system-dependent, it varies between $20\%$ ~\cite{Abt:2008ed} in $pA$ and $\sim 40\%$ in $pp$-collisions~\cite{Adare:2011vq} (see the detailed discussion in Ref.~\cite{Lansberg:2019adr}). Accounting for this contribution introduces additional uncertainties into the theoretical calculations of inclusive $J/\psi$ cross-sections. Hence, to provide additional constraints to production models, it is important to study production of $\chi_{cJ}$ and $\psi(2S)$ separately, through their decays $\chi_{cJ}\to \gamma J/\psi$ ($BF=0.014$, $0.343$ and $0.19$ for $J=0,1$ and 2) and $\psi(2S)\to J/\psi\ \pi^+ \pi^-$ ($BF=0.347$). The latter state is of special interest, because it is essentially free from feed-down contamination from higher charmonium states, due to the proximity of $D^0\overline{D}^0$-threshold. However, the separation of the $\chi_{c0,1,2}$ signals is a challenging experimental task due to the small mass difference between the states, which requires good energy resolution of the electromagnetic calorimeters for soft photons. The measurement of $\eta_c$-meson production cross-section at SPD NICA, using $\eta_c\to p\bar{p}$ decay mode ($BF=1.45\times 10^{-3}$) is also under discussion, but might be challenging.

 Besides, the task of accessing gluon distributions using heavy quarkonia is rather challenging also from the theoretical point of view. The heavy quark-antiquark pair couples directly to gluons from initial-state hadrons (Fig.~\ref{fig:diagrams-gg}(a)) and its production can be perturbatively calculated, because the hard scale of the process is limited from below by the heavy quark mass, providing the direct access to polarized and unpolarized gluon distributions. However, the process of the transition of the heavy quark-antiquark pair into a physical bound-state is presently not well understood~\cite{Lansberg:2019adr,Andronic:2015wma,Brambilla:2010cs,Lansberg:2006dh} and can become a source of significant theoretical uncertainties. We review modern status of the theory of quarkonium production in more detail in Sec.~\ref{quarkonia_th} to explain the latter point. 

Hence, quarkonium production can be used to study the structure of hadrons only with a great caution and only if the results consistent with other probes will eventually emerge. The study of the hadronic structure and the heavy quarkonium production mechanism should be treated as complementary analyses. The best strategy for quarkonium measurements at SPD is to study the yields and the polarization of different quarkonium states in a wide kinematic range, at various energies, both in polarized and non-polarized hadronic collisions. This would serve as an input to develop and to constrain the theory and to validate or exclude various models. When the theory of production of heavy quarkonia is firmly established, it will become an invaluable tool to study the details of hadronic structure.

\subsection{Open charm production\label{probes_och}}
\begin{figure}[!t]
  \begin{minipage}[ht]{0.48\linewidth}
      \center{\epsfig{file=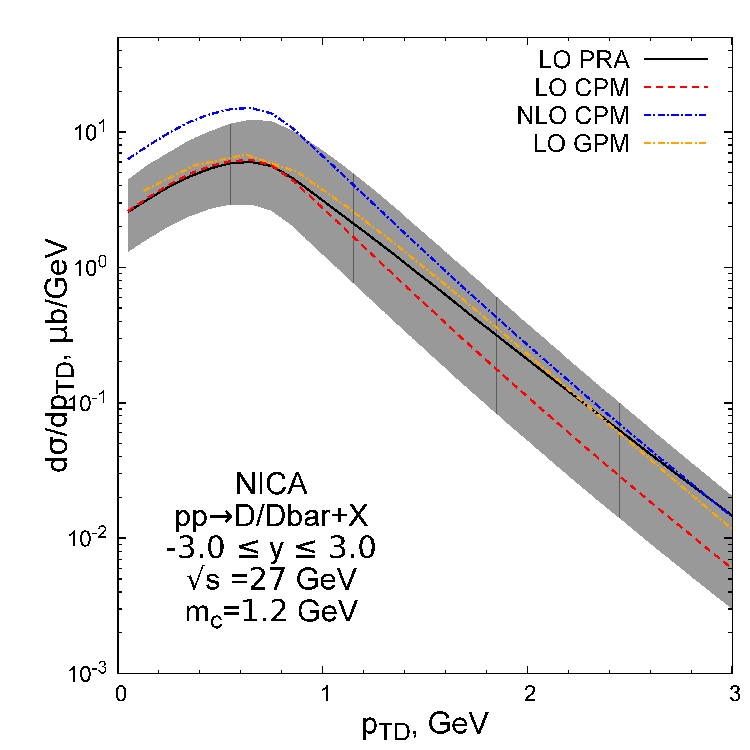,scale=0.6} \\ (a)}
  \end{minipage}
  \begin{minipage}[ht]{0.48\linewidth}
      \center{\epsfig{file=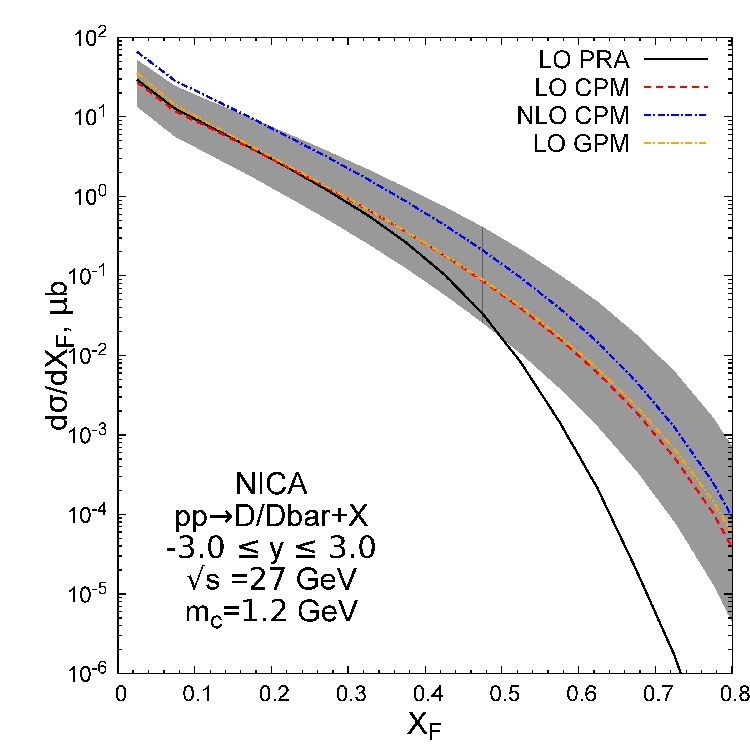,scale=0.6} \\ (b)}
  \end{minipage}
    \caption{Theoretical predictions for $p_T(D)$- (a) and $x_F$-differential (b) inclusiive $D$-meson production cross-section in the LO of CPM, PRA, GPM and NLO of CPM}
  \label{fig:NICA-D-pT-xF}  
\end{figure}

It is well known that the heavy flavor production offers direct probes of the gluon distributions in hadrons, see e.g.  \cite{Leveille:1978px} and references therein. The basic mechanism responsible for charm pair production in $pp$ collisions is the gluon fusion (GF, see Fig.~\ref{fig:diagrams-gg}(b)). In the framework of pQCD, the GF contributes to the hadron cross-section as ${\cal L}_{gg}\otimes {\hat \sigma}_{c\bar{c}}$, where the gluon luminosity ${\cal L}_{gg}$ is a convolution of the gluon densities in different protons, ${\cal L}_{gg}=g\otimes g$. At leading order in pQCD, ${\cal O}(\alpha_s^2)$, the partonic cross-section ${\hat \sigma}_{c\bar{c}}$ describes the process $gg\rightarrow c\bar{c}$. For different theoretical estimates of $p_T$ and Feynman variable $x_F$-differential $D$-meson hadroproduction cross-section an NICA energies see the Fig.~\ref{fig:NICA-D-pT-xF}. 

The GF contribution to the charmonia production in $pp$ collisions has the form ${\cal L}_{gg}\otimes {\hat \sigma}_{(c\bar{c})+X}\otimes W_{c\bar{c}}$. (For more details, see recent review \cite{Lansberg:2019adr}). At the Born level, the partonic cross-section ${\hat \sigma}_{(c\bar{c})+X}$ is of the order of $\alpha_s^3$ because its basic subprocess is $gg\rightarrow (c\bar{c})+g$ in the color-singlet model of charmonium production. Moreover, the quantity $W_{c\bar{c}}$, describing the probability for the charm pair to form a charmonium, imposes strong restrictions on the phase space of the final state.\footnote{To form a charmonium, the momenta of the produced quark and antiquark should be sufficiently close to each other.} For these two reasons, the $\alpha_s$-suppression and phase space limitation, the cross-sections for charmonia production are almost two orders of magnitude smaller than the corresponding ones for open charm, see Figs.~\ref{fig:kinematic} (b). 

%Production of $c\bar{c}$ pairs which then form two charmed hadrons, mainly $D$-mesons is related to the same hard processes as the production of charmonia. Only the difference is that a valence or sea quark from the nucleon can participate in formation of a final $D$-meson.
%Nevertheless, the cross-section of the open charm production is two orders of magnitude larger than the corresponding one for charmonia. 

To analyze the kinematics of a $DD$ pair, each of $D$-mesons has to be reconstructed. The decay modes $D^+\to \pi^+K^-\pi^+$ (BF=0.094) and $D^0\to K^-\pi^+$ (BF=0.04) can be used for that. In order to suppress the combinatorial background at SPD, the $D$-meson decay vertices that are about 100 $\mu$m away from the interaction point (the $c\tau$ values are 312 and 123 $\mu$m for the charged and neutral $D$-mesons, respectively) have to be selected. Identification of a charged kaon in the final state can be done using the time-of-flight system. The production and the decay of $D^*$-mesons can be used as an additional tag for open-charm events. Singe-reconstructed $D$-mesons also carry reduced but still essential information about gluon distribution that is especially important in the low-energy region with a lack of statistics.

\subsection{Prompt photon production} \label{probes_pp}
%{Prompt photons are photons that are produced by hard scattering of partons.
%Production of such photons in hadron collisions is the most direct way to access the gluon structure of hadrons.}
Photons emerging from the hard parton scattering subprocess, the so-called prompt photons, serve as a sensitive tool to access the gluon structure of hadrons in hadron-hadron collisions.
Inclusive direct photon production proceeds without fragmentation, i.e. the photon carries the information directly from the hard scattering process. Hence this process measures a combination of initial $\ktv$-effects and hard scattering twist–3 processes. There are two main hard processes for the production of direct photons: gluon Compton scattering, $gq(\bar{q})\rightarrow \gamma q(\bar{q})$ (Fig.~\ref{fig:diagrams-gg}(c)), which dominates, and  quark-antiquark annihilation, $q \bar{q} \rightarrow \gamma g$. Contribution of the latter process to the total cross-section is small. 

\begin{figure}[h!t]
  \begin{minipage}[ht]{0.50\linewidth}
        \center{\epsfig{file=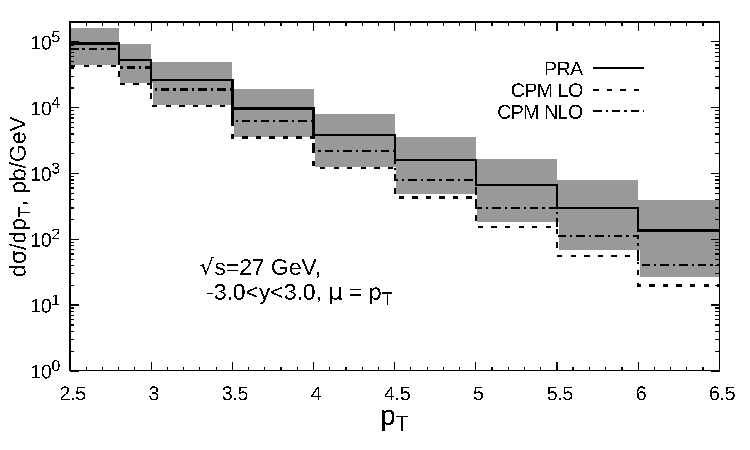,scale=0.6} \\ (a)}
  \end{minipage}
  \begin{minipage}[ht]{0.46\linewidth}
          \center{\epsfig{file=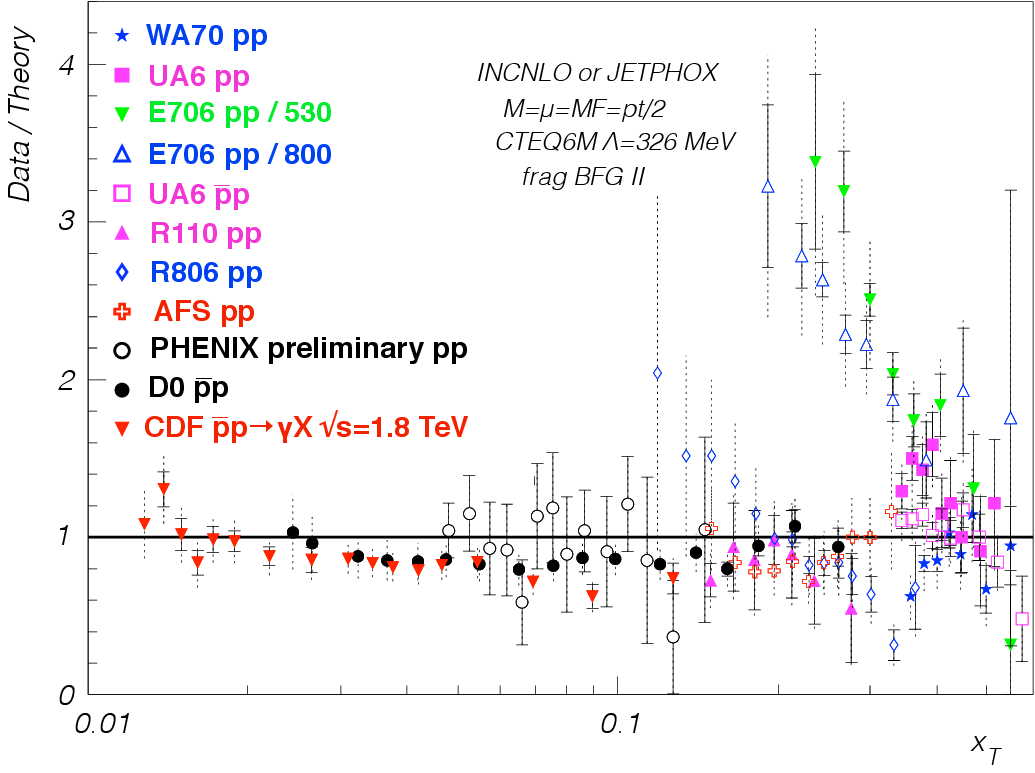,scale=0.45} \\ (b)}
  \end{minipage}
    \caption{(a) Prediction for prompt photon transverse momentum spectrum at $\sqrt{s}=27$ GeV obtained in LO (dashed line) and NLO (dash-dotted line) approximations of CPM and LO of PRA solid line). Uncertainty bands for PRA predictions are due to factorization/renormalization scale variation only. (b) Data-to-theory ratio for the fixed-target and collider experiments. Reprinted figure with permission from \cite{Aurenche:2006vj}  \textcopyright~(2006) by the American Physical Society.}
  \label{fig:NICA-photon-pT}  
\end{figure}

Theoretical predictions for transverse momentum spectrum for inclusive prompt photon production at the energy of $\sqrt{s}=27$ GeV are shown in Fig. \ref{fig:NICA-photon-pT}(a). Calculations are performed in LO and NLO approximations of  Collinear Parton Model (CPM), as well as in the Parton Reggeization Approach (PRA)~\cite{Karpishkov:2017kph}, which is a QCD and QED gauge-invariant version of $k_T$-factorization. They include direct and fragmentation contributions, the latter one is about 15-30 \%. The K-factor between LO and NLO calculations in the CPM is about $\sim$1.8 and slightly depends on $p_{T\gamma}$~\cite{Wong:1998pq}. LO prediction of PRA coincides with the result of NLO CPM calculation at moderate transverse momenta ($p_{T}<4$ GeV) while at higher $p_T$ PRA predicts somewhat harder $p_T$-spectrum. Figure~\ref{fig:NICA-photon-pT}(b)~\cite{Aurenche:2006vj} presents the comparison of the $p_T$ spectra ($x_T=2p_T/\sqrt{s}$) measured in a wide kinematic range of $\sqrt{s}$ in different fixed-target and collider experiments and the theoretical NLO calculations performed within the JETPHOX package~\cite{Binoth:1999qq}.  While high-energy collider results exhibit rather good agreement with the expectations, the situation at high-$x_T$ is complicated.
The results from E706 ($\sqrt{s}=31.6$ and $38.8$ GeV)~\cite{Apanasevich:1997hm} and R806 ($\sqrt{s}=63$ GeV) ~\cite{Anassontzis:1982gm} experiments diverge significantly from the theory and follow an evident trend, which could be an indication of possible systematic effects that have not been fully understood yet.

\begin{table}[h!]
\caption{Expected cross-section and event counts (without the acceptance correction) for each of the gluon probes per one year of SPD running ($10^7$ s).  For $\eta_c$ cross-section, the central LO PRA NRQCD (see Sec.~\ref{quarkonia_th}) estimate for $\sqrt{s}=24$ GeV is given, for other estimates see Fig.~\ref{fig:NICA-etac-dy}.}
\begin{center}
\begin{tabular}{|l|c|c|c|c|}
\hline \bigstrut 
                                &  $\sigma$\textsubscript{27\,GeV},  & $\sigma$\textsubscript{13.5\,GeV}, & $N$\textsubscript{27\,GeV}, & $N$\textsubscript{13.5\,GeV}  \bigstrut \\                    
   Probe                   &      nb ($\times$BF)                         &  nb   ($\times$BF)                             &   10$^6$             &     10$^6$       \bigstrut \\
\hline
\hline
Prompt-$\gamma$ ($p_T>3$ GeV/c) & 35          & 2 & 35 &  0.2 \bigstrut \\
\hline
\hline
$J/\psi$                    & 200          & 60 &  &   \bigstrut \\
~~~$\to\mu^+\mu^-$ & 12         & 3.6 & 12 & 0.36  \bigstrut \\
\hline
 $\psi(2S)$    &  25      &  5     &  &   \bigstrut \\
~~~$\to J/\psi \pi^+\pi^- \to \mu^+\mu^- \pi^+\pi^-$            &  0.5      &  0.1     & 0.5 & 0.01  \bigstrut \\
~~~$\to\mu^+\mu^-$                  &  0.2      &  0.04     & 0.2 & 0.004  \bigstrut \\
\hline
$\chi_{c1}$ + $\chi_{c2}$ & 200 & & & \bigstrut \\
% assuming 12.5% + 12.5% feed-down
~~~$\to \gamma J/\psi \to \gamma \mu^+\mu^-$ & 3.0 & & 3.0 & \bigstrut \\
\hline
% From the talk by Maxim, c.-s. at 24 GeV
$\eta_c$ & 400 & & & \bigstrut \\
~~~$\to p\bar p$ & 0.6 & & 0.6 & \bigstrut \\
\hline
\hline
% The cross-sections for sqrts = 27 GeV are taken from AguilarBenitez:1987rc.
Open charm: $D\bar{D}$ pairs       			     & $14000$& 1300&   & \bigstrut \\
Single $D$-mesons       			     & & &  & \bigstrut \\
~~~$D^{+}\to K^-2\pi^+$ ($D^{-}\to K^+2\pi^-$)           &   520            &   48      & 520 & 4.8  \bigstrut \\
~~~$D^{0} \to K^- \pi^+$ ($\bar{D}^{0} \to K^+ \pi^-$)           &   360             &    33     & 360 & 3.3 \bigstrut \\

\hline
\end{tabular}
\end{center}
\label{tab:crossections}
\end{table}%

In experiments prompt photons are detected alongside with a much larger number of photons from decays of secondary $\pi^0$ and $\eta$ mesons (minimum-bias photons). The main challenge is to subtract these decay contributions and filter the photons directly emitted from hard collisions. This kind of background is especially important at small transverse momenta of produced photons ($p_T$) and gives the lower limit of the accessible $p_T$ range. Therefore the prompt-photon contribution with $p_T\leq 2-3$ GeV is usually unreachable in the experiment~\cite{Vogelsang:1997cq}. 

%A pair of photons can be produced in hadronic interactions in the $q\bar{q}$ annihilation and gluon-gluon fusion hard processes (at the tree level and via quark loop, respectively).
 A pair of prompt photons can be produced in hadronic interactions in $q\bar{q}$ annihilation, quark-gluon scattering, and gluon-gluon fusion hard processes (at the leading, next-to-leading, and next-to-next-leading orders, respectively).
The double prompt photon production in nucleon interactions at low energies is not yet well studied experimentally. The production cross-section for proton-carbon interaction at $\sqrt{s}=19.4$ GeV/$c$ has been measured by the CERN NA3 experiment~\cite{Badier:1985vm}. Based on this result one can expect the cross-section of the double photon production with $p_T>2$~GeV/$c$ for each photon to be at the level of about 0.5~nb.

%\subsection{Neutral pions}
%\subsection{Central production}

Estimations of the expected event rates are evaluated for $p$-$p$ collisions at $\sqrt{s}=27$ and $13.5$~GeV for the projected integrated luminosity 1.0  and 0.1~fb$^{-1}$, respectively that corresponds effectively to one year of data taking (10$^7$ s).  The results are listed in Tab.~\ref{tab:crossections}.

\section{Theoretical motivation}\label{sec:ThIntro}
\subsection{Gluon TMDs\label{sec:gluon_TMD_intro}}

%\MGE{Add here the piece of Pavia guys, just a general text on current status of gluon TMDs. 
%Classification, add the typical table, discussion of their properties, evolution, small-x, probes, current knowledge (if any), models (e.g. spectator model)...}

The full list of leading-twist polarized gluon TMDs was first introduced in Ref.~\cite{Mulders:2000sh}. Tab.~\ref{tab:gluon_TMDs} contains the eight leading-twist gluon TMDs that are defined for a spin-1/2 hadron, using the naming scheme proposed in Ref.~\cite{Meissner:2007rx}, analogous to that of quark TMDs (see also~\cite{Lorce:2013pza}). In Ref.~\cite{Boer:2016xqr}, the study of gluon TMDs was extended to spin-1 hadrons, leading to the definition of 11 new functions.

The TMDs depend on the light-cone momentum fraction $x$ and the parton transverse momentum $\ktv$. In Tab.~\ref{tab:gluon_TMDs} they are listed in terms of both the polarization of the gluon itself and of its parent spin-1/2 hadron.
The two gluon TMDs on the diagonal of the table have the simplest physical interpretation: $f_1^g(x,\ktsq)$ is the distribution of unpolarized gluons inside an unpolarized hadron, and $g_1^g(x,\ktsq)$ is the distribution of circularly polarized gluons inside a longitudinally polarized hadron. Upon integration over ${\bf k}_T$ all TMD PDFs for spin-1/2 hadron vanish, except $f_1^g(x,\ktsq)$ and $g_1^g(x,\ktsq)$, which correspond to the well-known collinear unpolarized $f^{g}(x)$ and helicity $g_1^g(x)$ gluon PDFs, respectively. The collinear (${\bf k}_T$-integrated) gluon transversity PDF, $h_1^g(x)$, which is equal to zero for spin-1/2 hadrons, may be non-zero for spin-1 case (see Sec.~\ref{trans} of this review). The collinear unpolarized gluon PDF, $f^g(x)$, is at present the only gluon function that is known to a good extent, while we have a fair knowledge of the helicity PDF $g_1^g(x)$, but still with large uncertainties~\cite{Ethier:2020way} (see Sec.~\ref{hel} for further discussion).

As for quarks TMDs, gluon TMDs receive contributions from the resummation of logarithmically-enhanced terms in perturbative calculations. They could be called the "perturbative part" of the TMDs. Much is known about them~\cite{Bozzi:2003jy,Catani:2010pd,Echevarria:2015uaa}, but very little is known about the nonperturbative components.

The distribution of linearly polarized gluons in an unpolarized nucleon, $h_1^{\perp g}(x,\ktsq)$, \textit{i.e.} the gluonic counterpart of the Boer-Mulders function, is particularly interesting because it gives rise to spin effects even in collisions of unpolarized hadrons~\cite{Boer:2010zf,Sun:2011iw,Boer:2011kf,Pisano:2013cya,Dunnen:2014eta,Lansberg:2017tlc}. Analogous effects are generated at high transverse momentum by perturbative QCD: part of these contributions can be resummed and represent the perturbative part $h_1^{\perp g}(x,\ktsq)$~\cite{Catani:2010pd} (for further discussion see Sec.~\ref{un_BM}). 

The Sivers function, $f_{1T}^{\perp g}(x,\ktsq)$, which encodes the distribution of unpolarized gluons in a transversely polarized nucleon, has a very important role in the description of transverse-spin asymmetries (see Sec.~\ref{tmd_t3}).

%%%%%%%%%%%%%%  Tab. I %%%%%%%%%%%%%%%%%
{
\renewcommand{\arraystretch}{1.7}

 \begin{table}
\centering
 \hspace{1cm} gluon pol. \\ \vspace{0.1cm}
 \rotatebox{90}{\hspace{-1cm} nucleon pol.} \hspace{0.1cm}
 \begin{tabular}[c]{|m{0.7cm}|c|c|c|}
 \hline
 & $U$ & circular & linear \\
 \hline
 $U$ & $f_{1}^{g}$ & & \textcolor{blue}{$h_{1}^{\perp g}$} \\
 \hline	
 $L$ & & $g_{1}^{g}$ & \textcolor{red}{$h_{1L}^{\perp g}$} \\
 \hline	
 $T$ & \textcolor{red}{$f_{1T}^{\perp g}$} & \textcolor{blue}{$g_{1T}^{g}$} & \textcolor{red}{$h_{1}^{g}$}, \textcolor{red}{$h_{1T}^{\perp g}$} \\
 \hline
  \end{tabular}
 \caption{Nucleon gluon TMD PDFs at twist-2.
 $U$, $L$, $T$ describe unpolarized, longitudinally polarized and transversely polarized nucleons.
 $U$, `circular', `linear' stand for unpolarized, circularly polarized and linearly polarized gluons.
 Functions \textcolor{blue}{$h_{1}^{\perp g}$} and \textcolor{blue}{$g_{1T}^{g}$} are $T$-even.
 Functions $f_{1}^{g}$ and $g_{1}^{g}$ are $T$-even and survive integration over the transverse momentum.
 Functions \textcolor{red}{$f_{1T}^{\perp g}$}, \textcolor{red}{$h_{1}^{g}$}, \textcolor{red}{$h_{1T}^{\perp g}$} and  \textcolor{red}{$h_{1L}^{\perp g}$} are $T$-odd. For brevity functional dependence on $x$ and $\ktsq$ as well as hard-scale dependence is omitted.}
 \label{tab:gluon_TMDs}
 \end{table}

}
 %%%%%%%%%%%%%%%%%%%%%%%%%%%%%%%%%%

The precise definition of gluon TMDs should also take into account the fact that they can contain (at least) two different gauge-link configurations. This leads to the distinction between $f$-type and $d$-type gluon TMDs, also known as Weisz\"acker-Williams (WW) and dipole TMDs~\cite{Kharzeev:2003wz,Dominguez:2010xd,Dominguez:2011wm}.
In practice, each gluon TMD in Tab.~\ref{tab:gluon_TMDs} actually represents two distinct TMDs.

The WW TMDs contain either $[+,+]$ or $[-,-]$ Wilson line gauge links\footnote{Here $+$ ($-$) indicates the direction of the future- (past-) pointing Wilson lines corresponding to final (initial) state interactions.}, while the dipole TMDs contain either $[+,-]$ or $[-,+]$ gauge links. The WW type gluon TMDs occur in processes where the gluon interacts with a color-singlet initial particle (e.g., a photon in a DIS process) producing two colored final states (e.g., two jets). In processes where a gluon interacts with another gluon (color-octet state) and produces a color-singlet state (e.g., Higgs production), the relevant gluon TMDs have a $[-,-]$ gauge link structure. TMD factorization is expected to work in all these processes, and these relations are expected to follow from time-reversal invariance
\begin{align}
  f_1^{g\,[+,+]}(x,\ktsq) &= f_1^{g\,[-,-]}(x,\ktsq) \; && \text{(T-even),} \\
 f_{1T}^{g\,\perp[+,+]}(x,\ktsq) &= -f_{1T}^{g\,\perp[-,-]}(x,\ktsq)  && \text{(T-odd)}.
\end{align}

The dipole gluon TMDs, instead, occur when a gluon interacts with a colored initial particle and produces a colored final particle as, e.g., in photon-jet production in $pp$ collisions. In this case, TMD factorization has not been proven to work and may be affected by color-entanglement problems~\cite{Rogers:2013zha}. More complicated gauge-link structure are involved in processes where multiple color states are present both in the initial and final state~\cite{Bomhof:2006dp}. In these cases, TMD factorization can be even more questionable.

Experimental information on gluon TMDs is very scarce.
First attempts to perform phenomenological studies of the unpolarized gluon TMD have been presented in Refs.~\cite{Lansberg:2017dzg,Gutierrez-Reyes:2019rug,Scarpa:2019fol}. Experimental and phenomenological results related to the gluon Sivers function can be found in Refs.~\cite{Adolph:2017pgv,Szabelski:2016wym, DAlesio:2017rzj,DAlesio:2018rnv,DAlesio:2019qpk}.

%Several processes have been proposed to access gluon TMDs. Of relevance for the NICA project, we mention in particular $pp \to \eta_{c,b} X$.

At high transverse momentum and at low-$x$, the unpolarized and linearly polarized gluon distributions $f^g_1(x,\ktsq)$ and $h^{\perp g}_1(x,\ktsq)$ are connected~\cite{Dominguez:2011wm} to the Unintegrated Gluon Distribution (UGD), defined in the BFKL approach~\cite{Fadin:1975cb,Kuraev:1976ge,Kuraev:1977fs,Balitsky:1978ic} (see Refs.~\cite{Hentschinski:2012kr,Besse:2013muy,Bolognino:2018rhb,Bolognino:2018mlw,Bolognino:2019bko,Bolognino:2019pba,Celiberto:2019slj,Brzeminski:2016lwh,Bautista:2016xnp,Garcia:2019tne,Celiberto:2018muu} for recent applications).
%The UGD has been probed via distinct channels, as the DIS~\cite{Hentschinski:2012kr,Hentschinski:2013id} and the exclusive electroproduction of $\rho$ and $\phi$ mesons~\cite{Anikin:2009bf,Anikin:2011sa,Besse:2013muy,Bolognino:2018rhb,Bolognino:2018mlw,Bolognino:2019bko,Bolognino:2019pba,Celiberto:2019slj} at HERA, or the forward inclusive Drell--Yan production~\cite{Motyka:2014lya,Brzeminski:2016lwh,Motyka:2016lta,Celiberto:2018muu} at LHCb.

Since the information on gluon TMDs is at present very limited, it is important to study some qualitative features using relatively simple models. Pioneering work in this direction was done in the so-called \textit{spectator-model} approach~\cite{Lu:2016vqu,Mulders:2000sh,Pereira-Resina-Rodrigues:2001eda}.
Originally developed for studies in the quark-TMD sector~\cite{Bacchetta:2008af,Bacchetta:2010si,Gamberg:2005ip,Gamberg:2007wm,Jakob:1997wg,Meissner:2007rx}, this family of models relies on the assumption that the struck nucleon emits a gluon, together with remnants that are treated as a single, on-shell particle.
With this model, it is possible to generate all TMD densities at twist-2~(Table~\ref{tab:gluon_TMDs}).
A recent calculation for $T$-even distributions has been presented in Ref.~\cite{Bacchetta:2020vty}.
In that work, at variance with previous studies, the spectator mass is allowed to take a continuous range of values weighted by a spectral function, which provides the necessary flexibility to reproduce both the small- and the moderate-$x$ behaviour of gluon collinear PDFs.

\begin{figure}[htbp!]
  \begin{minipage}[ht]{0.50\linewidth}
        \center{\epsfig{file=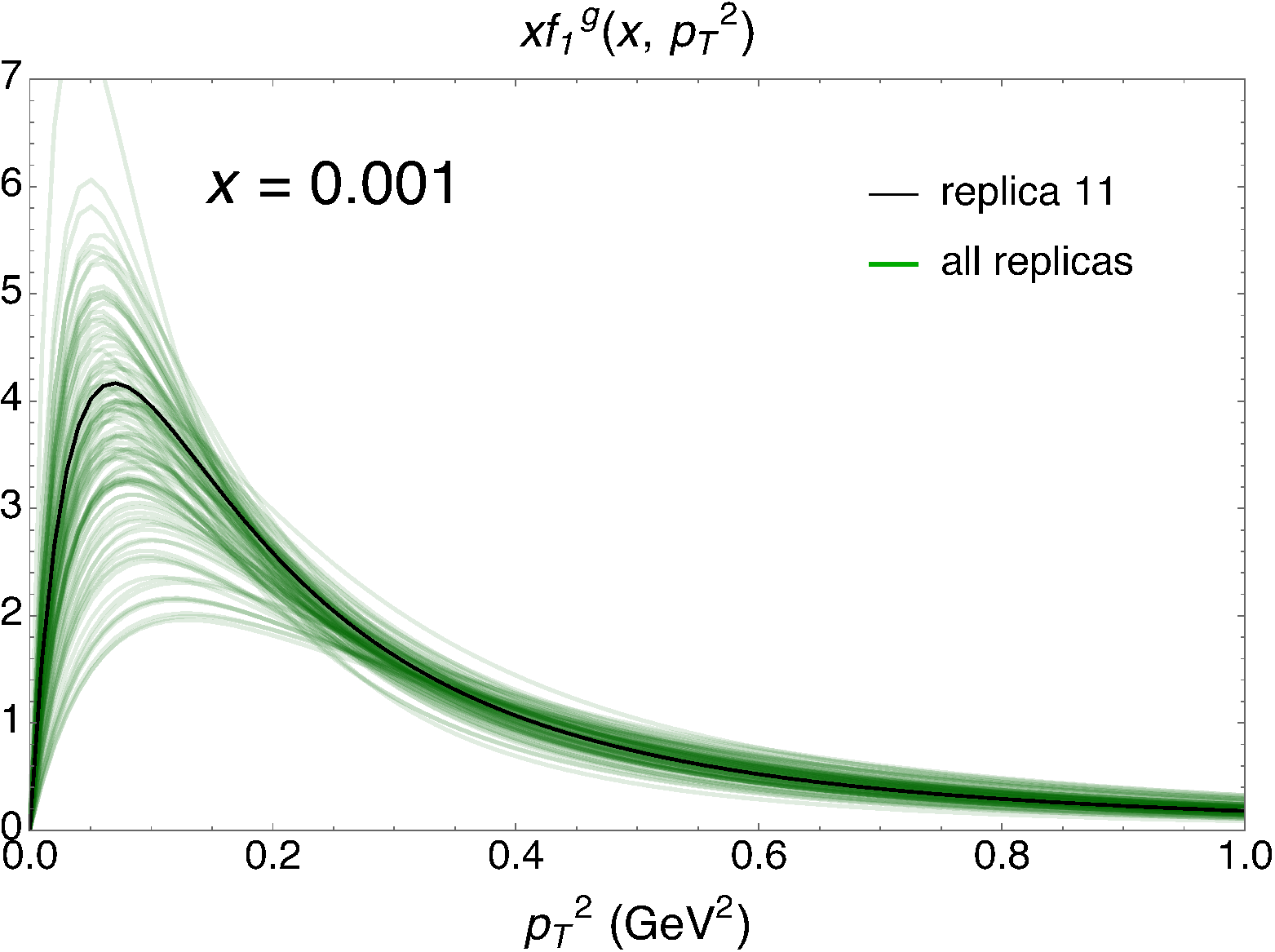,scale=0.29} \\ (a)}
  \end{minipage}
    \begin{minipage}[ht]{0.50\linewidth}
        \center{\epsfig{file=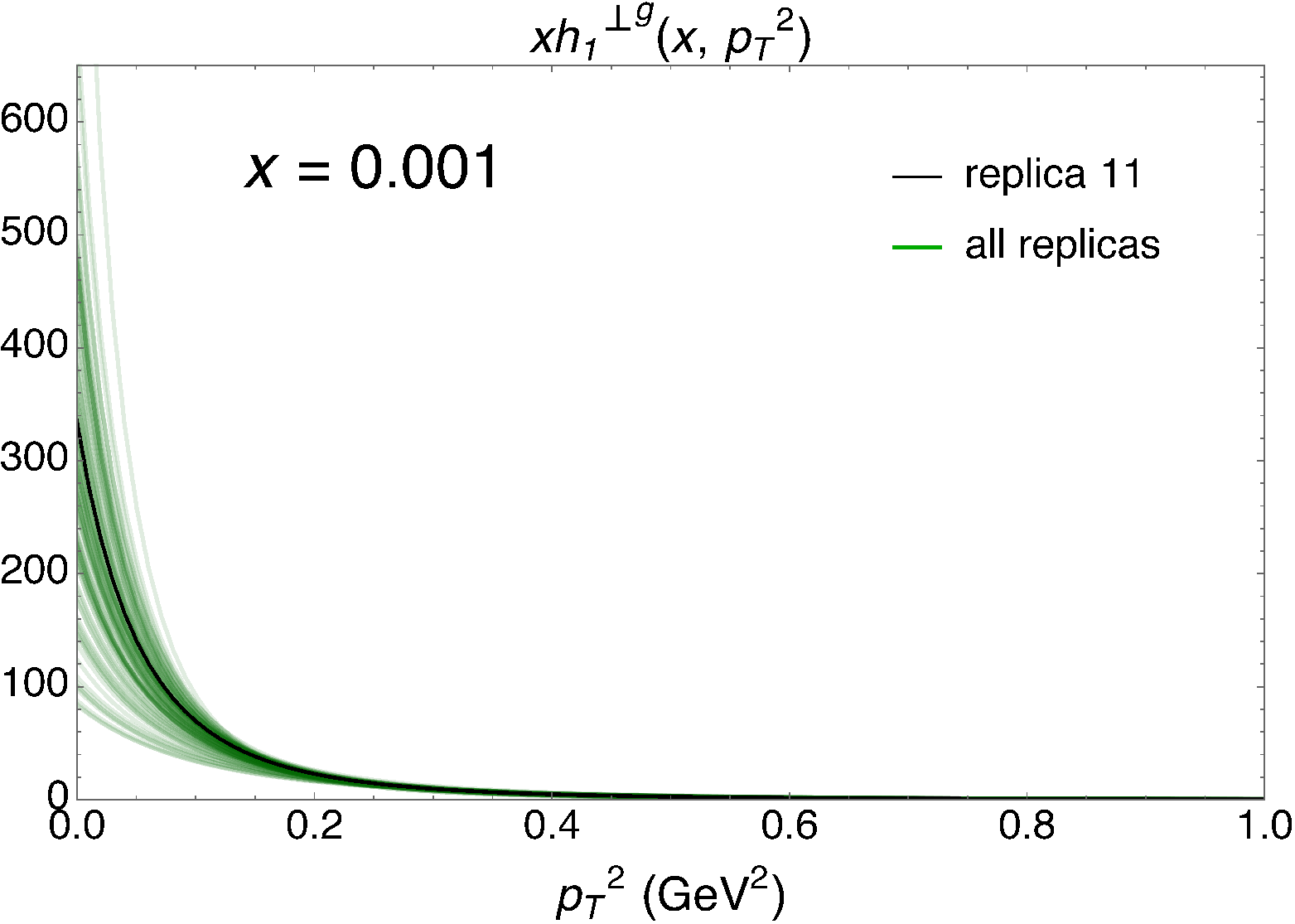,scale=0.30} \\ (b)}
  \end{minipage}
\caption{Examples of model calculations of the unpolarized(a) and Boer-Mulders (b) gluon TMDs as functions of $\ktsq$, for $x=10^{-3}$ and at the initial scale, $Q_0 = 1.64$ GeV. The figures are adapted from Ref.~\cite{Bacchetta:2020vty}~\textcopyright (2020) by The European Physical Journal.}
\label{fig:SM_gluon_TMDs}
\end{figure}

Predictions obtained in Ref.~\cite{Bacchetta:2020vty} for the unpolarized gluon TMD, $x f_1^g(x,\ktsq)$, and for the linearly-polarized gluon TMD, $x h_1^{\perp g}(x,\ktsq)$, are shown in Fig.~\ref{fig:SM_gluon_TMDs} as functions of the transverse momentum squared, $\ktsq$, for $x=10^{-3}$ and at the initial scale, $Q_0 = 1.64$ GeV (\textit{i.e.}, without the application TMD evolution). Predictions are given as a set of 100 replicas, which are statistically equivalent and reproduce well the collinear PDFs $f^g(x)$ and $g_1^g(x)$. Each line in the plot shows a single replica, with the black solid line corresponding to the most representative replica (n. 11), which has the minimal quadratic distance to the mean values of parameters obtained in the fit. Apart from details, it is important to note that: i) even if all replicas reproduce similar collinear PDFs, they predict very different results for the TMDs, ii) each TMD exhibits a peculiar trend both in $x$ and $\ktsq$. For instance, the unpolarized function presents a clearly non-Gaussian shape in $\ktsq$, and goes to a small but non-vanishing value for $\ktsq \to 0$.  The linearly-polarized gluon TMD is large at small $\ktsq$ and decreases fast. Both of them are increasingly large at small x.

\subsection{Hadron structure and mechanisms of charmonium hadroproduction\label{quarkonia_th}}

In this section, we give a short overview of the current status of the theory of heavy-quarkonium production with an emphasis on possible applications of heavy-quarkonium measurements for studies of the gluon content of hadrons. 

The hadroproduction of heavy quarkonia proceeds in two stages: firstly, a heavy quark-antiquark pair is produced at short distances, via gluon-gluon fusion but also with a non-negligible contribution of $q\bar{q}$ and $qg$-initiated subprocesses, depending on the collision energy. The second stage is represented by the hadronization process of the quark-antiquark pair into a physical quarkonium state, which happens at large distances (low energy scales/virtualities) and is accompanied by a complicated rearrangement of color via exchanges of soft gluons between the heavy quark-antiquark pair and other colored partons produced in the collision. At present, two approaches aimed to describe hadronization stage are most well-explored: the Non-Relativistic QCD factorization (NRQCD-factorization)~\cite{Bodwin:1994jh} formalism and the (Improved-)Color-Evaporation Model (CEM)~\cite{Fritzsch:1977ay,Halzen:1977rs,Barger:1979js, Barger:1980mg, Gavai:1994in, Ma:2016exq, Cheung:2017osx, Cheung:2018tvq, Maciula:2018bex,Lansberg:2020rft}, see e.g. recent reviews~\cite{Lansberg:2006dh,Brambilla:2010cs,Andronic:2015wma,Lansberg:2019adr}. 

Let us first discuss the conceptually simpler CEM. In this model, inspired by the idea of (local) quark-hadron duality, all color and angular-momentum states of $c\bar{c}$-pair with invariant masses below the threshold of production of open-charmed meson pair contribute to the production of charmonium-state ${\cal H}=\eta_c,$ $J/\psi$, $\chi_c$ with some constant probability $F_{\cal H}$, i.e. the ICEM cross-section is:
\begin{equation}
d\sigma(p+p\to {\cal H} + X) = F_{\cal H} \int\limits_{m_{\cal H}}^{2m_{D}}dM_{c\bar{c}}\ \frac{d\sigma(p+p\to c+\bar{c}+X)}{dM_{c\bar{c}}},\label{eq:ICEM-CS}
\end{equation}
where the cross-section $d\sigma(p+p\to c+\bar{c}+X)$ can be computed to any order in QCD perturbation theory and particular lower-limit of integration $M_{c\bar{c}}^{(\min)}=m_{\cal H}$ corresponds to ICEM~\cite{Ma:2016exq, Cheung:2017osx, Cheung:2018tvq}, while in traditional CEM~\cite{Barger:1979js, Barger:1980mg, Gavai:1994in} it has been set to $2m_c$. Also, due to $M_{c\bar{c}}\neq m_{\cal H}$, the three-momentum of $c\bar{c}$-pair is different form the momentum of produced charmonium, which in some approximation incorporates effects of soft-gluon emissions at hadronisation stage. This recoil effect is usually~\cite{Ma:2016exq, Maciula:2018bex} taken into account using simple rescaling $\bm{p}_{\cal H}=\bm{p}_{c\bar{c}}\times m_{\cal H}/M_{c\bar{c}} $, which is the correct prescription in the limit $M_{c\bar{c}}-m_{\cal H}\ll m_{\cal H}\ll |\bm{p}_{\cal H}|$ and assuming isotropic gluon emission, while in some studies the soft-gluon recoil is neglected~\cite{Cheung:2017osx, Cheung:2018tvq}. Due to steep decrease of the $p_T$-spectra, this kind of small momentum shifts are important and may lead to $O(10\%)$ change of probabilities $F_{\cal H}$.   

  Despite its simplicity, the ICEM is remarkably successful in describing $p_T$-dependence of ratios of yields of different charmonium states~\cite{Ma:2016exq}, the general shape of $p_T$-spectra of prompt $J/\psi$ and $\psi(2S)$ production at moderate transverse-momenta and even describes the unpolarized production of $J/\psi$ at high-$p_T$ in agreement with experimental data~\cite{Cheung:2018tvq}. However the detailed shape of $p_T$ spectra, in particular at large $p_T$, can not be reproduced in this model even at NLO in $\alpha_s$ for the short-distance cross-section $d\sigma(p+p\to c+\bar{c}+X)$, as it was shown in recent studies~\cite{Lansberg:2016rcx,Lansberg:2020rft}. Moreover, when the CEM is applied to the charmonium pair-production with hadronisation-probabilities, $F_{\cal H}$ fitted to the prompt single-production data, it underestimates the total and differential cross-sections in whole phase-space by up to two orders of magnitude~\cite{Lansberg:2020rft}. A similar problem arises with the estimation of $e^+e^-\to J/\psi + c\bar{c}$ cross-section in CEM~\cite{Lansberg:2019adr}.  We will discuss implications of latter findings in the context of NRQCD factorization below.   

 The NRQCD-factorization framework is based on the Non-Relativistic QCD (NRQCD) effective field theory which consists in a systematic expansion of usual QCD Lagrangian in powers of squared velocity of constituent heavy-quarks in a bound state -- $v^2$. Simple potential-model estimates lead to $v^2\simeq 0.3$ for charmonia and $\simeq 0.1$ for bottomonia (see e.g. Sec.~A in Ref.~\cite{Bodwin:1994jh}), so $v^2$ is assumed to be a good expansion parameter. The degrees of freedom of NRQCD are fields (field-operators) of non-relativistic quarks (annihilation-operator) $\psi(x)$ and anti-quarks (creation-operator) $\chi(x)$, as well as fields $A_{\mu}(x)$ of long-wavelength gluons with virtualities much smaller than $m_c$. All the effects caused by short-wavelength gluons and light-quarks are incorporated into matching-coefficients in front of local operators forming the NRQCD Lagrangian. The latter ones are perturbatively computable  to any order in $\alpha_s$. The production of $c\bar{c}$-pair at short distance is also represented as a perturbative matching coefficient, while the subsequent evolution of a $c\bar{c}$ state into a quarkonium is described by Long-Distance Matrix-Elements (LDMEs) of local operators, constructed from the NRQCD fields $\psi(x)$, $\chi(x)$ and $A_{\mu}(x)$.     

 Inclusive heavy-quarkonium production cross-section depends on LDMEs of operators~\cite{Bodwin:1994jh}:
\begin{equation}
 \sum\limits_{X} \left\vert \left\langle X+{\cal H} \right\vert \psi^\dagger(0)\kappa_i \chi(0) \left\vert 0 \right\rangle \right\vert^2 = \left\langle 0 \right\vert {\cal O}^{({\cal H})}_i(0) \left\vert 0 \right\rangle=\left\langle {\cal O}_{\cal H}[i] \right\rangle, \label{eq:Prod-LDME}
\end{equation}
where ${\cal O}_i^{({\cal H})}(x)=\chi^\dagger(0)\kappa^\dagger_i \psi(0) a^\dagger_{\cal H} a_{\cal H} \psi^\dagger(0)\kappa_i \chi(0)$ with $a^\dagger_{\cal H}$($a_{\cal H}$) being the creation(annihilation) operators for the physical state ${\cal H}$. The operators $\kappa_i$ are polynomials in Pauli matrices, covariant derivatives and QCD field-strength tensors, which project out a particular color-singlet (CS,$^{(1)}$) or color-octet (CO,$^{(8)}$) Fock-state of $c\bar{c}$-pair $i={}^{2S+1}L_J^{(1,8)}$ where the spectroscopic notation for spin -- $S$ and orbital-momentum -- $L$ has been used. To ensure the gauge-invariance of the LDMEs for color-octet states and the expected factorization properties of infra-red divergences in NRQCD perturbation theory, Wilson-line factors depending on $A_{\mu}(x)$ should be added~\cite{Nayak:2005rt,Nayak:2006fm} to the operators  ${\cal O}^{({\cal H})}_i(x)$, making them non local. Further complications in the definition of NRQCD operators arise when more than one heavy-quarkonium state is considered in the final-state~\cite{He:2018hwb}. 

In contrast to Eq.~(\ref{eq:Prod-LDME}), the inclusive decay rate of quarkonium ${\cal H}$ depends on different kind of LDMEs:
 \begin{equation}
\sum\limits_{X} \left\vert \left\langle X \right\vert \chi^\dagger(0)\kappa^\dagger_i \psi(0) \left\vert {\cal H} \right\rangle \right\vert^2 = \left\langle 0 \right\vert a_{\cal H} \chi(0)\kappa_i \psi^\dagger(0) \chi^\dagger(0)\kappa^\dagger_i \psi(0) a^\dagger_{\cal H} \left\vert  0 \right\rangle , \label{eq:Decay-LDME}
\end{equation}
which, for the case of color-octet states, cannot be related to LDMEs~(\ref{eq:Prod-LDME}) without further approximations. 

The velocity-scaling rules (VSRs)~\cite{Lepage:1992tx, Bodwin:1994jh} for LDMEs allow one to truncate to a desired order in $v^2$ the infinite tower of local NRQCD-operators which otherwise might contribute to the production of a state ${\cal H}$. As such, they are the cornerstone of phenomenological applications of NRQCD factorization. These scaling-rules assign specific scaling power in $v$ to the NRQCD fields $\psi(x)$, $\chi(x)$, $A_{\mu}(x)$ and to the covariant-derivative $D_{\mu}=\partial_\mu+ig_s A_\mu(x)$ (see the Tab. 1 in Ref.~\cite{Lepage:1992tx}). The latter one is particularly important, because one factor of ${\bm D}$ enters into operator $\kappa_i$ for each unit of orbital momentum $L$, leading to $(v^2)^L$-suppression of LDMEs of states with high values of $L$.    Another key ingredient of the derivation of VSRs for LDMEs is the $v^2$-expansion for the physical quarkonium states, originating from the physical picture of non-relativistic bound-state in Coulomb gauge for gluon fields. The latter expansion e.g. for $J/\psi$ has the form (see the discussion in Sec.~D of Ref.~\cite{Bodwin:1994jh}):
\begin{eqnarray*}
\left\vert J/\psi \right\rangle = a^\dagger_{J/\psi} \left\vert 0 \right\rangle &=& O(1) \left\vert c\bar{c}\left[{}^3S_1^{(1)} \right] \right\rangle + O(v) \left\vert c\bar{c}\left[{}^3P_J^{(8)} \right]g \right\rangle \\
&+& O(v^{3/2}) \left\vert c\bar{c}\left[{}^1S_0^{(8)} \right]g \right\rangle+O(v^2) \left\vert c\bar{c}\left[{}^3S_1^{(8)} \right]gg \right\rangle+\ldots ,
\end{eqnarray*}
where, by $O(v^n)$, we denote the order of velocity-suppression of the corresponding Fock-state.  Therefore, an additional $v^2$ suppression, besides the $v^2$-factor coming from the structure of operators $\kappa_i$, comes in for decay-LDMEs~(\ref{eq:Decay-LDME}) resulting into the intricate structure of $v^2$-suppression for LDMEs with various labels $i$, shown in the Tab.~\ref{tab:LDME-VSRs}. It is further assumed that production LDMEs~(\ref{eq:Prod-LDME}) follow the same pattern of $v^2$-suppression as the decay-LDMEs (\ref{eq:Decay-LDME}) and this assumption is supported by the results of higher-order perturbative calculations~\cite{Nayak:2005rt,Nayak:2006fm}. The color-singlet LDMEs in the LO of $v^2$-expansion are proportional to the modulus-squared of the radial part of the wave-function of potential model -- $|R(0)|^2$ (or $|R'(0)|^2$ for $P$-wave states), thus connecting the NRQCD-factorization with the original {\it Color-Singlet Model} of heavy-quarkonium production~\cite{Gastmans:1986qv}. The numerical smallness of color-octet LDMEs relatively to color-singlet also has been observed in global LDME-fits for charmonia~\cite{Butenschoen:2010rq,Butenschoen:2011yh,Butenschoen:2012px,Butenschoen:2012qr,Chao:2012iv,Gong:2012ug,Bodwin:2014gia} and bottomonia~\cite{Gong:2010bk,Wang:2012is,Gong:2013qka,Nefedov:2013qya}. 

\begin{table}[htp]
\caption{Velocity-scaling rules for LDMEs~\cite{Bodwin:2005hm} in the NRQCD-factorization formalism. \label{tab:LDME-VSRs}}
\begin{center}
\begin{tabularx}{0.9\textwidth}{|X|XXXX|XXXXXXXX|}
\hline
           & ${}^1S_0^{(1)}$ & ${}^{3}S_1^{(1)}$ & ${}^1S_0^{(8)}$ & ${}^3S_1^{(8)}$ & ${}^1P_1^{(1)}$ & ${}^3P_0^{(1)}$ & ${}^3P_1^{(1)}$ & ${}^3P_2^{(1)}$ & ${}^1P_1^{(8)}$ & ${}^3P_0^{(8)}$ & ${}^3P_1^{(8)}$ & ${}^3P_2^{(8)}$ \bigstrut \\
\hline
$\eta_c$   & $1$             &                   & $v^4$           & $v^3$           &                 &                 &                 &                 & $v^4$           &                 &                 &                 \bigstrut \\
$J/\psi$   &                 & $1$               & $v^3$           & $v^4$           &                 &                 &                 &                 &                 & $v^4$           & $v^4$           & $v^4$           \bigstrut \\
\hline
$h_c$      &                 &                   & $v^2$           &                 & $v^2$           &                 &                 &                 &                 &                 &                 &                 \bigstrut \\
$\chi_{c0}$&                 &                   &                 & $v^2$           &                 & $v^2$           &                 &                 &                 &                 &                 &                 \bigstrut \\
$\chi_{c1}$&                 &                   &                 & $v^2$           &                 &                 & $v^2$           &                 &                 &                 &                 &                 \bigstrut \\
$\chi_{c2}$&                 &                   &                 & $v^2$           &                 &                 &                 & $v^2$           &                 &                 &                 &                 \bigstrut\\
\hline
\end{tabularx}
\end{center}
\end{table}

  The description of detailed shapes of $p_T$ spectra of prompt charmonia~\cite{Butenschoen:2010rq,Butenschoen:2011yh,Butenschoen:2012px,Butenschoen:2012qr,Butenschoen:2012qr,Chao:2012iv,Gong:2012ug,Bodwin:2014gia} and bottomonia~\cite{Gong:2010bk,Wang:2012is,Gong:2013qka} with $p_T>3-5$ GeV, produced in $pp$ and $p\bar{p}$ collisions at Tevatron and LHC in the framework of global NRQCD fits at NLO in CPM is one of the main phenomenological achievements of NRQCD factorization so far. The global NLO fit~\cite{Butenschoen:2010rq,Butenschoen:2011yh,Butenschoen:2012px,Butenschoen:2012qr} which attempts to describe charmonium production cross-section in $e^+e^-$, $pp$ and $ep$ collisions leads to the dominance of ${}^3S_1^{(8)}$ and ${}^3P_J^{(8)}$ states at $p_T\gg M_{J/\psi}$. At high-$p_T$ the ${}^3S_1^{(8)}$ state is typically produced in a fragmentation of an almost on-shell gluon $g\to c\bar{c}\left[ {}^3S_1^{(8)} \right]$, resulting into the predominantly-transverse polarization of this state in the helicity frame. The ${}^3P_{J=1,2}^{(8)}$-states also give significant contribution to the transverse polarization of final quarkonium at NLO. There is no mechanism to de-polarize this states at leading-order in $v$, since the polarization is carried by heavy-quark spin, so the transverse polarization of produced quarkonium at high-$p_T$ for those fits in which ${}^3S_1^{(8)}$ and ${}^3P_J^{(8)}$-states dominate at high $p_T$ is one of the most solid predictions of NRQCD.

 Unfortunately, this prediction is by now clearly disfavored by experimental measurements of $p_T$ dependence of charmonium~\cite{Chatrchyan:2013cla} and bottomonium~\cite{Chatrchyan:2012woa} polarization parameters at the LHC, see the Ref.~\cite{Andronic:2015wma} for a global survey of heavy-quarkonium polarization data. All experimental data are consistent with unpolarized production at high-$p_T$, with $\psi(2S)$ data even consistent with small longitudinal polarization, in contradiction with predictions of the fit~\cite{Butenschoen:2010rq,Butenschoen:2011yh,Butenschoen:2012px,Butenschoen:2012qr}. On the other hand, if the consistency with unpolarized production is imposed, then agreement of the fit with $e^+e^-$ and photo-production data is lost~\cite{Butenschoen:2012px,Butenschoen:2012qr}. In the literature, this situation is called ``heavy-quarkonium polarization puzzle''.  

 Another prediction, based on the standard picture of VSRs in the NRQCD factorization is the relation between LDMEs of Fock states different by the flipping of the spin of a heavy-quark -- the so-called Heavy-Quark Spin Symmetry (HQSS) relations, for example:
\begin{eqnarray*}
\left\langle {\cal O}^{\eta_c} \left[ {}^1S_0^{(1/8)} \right] \right\rangle &=& \frac{1}{3} \left\langle {\cal O}^{J/\psi} \left[ {}^3S_1^{(1/8)} \right] \right\rangle + O(v^5), \\
\left\langle {\cal O}^{\eta_c} \left[ {}^3S_1^{(8)} \right] \right\rangle &=& \left\langle {\cal O}^{J/\psi} \left[ {}^1S_0^{(8)} \right] \right\rangle + O(v^4), \\
\left\langle {\cal O}^{\eta_c} \left[ {}^1P_1^{(8)} \right] \right\rangle &=& 3\left\langle {\cal O}^{J/\psi} \left[ {}^3P_0^{(8)} \right] \right\rangle + O(v^5), \\
\left\langle {\cal O}^{h_c} \left[ {}^1P_1^{(1)}/{}^1S_0^{(8)} \right] \right\rangle &=& 3\left\langle {\cal O}^{\chi_{c0}} \left[ {}^3P_0^{(1)}/{}^{3}S_1^{(8)} \right] \right\rangle + O(v^3).
\end{eqnarray*}
These relations had been confronted with phenomenology in Ref.~\cite{Butenschoen:2014dra} by taking CO LDMEs from several available NRQCD-fits of $J/\psi$ production, converting $J/\psi$ LDMEs to $\eta_c$-ones and predicting the $\eta_c$ transverse-momentum spectrum, which can be compared with the measurement by LHCb~\cite{Aaij:2014bga,Aaij:2019gsn}. The firm result of this investigation is, that LHCb data are perfectly compatible with predominantly direct production of $\eta_c$ via CS ${}^{1}S_0^{(1)}$-state. The feed-down from $h_c$ has been found to be negligible and addition of CO-contributions leads to overestimation of high-$p_T$ cross-section by more than order of magnitude above experimental data. This findings suggest the significant violation of VSRs for $\eta_c$ LDMEs, since CO contributions for $\eta_c$ are of the same order as for the $J/\psi$ (Tab.~\ref{tab:LDME-VSRs}) and corrections to the HQSS relations are of higher-order in $v$. Nevertheless, in Refs.~\cite{Zhang:2014ybe,Han:2014jya} it was shown that it is possible to improve the description of the $J/\psi$ and $\eta_c$ cross-sections as well as $J/\psi$-polarization in common LDME fits compatible with the HQSS-relations if one discards the data with $p_T<7$ GeV from the fit. However, besides the fact that there is no explanation for such a high value of the $p_T$-threshold within the NRQCD-factorization formalism, tensions in the data description either remain or are newly introduced: The CS LDME values fitted in Ref.~\cite{Zhang:2014ybe} differ from those obtained from potential models or decay widths, while the LDME fit of Ref.~\cite{Han:2014jya} still leads to significant transverse $J/\psi$ polarization at higher $p_T$, see also related Ref.~\cite{Shao:2014yta}.

Comparing the CEM and NRQCD-factorization formalisms to each-other, one finds~\cite{Bodwin:2005hm}, that CEM can be understood as NRQCD-factorization without $v^2$-suppression between $S$-wave CS and CO LDMEs. This suppression is characteristic for NRQCD, see Tab.~\ref{tab:LDME-VSRs}. The absence of the suppression is also clearly disfavored by recent studies of $J/\psi$ pair-production in NRQCD~\cite{Lansberg:2019fgm,He:2019qqr,He:2015qya} and CEM~\cite{Lansberg:2020rft}. These studies show that the bulk of the double-$J/\psi$ cross-section is explained by double-${}^3S_1^{(1)}$ CS-state production~\cite{Lansberg:2014swa,Lansberg:2013qka}, while the CEM prediction, dominated by CO states of $c\bar{c}$-pair, fails to describe the data. 

Concluding our mini-review of heavy-quarkonium production theory we emphasize, that neither CEM, nor NRQCD-factorization in its pure form can describe all available data on heavy-quarkonium production and polarization, and probably some hybrid of this two models should be constructed.  Presently, the study of the heavy-quarkonium production mechanism is an active field of research, with new approaches, such as subleading-power fragmentation~\cite{Kang:2011mg} and soft-gluon factorization~\cite{Ma:2017xno,Li:2019ncs,Chen:2020yeg}, being proposed recently.

   \begin{figure}[!h]
  \begin{minipage}[ht]{0.50\linewidth}
        \center{\epsfig{file=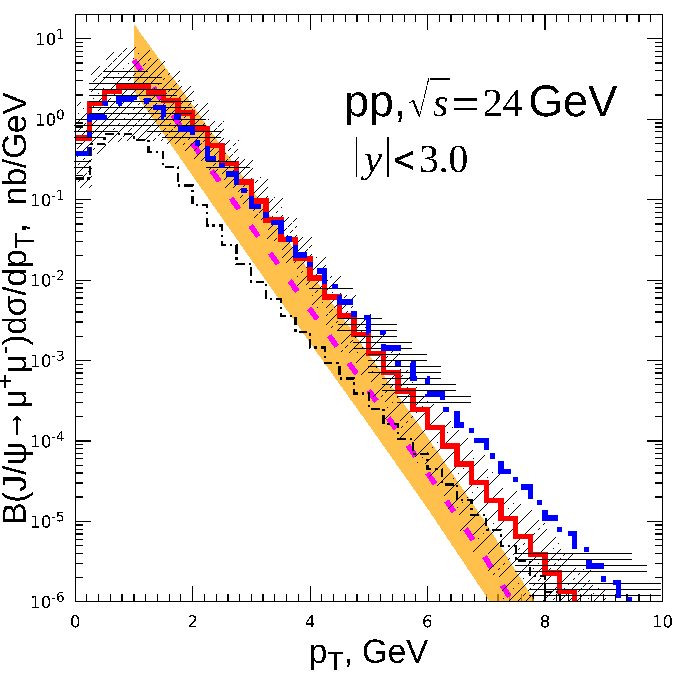,scale=0.63} \\ (a)}
  \end{minipage}
  \begin{minipage}[ht]{0.50\linewidth}
        \center{\epsfig{file=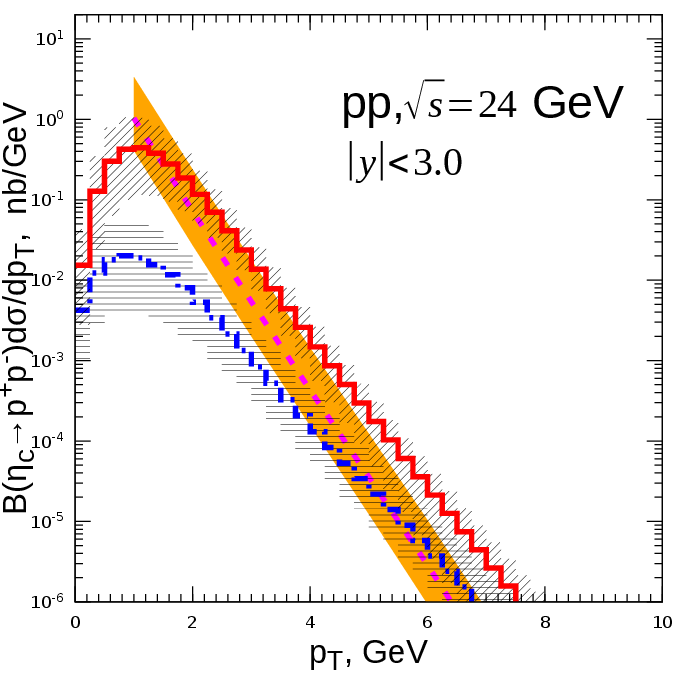,scale=0.63} \\ (a)}
  \end{minipage}
     \caption{Theoretical predictions for the inclusive $J/\psi$ $p_T$-spectrum (a) and the $\eta_c$ $p_T$-spectrum (b) in various models: NLO of CPM + NRQCD-factorization (thick dashed line with solid uncertainty band)~\cite{Butenschoen:2010rq,Butenschoen:2011yh}, LO of PRA + NRQCD-factorization (thick solid histogram with diagonally-shaded uncertainty band)~\cite{Saleev:2012hi,Karpishkov:2020wwe}, and LO PRA~\cite{Karpishkov:2017kph} + Improved Color Evaporation Model (thick dash-dotted histogram with horizontally-shaded uncertainty band)~\cite{Cheung:2018tvq}. The contribution of the $q\bar{q}$-annihilation channel to the central ICEM prediction is depicted by the thin dash-dotted histogram. Uncertainty bands are due to the factorization/renormalization scale variation only. }
     \label{fig:NICA-Jpsi-etac-pT}
 \end{figure}

   \begin{figure}[!h]
%\begin{minipage}[ht]{0.50\linewidth}
      \center{\epsfig{file=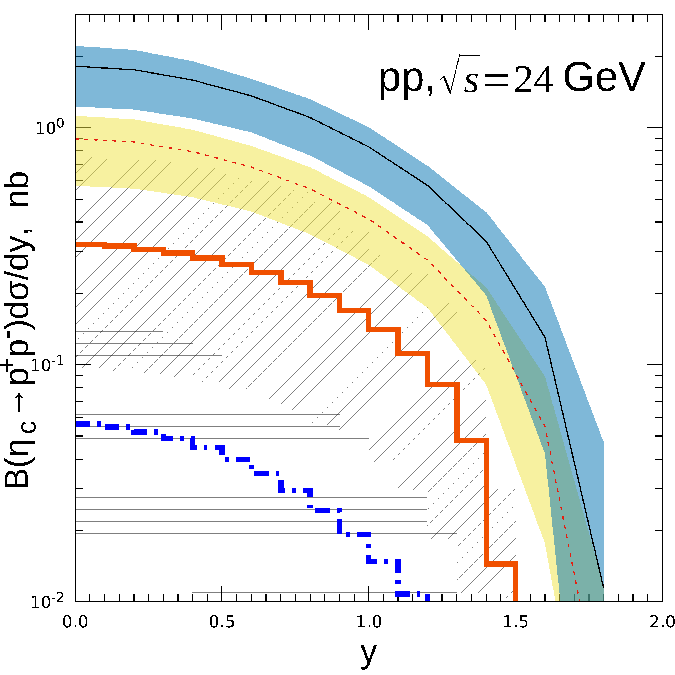,scale=0.63}}
%\end{minipage}
     \caption{Theoretical predictions for the rapidity-differential cross-section of $\eta_c$-production. Thin dashed and solid lines -- LO and NLO CPM + NRQCD predictions  with optimized factorization-scale choice~\cite{Feng:2015cba,Lansberg-Ozcelik}, uncertainty bands -- PDF and renormalization-scale uncertainties added in quadratures. Thick solid line -- LO PRA + NRQCD, dash-dotted line -- LO PRA + ICEM, full uncertainty bands. Last two predictions performed with the same model as predictions in the Fig.~\ref{fig:NICA-Jpsi-etac-pT}(b).  }
     \label{fig:NICA-etac-dy}
 \end{figure}

   \begin{figure}[!h]
%\begin{minipage}[ht]{0.50\linewidth}
       \center{\epsfig{file=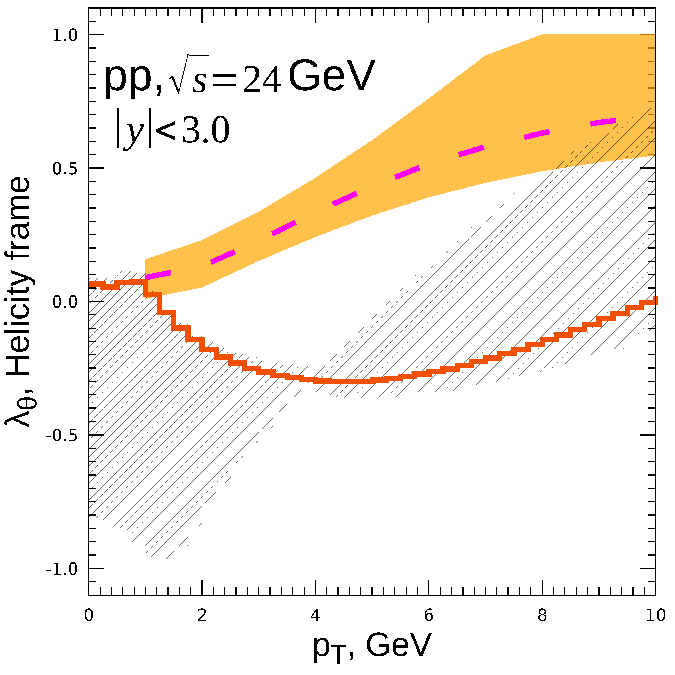,scale=0.7}}
     \caption{Theoretical predictions for the $p_T$-dependence of polarization parameter $\lambda_\theta$ in various models: NLO of Collinear Parton Model + NRQCD-factorization (dashed line with solid uncertainty band)~\cite{Butenschoen:2010rq,Butenschoen:2011yh}, LO of PRA + NRQCD-factorization (solid histogram with shaded uncertainty band)~\cite{Saleev:2012hi,Karpishkov:2020wwe}. Uncertainty bands are due to the factorization/renormalization scale variation and the variation of LDMEs within their fit-uncertainties. }
     \label{fig:NICA-Jpsi-pol}
 \end{figure}
  
Due to the theoretical problems described above and multitude of competing theoretical approaches and models available on the market, our lack of quantitative understanding of the mechanism of hadronization of charm quarks into charmonium can become a source of significant theoretical uncertainties if charmonium production is to be used as a tool to study the proton structure. The Fig.~\ref{fig:NICA-Jpsi-etac-pT} provides an insight on this situation at NICA SPD. In this figure, predictions of three models for the $p_T$ spectra of $J/\psi$ (Fig.~\ref{fig:NICA-Jpsi-etac-pT}(a)) and $\eta_c$ mesons (Fig.~\ref{fig:NICA-Jpsi-etac-pT}(b)) are compared. Furthermore, in the Fig.~\ref{fig:NICA-Jpsi-pol}, predictions for the $p_T$-dependence of the polarization parameter $\lambda_\theta$ in the $\hat{s}$-channel helicity frame are presented. The dashed curve with solid uncertainty bend in Fig.~\ref{fig:NICA-Jpsi-etac-pT}(a) represents the NLO calculation in collinear parton model (with LO being $O(\alpha_s^3)$, see Fig.~\ref{fig:diagrams-gg}(a)) for the short-distance part of the cross-section matched to the NRQCD-factorization formalism for the long-distance part, with LDMEs of the latter tuned to charmonium production data in hadronic collisions, DIS and $e^+e^-$-annihilation~\cite{Butenschoen:2010rq,Butenschoen:2011yh,Butenschoen:2012px,Butenschoen:2012qr}.  For the prediction represented by solid histogram in the Fig.~\ref{fig:NICA-Jpsi-etac-pT}(a), the short-distance part of the cross-section is calculated in the LO ($O(\alpha_s^2)$ for color-octet and $P$-wave contributions and $O(\alpha_s^3)$ for color-singlet $S$-wave ones) of PRA~\cite{Karpishkov:2017kph}, while  LDMEs in this calculation had been fitted to the charmonium hadroproduction data from RHIC, Tevatron and LHC~\cite{Saleev:2012hi,Karpishkov:2020wwe}. The thick dash-dotted histogram is calculated in the LO ($O(\alpha_s^2)$) of PRA with the same unintegrated PDFs as for the LO PRA+NRQCD prediction, but interfaced with an improved Color-Evaporation Model (ICEM) of Ref.~\cite{Cheung:2018tvq} for description of hadronization. Non-perturbative parameters of the ICEM had been taken from the Ref.~\cite{Cheung:2018tvq} where they had been fitted to charmonium hadroproduction data at Tevatron and LHC energies. Predictions of all three models for the inclusive $J/\psi$ $p_T$ spectrum at NICA SPD appear to be consistent within their uncertainty bands, see Fig.~\ref{fig:NICA-Jpsi-etac-pT}(a). However, the structure of this predictions is significantly different, with NRQCD-based predictions being dominated by gluon-gluon fusion subprocess, while ICEM prediction containing significant contamination from $q\bar{q}$-annihilation, shown with the thin dash-dotted histogram in the Fig.~\ref{fig:NICA-Jpsi-etac-pT}(a). The latter contribution reaches up to 50\% at low $p_T<1$ GeV and contributes up to 10\% at higher $p_T>3$ GeV. Also ICEM tends to predict a significantly harder $p_T$ spectrum at $p_T>5$ GeV, than NRQCD-based PRA prediction which had been performed with the same unintegrated PDFs.

 The above discussion  shows that the $J/\psi$ $p_T$-spectrum can be reliably predicted only in the limited range of transverse momenta, approximately from 3 to 6 GeV at $\sqrt{s}=24$ GeV. At higher $p_T$ the shape of the spectrum becomes highly model-dependent and at lower $p_T<M_{J/\psi}$ the TMD-factorization effects (including possible violation of factorization, see~\cite{Echevarria:2019ynx,Fleming:2019pzj}) come into the game and the contribution of $q\bar{q}$-annihilation subprocess becomes uncertain. Nevertheless, predictions and measurements of the rapidity or $x_F$-differential cross-sections even in this limited $p_T$-range could help to further constrain the gluon PDF, e.g. to rule-out the extreme values of $L$ in the $x\to 1$ asymptotic limit of the PDF $\sim (1-x)^L$. 

  In the Fig.~\ref{fig:NICA-Jpsi-etac-pT}(b), we perform a similar comparison of model-predictions for $\eta_c$-production at SPD-NICA. Taking into account results of Ref.~\cite{Butenschoen:2014dra} we include only CS ${}^1S_0^{(1)}$-contribution to the NLO CPM %~\cite{KB_priv}
  and LO PRA NRQCD predictions and neglect the possible feed-down from $h_c$-meson decays. For comparison we also show the LO PRA ICEM prediction for $\eta_c$ transverse-momentum spectrum, with $F_{\eta_c}=(1.8\pm 0.8)\times 10^{-2}$ in Eq.~(\ref{eq:ICEM-CS}) tuned to describe LHCb data~\cite{Aaij:2014bga} in the same model. The LO PRA and NLO CPM predictions again agree nicely with each-other, however ICEM-prediction has different slope and magnitude at SPD-NICA, while it describes the LHCb-data as well as the NRQCD prediction. This comparison shows, that studies of $\eta_c$-production at SPD-NICA will be instrumental for better understanding of its production mechanism. If the CS-dominated NRQCD-prediction turn out to be correct, then $\eta_c$  production becomes a unique instrument to study the gluon content of the proton without introducing additional free-parameters, such as CO LDMEs, to the analysis. 
  
  Model predictions for rapidity-dependence of $p_T$-integrated cross-section are presented in the Fig.~\ref{fig:NICA-etac-dy}. Here the LO and NLO CPM NRQCD predictions are performed with the optimized value of factorization and renormalization scales: $\mu_R=\mu_F=m_{\eta_c} e^{-A_{\eta_c}}$ with $A_{\eta_c}=1/2$ dictated by the high-energy behavior of NLO partonic cross-section. This particular scale-choice allows one to avoid the problem of negative NLO cross-sections at high energies~\cite{Feng:2015cba,Lansberg-Ozcelik}, however at NICA energies it leads to predictions for $p_T$-integrated $\eta_c$-yield significantly higher than predictions of other models. PRA predictions in the Fig.~\ref{fig:NICA-etac-dy} are based on the same model as PRA predictions in the Fig.~\ref{fig:NICA-Jpsi-etac-pT}(b). The latter model allows one to calculate rapidity-differential cross-section, because PRA cross-section is finite in $p_T\to 0$ limit due to $p_T$-spearing from unintegrated gluon PDFs. We have used the same scale-choice $\mu_R=\mu_F=\sqrt{m_{\eta_c}^2+p_T^2}$ (with $\langle p_T \rangle > 1$ GeV, see Fig.~\ref{fig:NICA-Jpsi-etac-pT}(b)) for PRA predictions in both Figs.~\ref{fig:NICA-Jpsi-etac-pT}(b) and \ref{fig:NICA-etac-dy}. The PRA ICEM cross-section in the Fig.~\ref{fig:NICA-etac-dy} is almost order of magnitude smaller than PRA NRQCD prediction, consistently with Fig.~\ref{fig:NICA-Jpsi-etac-pT}(b), thus even measurement of the total cross-section of $\eta_c$-production at SPD NICA will allow to exclude ICEM for this state. 
  
  Predictions of NLO CPM and LO of PRA for $J/\psi$-polarization parameter $\lambda_\theta$ (see the Fig.~\ref{fig:NICA-Jpsi-pol}) are significantly different, with PRA predicting mostly unpolarized production ($\lambda_\theta\simeq 0$) while CPM predicts transverse polarization ($\lambda_\theta=+1$) at high $p_T$. Disagreement of the predictions for polarization parameters  mostly reflects the difference of LDMEs obtained in two fits and their large uncertainty bands are due to significant uncertainties of LDMEs. Measurements of heavy quarkonium polarization at NICA energies will provide additional constraints on models, however due to well-known problems with description of polarization at high energies~\cite{Butenschoen:2012px,Butenschoen:2012qr} constraints coming from the polarization measurements should be interpreted with great care and one should try to disentangle conclusions for gluon PDF from the results related to heavy quarkonium polarization.

%\JPL{Good. You can probably add that at SPD energies there is a simplification because of the absence of $b$ feed-down. You can certainly also add some discussion for $\psi'$ and then $\chi_c$. The physics case for $\eta_c'$ studies at the LHC has been detailed here: \cite{Lansberg:2017ozx}. Most likely $Br(\eta_c' \to pp)/Br(\eta_c \to pp) \simeq 0.1$ with possibly a smaller background.}

\subsection{TMD factorization with gluon probes \label{un_tmd}}
The description of hard processes which involve hadrons is based on factorization theorems. In particular, formulation of factorization theorems in terms of the TMD PDFs (see e.g. the monograph~\cite{Collins:2011zzd} or the review~\cite{Angeles-Martinez:2015sea}) of quarks and gluons is the most important step towards studying the 3D structure of hadrons in momentum space and the nature of their spins.

The field of TMDs has taken a big step forward in the last years.
Both the theoretical framework~\cite{GarciaEchevarria:2011rb,Echevarria:2012pw,Echevarria:2012js,Echevarria:2014rua,Collins:2011zzd,Echevarria:2015uaa,Scimemi:2018xaf}, mainly focused on the proper definition of TMDs and their properties, and the phenomenological analyses (see e.g. these recent extractions \cite{DAlesio:2014mrz,Echevarria:2014xaa,Bacchetta:2015ora,Bacchetta:2017gcc,Anselmino:2016uie,Scimemi:2017etj,Bertone:2019nxa}), have been developed, together with the appearance of new higher-order perturbative calculations (see e.g.~\cite{Gutierrez-Reyes:2019rug,Gutierrez-Reyes:2018iod,Vladimirov:2017ksc,Echevarria:2016scs,Echevarria:2015byo,Echevarria:2015usa,Bacchetta:2018dcq}).
However this progress has mainly been done in the quark sector, due to the difficulty to cleanly probe gluons in high-energy processes.

Several processes have been proposed to access gluon TMDs in lepton-hadron collisions, like open charm production~\cite{Boer:2016fqd,Boer:2011fh,Burton:2012ug} and dijet or high-$p_T$ charged dihadron production~\cite{delCastillo:2020omr,Zheng:2018awe}.
However in hadron-hadron collisions the production of a color neutral final state is a must for the TMD factorization not to be broken by Glauber gluon exchanges~\cite{Collins:2007nk,Collins:2007jp,Rogers:2010dm,Rogers:2013zha,Gaunt:2014ska,Schwartz:2018obd}.
%On the other hand, processes where TMD factorization does not hold could be an opportunity to quantify these effects.
An example of a process which probes directly gluon TMDs in hadron-hadron collisions is the Higgs boson production~\cite{Gao:2005iu,Chiu:2012ir, Echevarria:2015uaa,Neill:2015roa,Gutierrez-Reyes:2019rug}.  
However the extraction of gluon TMDs from its transverse momentum distribution is challenging due to the large mass of the boson and small available statistics.
In other words, genuine non-perturbative TMD effects are somehow hidden in the spectrum and are thus difficult to constrain.
All in all, quarkonium production processes seem to be the best tool at our disposal to probe gluon TMDs, and they have indeed attracted an increasing attention lately ~\cite{Boer:2011fh,Boer:2012bt,Ma:2012hh,Dunnen:2014eta,Zhang:2014vmh,Ma:2015vpt, Boer:2015uqa,Boer:2016fqd,Bain:2016rrv,Mukherjee:2015smo,Mukherjee:2016cjw,Lansberg:2017tlc,Lansberg:2017dzg,Bacchetta:2018ivt,DAlesio:2019qpk,Echevarria:2019ynx,Fleming:2019pzj,Scarpa:2019fol,DAlesio:2019gnu,DAlesio:2020eqo,Grewal:2020hoc,Boer:2020bbd,Echevarria:2020qjk}.

The NRQCD factorization formalism, which is briefly reviewed in Sec.~\ref{quarkonia_th}, can only be applied for transverse momentum spectra when the quarkonium state is produced with a relatively large transverse momentum compared to its mass, i.e. $p_T \sim 2 m_Q$, with $m_Q$ the mass of the heavy quark.
This is because the emissions of soft gluons from the heavy quark pair cannot modify the large transverse momentum of the bound state.
The hard process generates this $p_T$, while the infrared divergences are parameterized in terms of the well-known long-distance matrix elements (LDMEs) and integrated PDFs.

On the contrary, when quarkonium states are produced with a small transverse momentum, soft gluons can no longer be factorized.
Indeed, it was found that for quarkonium photo/lepto-production in the endpoint region~\cite{Beneke:1997qw, Fleming:2003gt, Fleming:2006cd}, processes which are sensitive to soft radiation and where NRQCD approach fails, the LDMEs need to be promoted to shape functions~\cite{Echevarria:2019ynx, Fleming:2019pzj}.
In the same way, in order to properly deal with soft gluon radiation at small $p_T$ in a TMD spectrum of quarkonium production, it has recently been found that one needs to promote the LDMEs to the so-called TMD Shape Functions (TMDShFs)~\cite{Echevarria:2019ynx, Fleming:2019pzj}, which encode the two soft mechanisms present in the process: the formation of the bound-state and soft gluon radiation.

As an example, let us consider the TMD factorization for single quarkonium $\cal H$ production in hadronic collisions, with mass $m_{\cal H}$ and rapidity $y$, which reads
\begin{align}
\label{eq:TMD-shape}
\frac{d\sigma}{dy dp_T} &=
\int\frac{d^2\bm{b}_\perp}{(2\pi)^2} e^{-i (\bm{p}_T\cdot \bm{b}_\perp)}
\sum_{i\in \{^1S_0^{[1]},\ldots \}} 
H^{(i)}(m_{\cal H},y,s;\mu)\,
F_{g/A}(x_A,\bm{b}_{\perp};\mu,\nu)\, 
F_{g/B}(x_B,\bm{b}_{\perp};\mu,\nu)\,
S^{(i)}_{\cal H}(\bm{b}_{\perp},\mu,\nu)
+{O}\Big(\frac{p_T}{m_{\cal H}}\Big)
\,,
\end{align}
where $y$ is the quarkonium rapidity, $x_{A,B}=m_{\cal H}e^{\pm y}/\sqrt{s}$ are the longitudinal momentum fractions, $F_{g/A(B)}$ are the gluon TMD PDFs, $H^{(i)}$ are the process-dependent hard-scattering coefficients and $S^{(i)}_{\cal H}$ are the polarization-independent quarkonium TMDShFs (for more details see Refs.~\cite{Echevarria:2019ynx, Fleming:2019pzj}).  
$\mu$ and $\nu$ are the factorization/resummation and rapidity scales, respectively,  $\bm{b}_{\perp}$ is the Fourier conjugate of $\bm{p}_T$.
The summation is performed over the various colour and angular momentum configurations. 
Similarly to LDMEs, the TMDShFs scale with the relative velocity, $v$, of the heavy quark-antiquark pair in the quarkonium rest frame. 
Therefore, the factorisation formula is a simultaneous expansion in the relative quark-pair velocity $v$ and $\lambda = p_T/m_{\cal H}$.

The factorization theorem contains three or more non-perturbative hadronic quantities at low transverse momenta: gluon TMD PDFs and the TMDShFs. 
Thus, the phenomenological extraction of gluon TMDs from quarkonium production processes is still possible, i.e., a robust factorization theorem can potentially be obtained in any particular case of heavy meson production. 
However one also needs in principle to model and extract the involved TMDShFs.

Finally, a couple of open questions remain with regard to the factorization formula just discussed.
On one hand, the double expansion in $\lambda$ and $v$ allows for, e.g., terms that are suppressed (i.e. sub-leading) in $\lambda$ but at the same time are numerically important due to the enhancement in $v$. 
Although this double expansion introduces theoretical challenges, it is also an opportunity to observe the contributions to the cross-section from sub-leading factorization or from the terms with higher twist.
On the other hand, Glauber gluon exchanges when summing over colour octet final-state channels can spoil the factorization (as in $P$-wave quarkonium production \cite{Ma:2014oha}).
However this actually represents an opportunity to quantify these effects in QCD, which connect long and short distance physics.
Both of these issues are present in studies of inclusive $J/\psi$ and $\Upsilon(1S)$ production in hadronic colliders. 
Anyhow, the production of $\eta_{c/b}$~\cite{Echevarria:2019ynx}, with leading contribution from the channel $^1S_0^{[1]}$, could provide a clean channel for the study of quarkonium TMDShFs.

 \begin{figure}[!h]
  \begin{minipage}[ht]{0.48\linewidth}
        \center{\epsfig{file=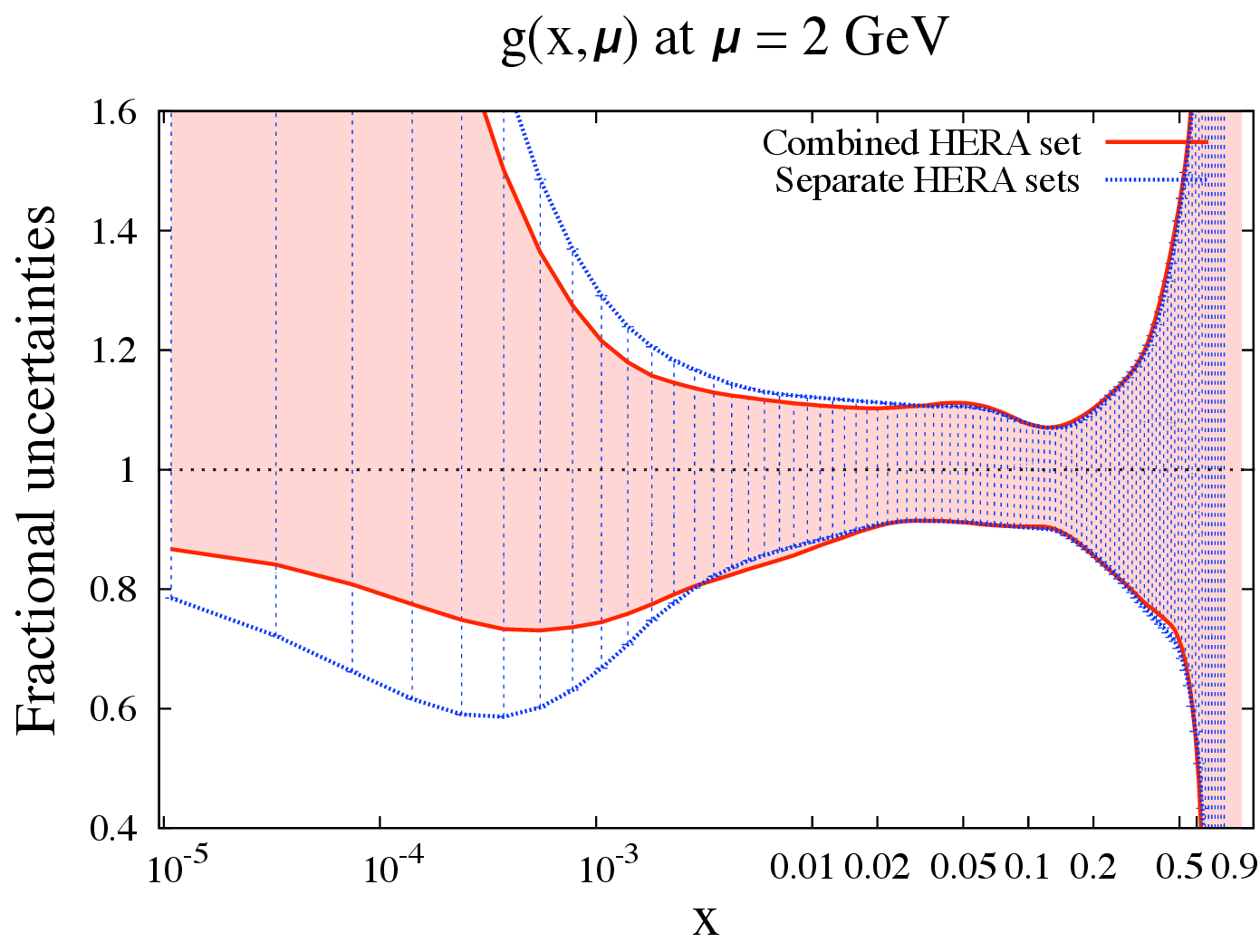,scale=0.31} \\ (a)}  
  \end{minipage}
  \hfill
  \begin{minipage}[ht]{0.48\linewidth}
          \center{\epsfig{file=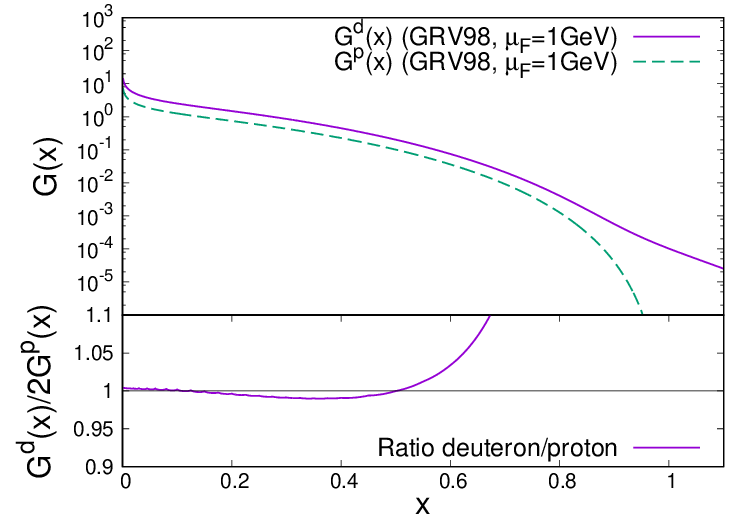,scale=0.63} \\ (b)}  
  \end{minipage}
  \caption{(a) Uncertainty of unpolarized gluon PDF based on HERA data ($\mu=2$ GeV). Reprinted figure with permission from Ref.~\cite{Lai:2010vv}  \textcopyright~(2010) by the American Physical Society. (b) Comparison of the model-prediction for gluon PDF in the deuteron in comparison with the nucleon. Reprinted figure from~\cite{Brodsky:2018zdh} \textcopyright~(2018) by Elsevier. }
  \label{fig:unpolarized} 
\end{figure}

\subsection{Gluon content of the proton spin from Lattice QCD  \label{lqcd}}

There are at least two different ways to define the gluon content in the nucleon~\cite{Jaffe:1989jz,Ji:1996ek}. One is
defined in the light-cone frame with the light-cone gauge and the other is frame-independent. They are both measurable 
but from different experimental observables. They are associated with the two different
decompositions of the proton spin and momentum in terms of the quark and
gluon contributions that can be defined from the matrix elements of the QCD energy-momentum tensor. 
There are, in principle, infinitely-many ways to define this decomposition. A meaningful
decomposition will depend on whether each component in the division can be
measured experimentally. As far as lattice calculations are concerned, it would
be desirable if the components can be defined by matrix elements of local operators.
 
The Jaffe-Manohar decomposition~\cite{Jaffe:1989jz} is given as follows:
\begin{equation}  \label{JM}
J = \frac{1}{2} \Delta \Sigma + L_q^{JM} + \Delta{G} + L_G,
\end{equation}
where $\frac{1}{2}\Delta\Sigma$ ($\Delta{G}$) is the quark (gluon) spin contribution, and
$L_q^{JM}$ ($L_G$) is the quark (gluon) Orbital Angular Momentum (OAM)
contribution. Contributions $\Delta \Sigma$ and $\Delta G$ are related to the corresponding helicity-PDFs as $\Delta \Sigma=\int_0^1 dx\ g_1^q (x)$ and $\Delta G=\int_0^1 dx\ \Delta g(x)$. This is derived from the canonical energy-momentum tensor in the
infinite-momentum frame with $A^0 = 0$ gauge (or light-cone frame with $A^+ =0$ gauge). Thus, this is superficially gauge-dependent. Furthermore, while $\Delta{G}$ can be extracted from high energy experiments, it had been thought that it cannot be obtained from a matrix
element based on a local operator. This has posed a challenge for the lattice approach for many years.

The Ji decomposition~\cite{Ji:1996ek} is:
\begin{equation}  \label{Ji}
J = \frac{1}{2} \Delta \Sigma + L_q^{Ji} + J_G,
\end{equation}
where $ \frac{1}{2} \Delta \Sigma$ is the same quark spin contribution as in
Eq.~(\ref{JM}), $L_q^{Ji}$ is the quark OAM, and $J_G$ is the gluon angular
momentum contribution. This is derived from the energy-momentum tensor in
the Belinfante form and each term in Eq.~(\ref{Ji}) is gauge invariant and can
be calculated on the lattice with local operators in the finite momentum frame.

The intriguing difference between these two decompositions and their respective
realization in experiments have perplexed the community for many years. 
The partonic picture of the gluon spin $\Delta G$ and OAM
are naturally depicted in the light-front formalism with $\Delta G$ extracted from high 
energy $pp$ collisions and OAM from GPDs and GTMDs. 
Unfortunately, the light-front coordinates are not accessible to lattice QCD calculation since the latter is based on Euclidean path-integral formulation. To bridge the gap between the light-front formulation
and the lattice calculation, it was shown that the matrix elements of appropriate equal-time local operator, when boosted to the infinite momentum frame, is the same as  those of the gauge-invariant but non-local operator on the light-cone.
The proof was first carried out for the gluon spin $\Delta G$~\cite{Ji:2013fga}. It is also proven
for the $L_q^{JM}$ and $L_G$ defined from the generalized transverse momentum
distribution (GTMD)~\cite{Zhao:2015kca}. After the usual continuum extrapolation of the lattice results at large but
finite momenta in the $\overline{MS}$ scheme at $\mu$, a large momentum effective field theory (LaMET)
~\cite{Ji:2013fga,Ji:2014gla,Ji:2014lra}, which takes care of the non-commuting UV and $P_z \rightarrow \infty$ limits, is suggested to match the matrix elements of local operators to those of the non-local operators measured on the light-front. 

     It is instructive to have a comparison with QED at this point. Many years of experimental study of paraxial light beam on matter has been able
to distinguish the different manifestation of the spin and OAM of the beam from the radiation-pressure force (a measure of OAM) and the torque (a measure of spin) on the probed dipole particle~\cite{Bliokh:2014ara}. The separation of spin and OAM is based on the canonical energy-momentum tensor in the physical Coulomb gauge. As we learned before, this separation is frame-dependent. Since the light is always in the light-front frame, it is the natural frame to define the spin and OAM of the optical beam. Likewise, boosting the proton to the infinite momentum frame makes the gluons
in the proton behave like the photons in the light beam. The transverse size of the proton, like the width of the light beam,
admits the existence of OAM of the gluon in the proton.
%
%
%The recent analyses~\cite{deFlorian:2014yva,Nocera:2014gqa} of the high-statistics data from STAR~\cite{Djawotho:2013pga} and PHENIX~\cite{Adare:2014hsq} experiments at RHIC, as well as the data from SIDIS-experiments such as COMPASS (see~\cite{Adolph:2015cvj} and the references therein) showed evidence of non-zero gluon helicity in the proton, $\Delta g$ (see the Sec.~\ref{hel} for more details). For $Q^2=10$ GeV$^2$, the gluon helicity distribution $\Delta g(x,Q^2)$ is found to be positive and away from zero in the momentum fraction region $0.05\leq x$ {(specifically $0.05\leq x\leq0.2$, the region in which RHIC can determine $\Delta g(x)$ much better than the other regions)}. However, the results have very large uncertainty in the region $x\le 0.05$.  
Recently, the experimental evidence for non-zero gluon helicity in the proton $\Delta g$ was found, see the Sec.~\ref{hel} of this review for detailed discussion and references. These experimental results support the interest in theoretical calculations of gluon contributions to the proton spin.

Following the suggestion of calculating $\Delta G$ on the lattice as discussed above,
a lattice calculation is carried out with the local operator $\bm{S}_G = \int d^3 x\, Tr (\bm{E}\times \bm{A}_{phys})$,
where $\bm{A}_{phys}$ transforms covariantly under the gauge transformation and satisfies the non-Abelian transverse condition, $\bm{D} \bm{A}_{phys} = 0$~\cite{Chen:2008ag}. Thus $\bm{S}_G$ is gauge invariant. It is shown on the lattice~\cite{Zhao:2015kca} that
$A_{phys}^i$ is related to $A_{c}^i$ in the Coulomb gauge via the gauge transformation, i.e. 
$A_{phys}^{\mu}(x) = g_c (x) A_c^{\mu} g_c^{-1}(x) + \mathcal{O}(a)$, where $g_c$ is the gauge transformation which
fixes the Coulomb gauge. As a result, the gluon spin operator
\begin{equation}
\bm{S}_G = \int d^3 x\, Tr (\bm{E}\times \bm{A}_{phys}) = \int d^3 x\, Tr (\bm{E}_c\times \bm{A}_c)
\end{equation}
can be calculated with both $\bm{E}$ and $\bm{A}$ in the Coulomb gauge.
A lattice calculation with the overlap fermion action is carried out on 5 lattices with 4 lattice spacings and several sea quark masses including one corresponding to the physical pion mass. The result, when extrapolated to the infinite momentum
limit, gives $\Delta G = 0.251(47)(16)$~\cite{Yang:2016plb} which suggests that the gluon spin contribute about 
half of the proton spin. However, there is a caveat. It was found that the finite piece in the one-loop large momentum effective theory (LaMET) matching coefficient is very large which indicates a  convergence problem for the perturbative series even
after one re-sums the large logarithms. Due to above-mentioned perturbative instability of the matching coefficient, the LaMET matching at current stage is not applicable. In this sense, the gluon helicity calculation on the lattice is not completed.

\begin{table}[tbp]
  \vspace*{0.3cm}
  \centering
  \renewcommand{\arraystretch}{1.4}
    \caption{Values of $J_g$ from different lattice studies.}
  \begin{tabular}{|c|c|c|c|}
  \hline
        Pub. & Lattice action  &  $J_g$ & Scale {$(Q^2)$} \bigstrut \\
        \hline
       ~\cite{Deka:2013zha}    & Quenched Fermionic action & 28(8)\% & $4$ GeV$^2$ \bigstrut \\
        \hline
        \multirow{2}{*}{\cite{Alexandrou:2017oeh}}
        & 2-flavor Twisted-mass (+clover term) &  \multirow{2}{*}{27(3)\%} & \multirow{2}{*}{$4$ GeV$^2$} \bigstrut \\
        &  at physical pion mass && \bigstrut\\
        \hline
        Preliminary
        & (2+1)-flavor overlap fermion on Domain Wall & \multirow{2}{*}{39(10)\%} & \multirow{2}{*}{$4$ GeV$^2$} \bigstrut \\
       ~\cite{Yang:2019dha} & sea fermion at $400$~MeV pion mass && \bigstrut \\
       \hline
  \end{tabular}  
  \label{tab:j_g}
\end{table}

\begin{figure}[htbp]
  \begin{minipage}[ht]{0.335\linewidth}
  \center{\epsfig{file=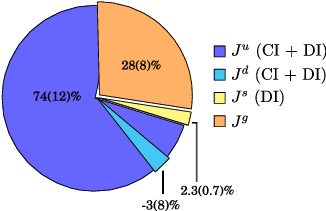,scale=1.} \\ (a)}  
  \end{minipage}
  \begin{minipage}[ht]{0.315\linewidth}
   \center{\epsfig{file=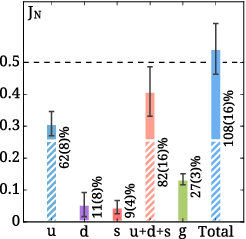,scale=1.} \\ (b)}  
  \end{minipage}
  \begin{minipage}[ht]{0.335\linewidth}
   \center{\epsfig{file=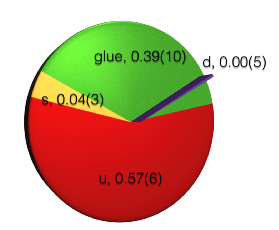,scale=1.} \\ (c)}  
  \end{minipage}
  \caption{The  angular  momentum  fractions  from  the quenched  clover  calculation  (a),  2 flavor Twisted-mass (+clover term) fermion calculation at physical pion mass (b), and the  non-perturbative renormalized 2+1 flavor overlap fermion on Domain Wall sea with 400 MeV pion mass (c). See Refs.~\cite{Deka:2013zha,Alexandrou:2017oeh,Yang:2019dha}. Reprinted figures with permission from Refs. ~\cite{Deka:2013zha,Alexandrou:2017oeh}  \textcopyright~(2015, 2017) by the American Physical Society.} 
  \label{fig:total_ang}
\end{figure}

Another approach is to calculate the polarized gluon distribution function $\Delta g(x)$  through the quasi-PDF approach~\cite{Fan:2018dxu} and then take the first moment to obtain $\Delta G$. 

To summarize existing LQCD studies regarding the gluon angular momentum contribution to the proton spin ($J_G$ in Eq.~(\ref{Ji})) we list obtained values of $J_G$ in Table~\ref{tab:j_g}. Also momentum fraction ($\langle x\rangle_g$ or $T_1(0)$) has been calculated; whereas, the gluon angular momenta are the sum of $T_1(0)$ and $T_2(0)$, the anomalous gravitomagnetic moment. In other words, $J_g = 1/2 \left( T_1(0) + T_2(0)\right)$~\cite{Ji:1996ek}. We also show the percentage of the
$J_G$ contribution to proton spin together with the quark contributions in Fig.~\ref{fig:total_ang}. It is worth emphasizing again that both $\langle x\rangle_g$ and $J_g$ are gauge invariant. 

An exploratory LQCD study of the gluonic structure of the $\phi$ meson is presented in Ref.~\cite{Detmold:2016gpy} that 
sets the stage for more complex studies of gluonic structure in light nuclei with an exotic gluon.

\section{Gluon content of unpolarized proton and deuteron\label{unpol_measurements}}

\subsection{Gluons at large x and perturbative QCD  \label{un_pdf}}

\begin{figure}[ht]
  \begin{minipage}[ht]{0.48\linewidth}
        \center{\epsfig{file=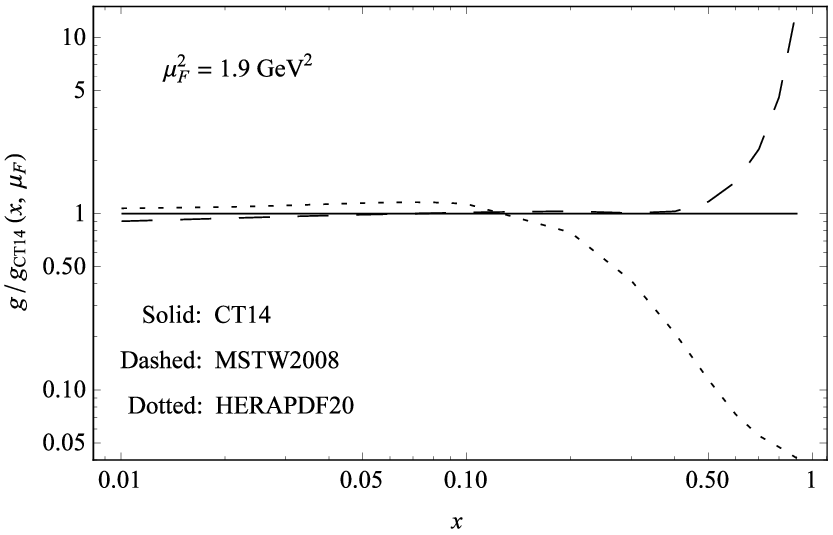,scale=0.53} \\ (a)}  
  \end{minipage}
    \begin{minipage}[ht]{0.48\linewidth}
        \center{\epsfig{file=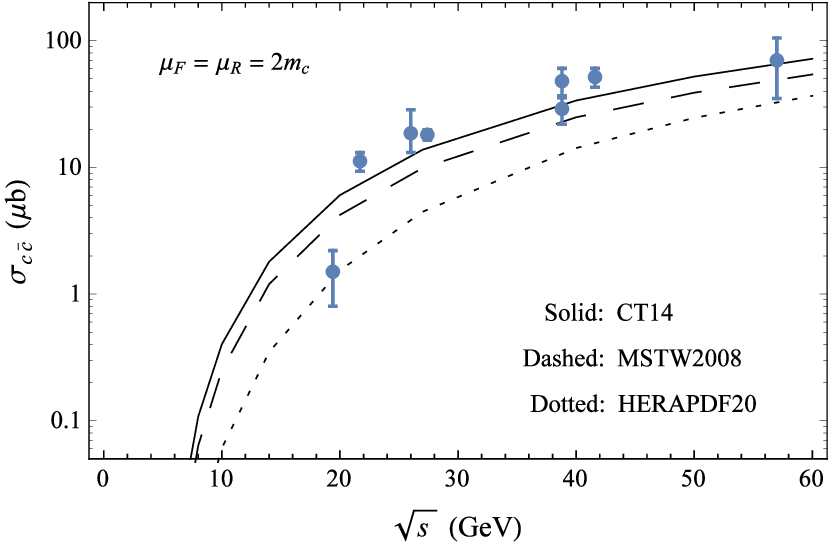,scale=0.53} \\ (b)}  
  \end{minipage}
\caption{(a) Ratios of the NLO MSTW2008 and HERAPDF20 predictions for the gluon density to the CT14 ones, $g/g_{{\rm CT}14}~(x,\mu_F)$, at $\mu_F^2=$1.9 GeV$^2$. (b) NLO QCD predictions for the total $pp\rightarrow c\bar{c}X$ cross-section as a function of $\sqrt{s}$ using CT14, MSTW2008 and HERAPDF20 sets of PDFs at $\mu_F=\mu_R=2m_c$.}
 \label{OCFg.1}
\end{figure}

The available data constrain weakly the gluon distribution function in the proton, $g(x,\mu_F)$\footnote{Which is also referred to as $f^{g} (x,\mu)$ }, for $x$ greater than 0.5 \cite{Abt:2020mnr,Sirunyan:2017azo}. In the high-$x$ region, the gluon density is usually parameterized as  $g(x,\mu_F)\sim (1-x)^L$, and values of $L$ extracted from global fits differ considerably from each other. In particular, obtained results for $L$ vary from 3 to 11 at $\mu_F^2=1.9$ GeV$^2$~\cite{Abdolmaleki:2019tbb}. In Fig.~\ref{OCFg.1}(a), the NLO gluon densities from the CT14, MSTW2008 and HERAPDF20 sets of PDFs are compared. One can see sizable difference between these predictions for large $x$. 
Note also that the uncertainty bands resulting from the PDF fits are quite large in the region $x>0.5$, see e.g. Fig.~\ref{fig:unpolarized}(a). 
%\textcolor{blue}{NI: Be careful, please. Fig.\ref{OCFg.1} demonstrates the ratios $g_A/g_B\sim 10$. Do you see such values in Fig.\ref{fig:unpolarized}(a)? I think, this last comment/addition should be rejected.}

To improve the situation with large $x$, one needs precise data on the heavy flavor production at energies not so far from the production threshold. Concerning the open charm production in $pp$ collisions, the corresponding cross-sections are poorly known for $\sqrt{s}<27$~GeV~\cite{Lourenco:2006vw,Accardi:2016ndt}.\footnote{On the contrary, the $J/\psi$ production cross-section is known well enough practically down to the threshold, see Fig.~\ref{fig:csec}.} 
In this region, only three hundred events on $D$ meson production in $pA$ collisions are presently available \cite{Lourenco:2006vw}. Unfortunately, these results have large uncertainties, and we can only estimate the order of magnitude of the $pp\rightarrow c\bar{c}X$ cross-section at $\sqrt{s}\approx 20$~GeV. For this reason, future studies of the open charm production at SPD in $pp$, $pd$ and $dd$ collisions for $\sqrt{s}\leq 27$~GeV are of special interest. In particular, they will allow to reduce significantly the present uncertainties in the gluon density (as well as in $\alpha_s$ and $c$-quark mass, $m_c$) at a GeV scale, especially for high $x$.

In the framework of pQCD, the inclusive $pp\rightarrow Q\bar{Q}X$ cross-section can be written as 
\begin{align} 
\sigma_{Q\bar{Q}}(\rho_h)&= \sum_{ij}\int_{\rho_h}^1\frac{{\rm d}z}{z}{\cal L}^{ij}(z,\mu_F)\,\hat{\sigma}_{ij}(\rho_h/z,\mu_F,\mu_R),  \label{OC1}\\
{\cal L}^{ij}(z,\mu_F)&=z\int_{0}^1\!{\rm d}x_1\!\int_{0}^1\!{\rm d}x_2\, f^{i}(x_1,\mu_F) f^{j}(x_2,\mu_F)\,\delta(x_1x_2-z), \label{OC1l}
\end{align}
where $i,j$ run over all the initial state partons, $\rho_h=4m^2/s$, $m$ is the heavy-quark mass, $\mu_R$ and $\mu_F$ are the renormalization and factorization scales, while $f^{i}(x,\mu_F)$ describes the parton $i$ density in the proton ($i=g,q,\bar{q}$). The partonic total cross-sections, $\hat{\sigma}_{ij}(\rho)$ with $\rho=4m^2/\hat{s}$ and  $\hat{s}=(p_i+p_j)^2=x_1x_2s$, are expanded in $\alpha_s\equiv\alpha_s(\mu_R)$ as follows:
\begin{align}
\hat{\sigma}_{ij}(\rho,\mu_F,\mu_R)=\frac{\alpha_s^2}{m^2}\Bigg\{ f^{(0)}_{ij}(\rho)&+\alpha_s\left[f^{(1)}_{ij}(\rho)+f^{(1,1)}_{ij}(\rho)\ln(\mu_F^2/m^2) \right] \label{OC2} \\ 
&+\alpha_s^2\left[f^{(2)}_{ij}(\rho)+f^{(2,1)}_{ij}(\rho)\ln(\mu_F^2/m^2)+f^{(2,2)}_{ij}(\rho)\ln^2(\mu_F^2/m^2) \right] +{\cal O}(\alpha_s^3) \Bigg\}. \nonumber  
\end{align}

Main partonic subprocesses of the heavy flavor production are:  gluon fusion, $gg\rightarrow Q\bar{Q}X$, $q\bar{q}$- annihilation, $q\bar{q}\rightarrow Q\bar{Q}X$, and quark-gluon fusion, $qg\rightarrow Q\bar{Q}X$. Presently, the total cross-sections of these processes are known to NNLO in QCD, ${\cal O}(\alpha_s^4)$, see Refs.~\cite{Mitov:2012qq,Mitov:2013qg,Mitov:2013gg}.\footnote{At NNLO, three more fermion–pair–initiated partonic channels contribute to heavy flavor production: $qq\rightarrow Q\bar{Q}X$, $qq^\prime\rightarrow Q\bar{Q}X$ and $q\bar{q}^\prime\rightarrow Q\bar{Q}X$ with $q^\prime\neq q$.  However, these contributions are usually ignored due to their numerically negligible sizes.} The NLO results for $\hat{\sigma}_{gg}$, $\hat{\sigma}_{q\bar{q}}$ and  $\hat{\sigma}_{qg}$ are presented in Ref.~\cite{Nason:1988exact}. At LO in $\alpha_s$, the  gluon fusion and $q\bar{q}$- annihilation only contribute:
\begin{align} \label{OC3}
f^{(0)}_{gg}(\rho)&=\frac{\pi\rho\beta}{192}\Bigg[\frac{1}{\beta}(\rho^2+16\rho+16)\ln\left(\frac{1+\beta}{1-\beta}\right)-28-31\rho \Bigg],& f^{(0)}_{q\bar{q}}(\rho)&=\frac{\pi\rho\beta}{27}(2+\rho),&  f^{(0)}_{qg}(\rho)&=0,
\end{align}
where $\beta=\sqrt{1-\rho}$.

Due to relatively low $c$-quark mass, the charm production cross-sections are usually calculated within pQCD at $\mu_F=\mu_R=2m_c$. Such a choice makes it possible to improve convergence of the  perturbative series for $\sigma_{c\bar{c}}$. Unfortunately, the NNLO predictions for the key gluon fusion process are presently available only for the case of $\mu_F=m$, i.e. the dimensionless coefficients $f^{(2,1)}_{gg}$ and $f^{(2,2)}_{gg}$ are unknown \cite{,Mitov:2013gg}. At the same time, it is well known that the cross-section $\sigma_{c\bar{c}}$ is very sensitive to the choice of the factorization scale in the region of a few GeV. For this reason, although the NNLO predictions \cite{Mitov:2012qq,Mitov:2013qg,Mitov:2013gg} are successfully used in the top quark phenomenology, applicability of these results for description of the charm production is presently questionable. So, we restrict ourselves by consideration of the NLO approximation only. 

In Fig. \ref{OCFg.1}(b), the NLO QCD predictions for the total $pp\rightarrow c\bar{c}X$ cross-section are shown as a function of $\sqrt{s}$ using the CT14, MSTW2008 and HERAPDF20 sets of PDFs at $\mu_F=\mu_R=2m_c$. The experimental data are taken from Ref.~\cite{Lourenco:2006vw}.\footnote{Note that $\sigma_{D\bar{D}}=(0.78\pm 0.03)\,\sigma_{c\bar{c}}$. For more details, see e.g. \cite{Lourenco:2006vw} and references therein.} We see essential differences  between these predictions, especially at low energies. These differences are due not only to the different threshold behavior of the PDF sets we compare. Main source of these uncertainties are different values of the $c$-quark mass used in various analyses of the world data. In particular, the mass $m_c$ varies from 1.3 GeV to 1.47 GeV in the PDF sets presented in Fig.~\ref{OCFg.1}(b).\footnote{In the recent edition of the Review of Particle Physics \cite{Zyla:2020zbs}, the following value for the $c$-quark $\overline{\rm MS}$ mass is presented: $m_c=1.27\pm 0.02$ GeV.} On the other hand, the quantities $\alpha_s(\mu_R)$ and $g(x,\mu_F)$ are very sensitive to the values of $\mu_R$ and $\mu_F$ at a GeV scale. As a result, the pQCD predictions for $\sigma_{c\bar{c}}$ are crucially dependent on the value of $c$-quark mass in use. Therefore future precise data on the total cross-section $\sigma_{c\bar{c}}$ from NICA SPD should reduce uncertainties in the $c$-quark mass and improve essentially description of the processes with charm production within pQCD.

To probe the PDFs at high $x$, one needs to measure the differential cross-sections of the charmed particle production in sufficiently wide region. In particular, the invariant mass of the $c\bar{c}$ pair, $M^2_{c\bar{c}}=(p_c+p_{\bar{c}})^2$, should be large enough because $M^2_{c\bar{c}}\big/s<x_1x_2<1$ for $x_1$ and $x_2$ in Eq.~\eqref{OC1l}. Therefore to scan high values of $x$, we need $M^2_{c\bar{c}}\big/s\sim x$.

Detailed information on the gluon distribution at large $x$ is very important for various phenomenological applications. %\textcolor{red}{[Also for estimates of production cross-section of various heavy ``New Physics'' particles at LHC... M.N.]} 
For instance, it is of current interest to estimate the $b\bar{b}$ pair production cross-section at NICA energies. Such predictions are however presently unreliable due to their strong dependence on the exponent $L$ which is poorly known. 
Likewise, the threshold behavior of PDFs could be useful in search for various heavy particles beyond the Standard Model.  Another example is the DGLAP evolution of the PDFs. Using precise data on $g(x,\mu_F)$ (and $\alpha_s$) at $\mu_F\sim 2m_c$ as boundary conditions in DGLAP equations, one could reduce essentially the uncertainties in evolution of PDFs for higher values of $\mu_F$.

%\textcolor{red}{[I think the discussion of threshold resummation and it's uncertainties should be extended. It is written below, that uncertainties are potentially large. M.N.]}
From the theoretical point of view, the threshold behavior of cross-sections is closely related to the so-called infrared renormalon problem. It is well known that radiative corrections to the production cross-sections contain mass (or threshold) logarithms whose contribution is expected to be sizable near the threshold. These logarithms are usually taken into account within the soft gluon resummation  formalism~\cite{Sterman:1986aj,Catani:1989ne,Contopanagos:1996nh,Kidonakis:1998nf, Ivanov:2001wm}. Formally resummed cross-sections are however ill-defined due to the Landau pole contribution, and few prescriptions have been proposed to avoid the renormalon ambiguities~\cite{Catani:1996yz,Berger:1996ad,Kidonakis:2000ui,Forte:2006mi}. Unfortunately, numerical predictions for heavy quark production cross-sections can depend significantly on the choice of resummation prescription. Undoubtedly, anticipated data from SPD on the charm production not so far from the production threshold could clarify the role of soft gluon resummation in the description of heavy flavor production.

Another interesting problem for NICA SPD is to probe the intrinsic charm (IC) content of the proton~\cite{Brodsky:1980pb,Brodsky:1981se}. The IC contribution to open charm production is expected to be sizable near the threshold because its PDF, $c(x,\mu_F)$, is predicted to be harder than the gluonic one. As a result, the IC density in the proton can be dominant at sufficiently large $x$ independently of its overall normalization~\cite{Ananikyan:2007ef}. To visualize the IC component, one needs to collect enough events like $D\bar{D}$ pair  produced in $pp\rightarrow D\bar{D}$  with a large overall $x_F$. That events are predicted to be very rare within the GF mechanism and would directly indicate the five-quark component in the proton, $|\,uudc\bar{c}\rangle$. %\JPL{A valence like IC, like BHPS, show a peak near $x_c=0.3$. This is in this region that it is most important to look, not much at $x\to 1$.I would suggest to rephrase. Maybe also highlight the complementarity with AFTER@LHC~\cite{Hadjidakis:2018ifr} where the determination of gluon at large $x$ is also very important like the IC. It is important to realise that pQCD computations of charm production show very large uncertainties. A complete PDF study taking into account such uncertainties is therefore needed. It is also important to have a large enough range in $y$ to scan over $x$; this can then be used to mitigate the effect of the scale uncertainties. In this sense, it is important to measure charm also at mid $x$ below the expected IC peak. Finally, and this is a strong case for SPD, the extraction of IC can only be done if gluon at large $x$ (above $x=0.1$) are relatively well known; multiple probes are needed. We inist on that for AFTER@LHC.}
 
Investigation of the open charm production in $pp$, $pd$ and $dd$ collisions might be one of the key points in the NICA SPD program. The motivation is twofold. On the one hand, production of $D$-mesons in $pp$ collisions is practically unmeasured at NICA energies. On the other hand, these presently unavailable data on open charm production rates are strongly necessary for determination of the gluon density $g(x,\mu_F)$ at large $x$ where this PDF is practically unknown. Future SPD measurements should also reduce uncertainties in the charm quark mass $m_c$ and $\alpha_s$ at a GeV scale.

Moreover, anticipated results on the open charm production are very important for many other current issues in particle physics: from infrared renormalon ambiguities in cross-sections to intrinsic charm content of the proton.

\subsection{Linearly polarized gluons in unpolarized nucleon \label{un_BM}}

\subsubsection{Parton-hadron correlators and TMD distribution functions}

Search for the polarized quarks and gluons in unpolarized hadrons is of special interest in studies of the spin-orbit couplings of partons and understanding of the proton spin decomposition. The corresponding intrinsic transverse momentum dependent distributions of the transversely polarized quarks, $h_{1}^{\perp q}(x,\ktsq)$, and linearly polarized gluons, $h_{1}^{\perp g}(x,\ktsq)$, in an unpolarized nucleon have been introduced in Refs.~\cite{Boer:1997nt} and \cite{Mulders:2000sh}. Unfortunately, both these functions are presently unmeasured.

Information about parton densities in unpolarized nucleon is formally encoded in corresponding TMD parton correlators. The parton correlators describe the nucleon $\rightarrow$ parton transitions and are defined as matrix elements on the light-front (LF) (i.e., $\lambda^{+}=\lambda\cdot n=n^2=0$, where $n$ is a light-like vector conjugate to the nucleon momentum, $P\cdot n=1$).
%\textcolor{red}{[M.N. Not sure that definition of light-front is correct here. I use the following decomposition for $P=P^+n^- + m_N n^+/(2P^+)$, where $(n^{\pm})^2=0, n^+n^-=1$. Then light-front is indeed a surface $\lambda^+=\lambda\cdot n^+=0$ and integration in Eq. (\ref{TMD1}) goes over $\lambda^-$.]}  \textcolor{blue}{[NI: Be sure, all is ok here. These notations are widely used. Replacing $P^+n^-\rightarrow n^+$ and $n^+/P^+\rightarrow n^-$, you will obtain yours results up to ${\cal O}(1/P^+)$].}
At leading twist and omitting process-dependent gauge links that ensure gauge invariance, the quark correlator is given by \cite{Boer:1997nt,Pisano:2013cya} 
\begin{eqnarray} \label{TMD1}
\Phi_q(x, \ktv)&=& \int \frac{{\rm d}(P\cdot\lambda){\rm d}^2\bm{\lambda}_T}{(2\pi)^3}e^{ik\cdot\lambda}\left <P|\bar{\psi}(0)\psi(\lambda) |P\right >_{\rm LF} \nonumber \\ 
&=&\frac{1}{2}\left\{f_{1}^{q}\big(x,\ktsq\big)\hat{P}+i h_{1}^{\perp q}\big(x,\ktsq\big)\frac{[\hat{k}_{T},\hat{P}]}{2m_N}\right\},
\end{eqnarray}
where $m_N$ and $k=x P+k_T+k^{-}n$.
%\textcolor{red}{[$k=xP+k_T+k^{-}n^+$ in my notation, M.N.]} \textcolor{blue}{[NI: Ok, see previous comment]}
are the nucleon mass and parton momentum, respectively. $f_{1}^{q}\big(x,\ktsq\big)$ denotes the TMD distribution of unpolarized quarks inside  unpolarized nucleon. Its integration over $\ktv$ reproduces the well-known collinear momentum distribution, $\int {\rm d}^2\bm{k}_T f_{1}^{q}\big(x,\ktsq\big)=q(x)$.
The function $h_{1}^{\perp q}\big(x,\ktsq\big)$, referred to as Boer-Mulders PDF, is time-reversal (T-) odd and describes the distribution of transversely polarized quarks
inside an unpolarized nucleon. Similarly, for an antiquark correlator we have:
\begin{eqnarray} \label{TMD2}
\Phi_{\bar{q}}(x, \ktv)&=& -\int \frac{{\rm d}(P\cdot\lambda){\rm d}^2\bm{\lambda}_T}{(2\pi)^3}e^{-ik\cdot\lambda}\left <P|\bar{\psi}(0)\psi(\lambda) |P\right >_{\rm LF} \nonumber \\ 
&=&\frac{1}{2}\left\{f_{1}^{\bar{q}}\big(x,\ktsq\big)\hat{P}+i h_{1}^{\perp \bar{q}}\big(x,\ktsq\big)\frac{[\hat{k}_{T},\hat{P}]}{2m_N}\right\}.
\end{eqnarray} 
Omitting process-dependent gauge links, the gluon-nucleon correlator has the form \cite{Mulders:2000sh}:
\begin{eqnarray} \label{TMD3}
\Phi_g^{\mu\nu}(x, \ktv)&=& \frac{n_{\rho}n_{\sigma}}{(P\cdot n)^2}\int \frac{{\rm d}(P\cdot\lambda){\rm d}^2\bm{\lambda}_T}{(2\pi)^3}e^{ik\cdot\lambda}\left <P|{\rm Tr}[F^{\rho\mu}(0)F^{\sigma\nu}(\lambda)] |P\right >_{\rm LF} \nonumber \\
&=& \frac{x}{2}\left\{- g_T^{\mu\nu}f_{1}^{g}\big(x,\ktsq\big)+\left(g_T^{\mu\nu}-
2\frac{k_T^\mu k_T^\nu}{k_T^2}\right)\frac{\ktsq}{2m^2_N}h_{1}^{\perp g}\big(x,\ktsq\big)\right\},
\end{eqnarray}
with the gluon field strength $F^{\mu\nu}_a(x)$ and a transverse metric tensor $g_T^{\mu\nu}=g^{\mu\nu}- \frac{P^\mu n^\nu+P^\nu n^\mu}{P\cdot n}$. 

The TMD PDF $f_{1}^{g}\big(x,\ktsq\big)$ describes the distribution of unpolarized gluons inside an unpolarized nucleon, and, integrated over $\ktv$, gives the familiar gluon density, $\int {\rm d}^2\bm{k}_T f_{1}^{g}\big(x,\ktsq\big)=g(x)$.
The distribution $h_{1}^{\perp g}\big(x,\ktsq\big)$ is gluonic counterpart of the Boer–  Mulders function; it describes the linear polarization of gluons inside unpolarized nucleon. The degree of their linear polarization is determined by the quantity $r=\frac{\ktsq h_{1}^{\perp g}}{2m^2_N f_{1}^{g}}$. In particular, the gluons are completely polarized along the $\ktv$ direction at $r=1$. Note also that the TMD densities  under consideration have to satisfy the positivity bound \cite{Mulders:2000sh}: 
\begin{equation} \label{TMDbound}
\frac{\ktsq}{2m^2_N}\big|h_{1}^{\perp g}(x,\ktsq)\big|\leq f^g_{1}(x,\ktsq).
\end{equation} 

To probe the TMD distributions, the momenta of both heavy quark and anti-quark, $\bm{p}_{Q}$ and $\bm{p}_{\bar{Q}}$, should be measured (reconstructed). For further analysis, the sum and difference of the transverse heavy quark momenta are introduced,
\begin{align} \label{TMD4}
\bm{K}_{\perp}&=\frac{1}{2}\left(\bm{p}_{Q\perp}-\bm{p}_{\bar{Q}\perp}\right), &
\bm{q}_{T}&=\bm{p}_{Q\perp}+\bm{p}_{\bar{Q}\perp},
\end{align}
in the plane orthogonal to the collision axis. The azimuthal angles of $\bm{K}_{\perp}$ and $\bm{q}_{T}$ are denoted by $\phi_{\perp}$ and $\phi_{T}$, respectively.

\subsubsection{Probing the density $h_{1}^{\perp g}$ with heavy flavor production}

Contrary to its quark version, the TMD density $h_{1}^{\perp g}$ is $T$- and chiral-even, and thus can directly be probed in certain experiments. Azimuthal correlations in heavy-quark pair production in unpolarized $ep$ and $pp$ collisions as probes of the linearly polarized gluons have been considered in Refs.~\cite{Boer:2010zf,Pisano:2013cya}. For the case of DIS,
the complete angular structure of the pair production cross-section has been obtained in terms of seven azimuthal modulations containing the angles $\phi_{\perp}$ and $\phi_{T}$. However, in the considered kinematics, only two of these modulations are really independent; they can be chosen as the $\cos \varphi$ and $\cos 2\varphi$ distributions, where $\varphi$ is the heavy quark azimuthal angle \cite{Efremov:2017ftw}.  Integrating over the anti-quark azimuth, the following result for the reaction $l(\ell)+N(P)\rightarrow l^{\prime}(\ell -q)+Q(p_Q)+\bar{Q}(p_{\bar{Q}})+X$ was obtained:
\begin{equation} \label{TMD5}
{\rm d}\sigma_{lp}\propto\left[1+(1-y)^2 \right]{\rm d}\sigma_{2}-y^2{\rm d}\sigma_{L} 
+2(1-y)\,{\rm d}\sigma_{A}\cos2\varphi +(2-y)\sqrt{1-y}\,{\rm d}\sigma_{I}\cos\varphi, 
\end{equation}
where $\sigma_2=\sigma_T+\sigma_L$, $x,y,Q^2$ are usual Bjorken variables, $m$ is the heavy-quark mass and $z=\frac{p_Q\cdot P}{q\cdot P}$. The contributions of the linearly polarized gluons to ${\rm d}\sigma_{i}$ $(i=2,L,A,I)$ were analyzed in Refs. \cite{Ivanov:2018tvg,Efremov:2018myn}.

\begin{figure}[ht]
  \begin{minipage}[ht]{0.48\linewidth}
        \center{\epsfig{file=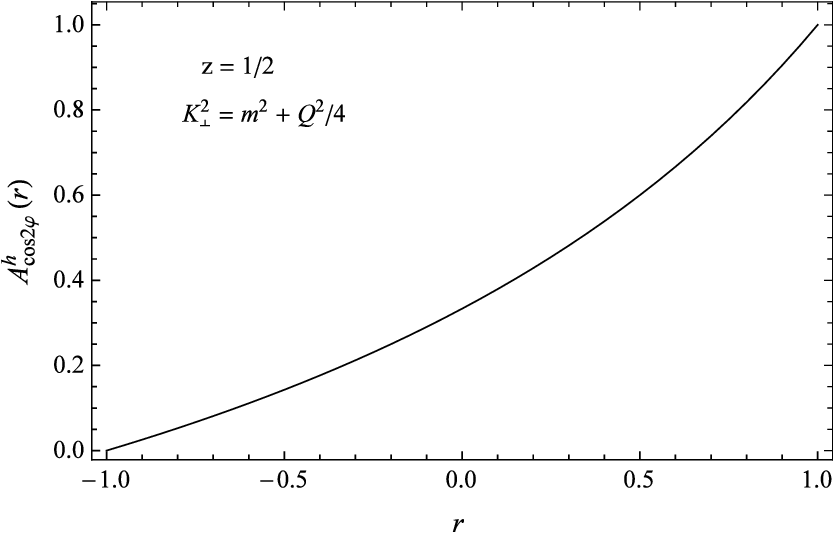,scale=0.53} \\ (a)}  
  \end{minipage}
    \begin{minipage}[ht]{0.48\linewidth}
        \center{\epsfig{file=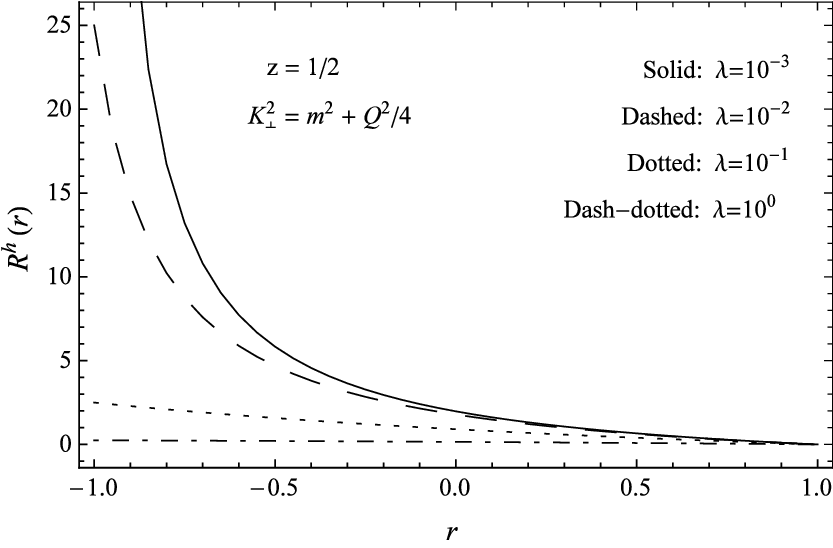,scale=0.53} \\ (b)} 
  \end{minipage}
\caption{(a) Maximum value of the $\cos2\varphi$ asymmetry with the contribution of linearly polarized gluons, $A^h_{\cos2\varphi}(r)$, as a function of $r$. (b) Maximum value of the Callan-Gross ratio, $R^h(r)$, with the contribution of linearly polarized gluons as a function of $r$. The figures are reprinted from Refs. \cite{Ivanov:2018tvg,Efremov:2018myn} \textcopyright~(2018) by Elsevier.}
\label{TMDFg.1}
\end{figure}

First, the LO predictions for the azimuthal asymmetries ($A_{\cos\varphi}={\rm d}\sigma_I/{\rm d}\sigma_2$ and  $A_{\cos2\varphi}={\rm d}\sigma_A/{\rm d}\sigma_2$) and  Callan-Gross ratio ($R={\rm d}\sigma_L/{\rm d}\sigma_T$) were discussed in the case when the unpolarized gluons only contribute, i.e. for $r=\frac{\bm{q}_{T}^2 h_{1}^{\perp g}}{2m^2_N f_{1}^{g}}=0$. It was shown that the maximal values of these quantities allowed by the photon-gluon fusion with unpolarized initial gluons are sizable: $A_{\cos 2\varphi}\le 1/3$, $|A_{\cos\varphi}|\le (\sqrt{3}-1)/2\simeq 0.366$ and $R\le \frac{2}{1+12\lambda}$ with $\lambda=m^2/Q^2$. 
Then the contributions of the linearly polarized gluons to the above distributions were considered. As shown in Refs. \cite{Ivanov:2018tvg,Efremov:2018myn}, the maximal values of the discussed quantities, $A^h_{\cos 2\varphi}$, $A^h_{\cos \varphi}$ and $R^h$, are very sensitive to the gluon density $h_{1}^{\perp g}$. In particular,  $A^h_{\cos 2\varphi}(r)=\frac{1+r}{3-r}$ and $R^h(r)=\frac{2(1-r)}{1+r+4\lambda (3-r)}$. The functions  $A^h_{\cos2\varphi}(r)$ and $R^h(r)$ are depicted in  Figs.~\ref{TMDFg.1}(a) and \ref{TMDFg.1}(b) where their strong dependence on $r$ is seen. Note also an unlimited growth of the Callan-Gross ratio with $Q^2$ at $r\rightarrow-1$. This is because the transverse cross-section, ${\rm d}\sigma_T$, vanishes at high $Q^2$ for $r\rightarrow-1$.

So, we see that the azimuthal distributions and Callan-Gross ratio in heavy-quark leptoproduction are predicted to be large and very sensitive to the gluon density $h_{1}^{\perp g}$.
Taking into account that the $\cos 2\varphi$ asymmetry and Callan-Gross ratio are well defined in pQCD (they are stable both perturbatively and parametrically \cite{Ivanov:2003dis,Ivanov:2009epj}), one can conclude that the considered quantities could be good probes of the linearly polarized gluons in unpolarized proton.

\subsubsection{Azimuthal distributions in heavy-quark hadroproduction}

%\JPL{I am sorry but this does not work. There is a detailed discussion in the previous section of what TMD factorisation implies, i.e. the introduction of soft functions, that one should be careful, etc., and now a process which DOES NOT TMD factorise is discussed. This is NOT coherent. Either the discussion in the beginning is made wider, introducing how processes which are a priori not factorisable can studied and used, or at the beginning of this section, BIG WARNINGS should be written. Otherwise this can only introduce confusion. {\blue NI: I could say more: the charm production IS NOT factorizable even in the collinear approximation (except for large $p_T$) due to small $c-$quark mass. Is this the reason to forbid a using of pQCD for estimates of the charm  cross-sections? It seems that the people involved in TMD business overestimate  seriously the role of factorization theorems in the colliders life...}}
In this Section, we discuss the contributions of the gluon density $h_{1}^{\perp g}$ to the azimuthal modulations anticipated in the charm quark pair production in $pp$ collisions at NICA SPD. Main goal is to estimate the upper bounds of theoretical expectations for these azimuthal distributions. In our analysis, the effects of initial and final state interactions (and, correspondingly, the problems with factorization) are ignored.\footnote{For the details related to TMD factorization, see Sec.~\ref{un_tmd}.} We expect that (possible) factorization breaking contributions are not able to affect essentially the parton model predictions for the order of magnitudes of the azimuthal asymmetries under consideration. 
%Even if this process is not TMD factorizable (for more details see section \ref{un_tmd}), we rely on the Generalized Parton Model in order to estimate the role of $h_{1}^{\perp g}$ and assess its measurability. Moreover, this process can be used to test TMD factorization breaking effects.

The angular structure of the $p(P_1)+p(P_2)\rightarrow Q(p_Q)+\bar{Q}(p_{\bar{Q}})+X$ cross-section has the following form \cite{Boer:2010zf,Pisano:2013cya}: 
\begin{equation} \label{TMD15}
{\rm d}\sigma_{pp}\propto A(\qtsq) + B(\qtsq) \qtsq \cos2(\phi_{\perp}-\phi_{T})+ C(\qtsq) \bm{q}_T^4 \cos4(\phi_{\perp}-\phi_{T}).
\end{equation}
Besides $\qtsq$, the terms $A$, $B$ and $C$ depend on other kinematic variables not explicitly shown: $\bm{K}_{\perp}=\frac{1}{2}\left(\bm{p}_{Q\perp}-\bm{p}_{\bar{Q}\perp}\right)$ and $M^2_{Q\bar{Q}}=(p_Q+p_{\bar{Q}})^2$. At LO in pQCD, ${\cal O}(\alpha_s^2)$, the quantities $A$ and $B$ are given by the sum of the gluon fusion, $gg\rightarrow Q\bar{Q}$, and $q\bar{q}$- annihilation, $q\bar{q}\rightarrow Q\bar{Q}$, subprocesses: $A=A^{gg}+A^{q\bar{q}}$ and $B=B^{gg}+B^{q\bar{q}}$. In particular,
\begin{eqnarray} \label{TMD16}
A^{q\bar{q}}&=&A_0\left(z,\frac{m^2}{m^2_{\perp}}\right)\left\{{\cal C}\left[f_1^q\, f_1^{\bar{q}}\right]+{\cal C}\left[ f_1^{\bar{q}}\, f_1^q \right] \right\}, \\
A^{gg}&=&A_1\left(z,\frac{m^2}{m^2_{\perp}}\right){\cal C}\left[f_1^g\, f_1^g\right]+ A_2\left(z,\frac{m^2}{m^2_{\perp}}\right){\cal C}\left[w_0\, h_1^{\perp g}\, h_1^{\perp g}\right], \nonumber
\end{eqnarray}
where the variables $m^2_{\perp}=m^2+\bm{K}^2_{\perp}$ and $z$ are related to $M^2_{Q\bar{Q}}$ as follows: $z(1-z)M^2_{Q\bar{Q}}=m^2_{\perp}$.

The TMD convolutions ${\cal C}\left[w\, f\, g\right]$ are defined as
\begin{equation} \label{TMD17}
{\cal C}\left[w\, f\, g\right]=\int{\rm d}^2\bm{k}_{1T}\int{\rm d}^2\bm{k}_{2T} \delta^2(\bm{k}_{1T}+\bm{k}_{2T}-\bm{q}_{T})w(\bm{k}_{1T},\bm{k}_{2T})f(x_1,\bm{k}^2_{1T},\mu_F)g(x_2,\bm{k}^2_{2T},\mu_F),
\end{equation}
where, at LO in $\alpha_s$, the light-cone momentum fractions $x_{1,2}$ are related to the rapidities $y_{1,2}$ and $s=(P_1+P_2)^2$ by $x_{1,2}=\frac{m_{\perp}}{\sqrt{s}}(e^{\pm y_1}+e^{\pm y_2})$. In the considered reactions, the following weights are used:
\begin{align}
w_0&=\frac{1}{m^4_N}\Big[ 2(\bm{k}_{1T}\cdot \bm{k}_{2T})^2-\bm{k}^2_{1T}\bm{k}^2_{2T} \Big],& w_1&=\frac{1}{m^2_N}\Big[ 2\frac{(\bm{q}_{T}\cdot \bm{k}_{1T})^2}{\qtsq}-\bm{k}^2_{1T} \Big],\label{TMD18} \\
w_3&=\frac{1}{m^2_N}\Big[2\frac{(\bm{q}_{T}\cdot \bm{k}_{1T})(\bm{q}_{T}\cdot \bm{k}_{2T})}{\qtsq}-(\bm{k}_{1T}\cdot \bm{k}_{2T}) \Big],& w_2&=\frac{1}{m^2_N}\Big[ 2\frac{(\bm{q}_{T}\cdot \bm{k}_{2T})^2}{\qtsq}-\bm{k}^2_{2T} \Big],\nonumber
\end{align}
and $w_4=2w^2_3-\bm{k}^2_{1T}\bm{k}^2_{2T}/m^4_N$.  Analogously to Eqs.(\ref{TMD16}), we have
\begin{eqnarray} \label{TMD19}
\qtsq B^{q\bar{q}}&=&B_0\left(z,\frac{m^2}{m^2_{\perp}}\right)\left\{{\cal C}\left[w_3\,h_1^{\perp q}\, h_1^{\perp\bar{q}}\right]+{\cal C}\left[w_3\,h_1^{\perp\bar{q}}\, h_1^{\perp q} \right] \right\}, \\
\qtsq B^{gg}&=&B_1\left(z,\frac{m^2}{m^2_{\perp}}\right)\left\{{\cal C}\left[w_1\, h_1^{\perp g}\, f_1^g\right]+{\cal C}\left[w_2\, f_1^{g}\, h_1^{\perp g}\right]\right\}. \nonumber
\end{eqnarray}
Finally, the $\cos4(\phi_{\perp}-\phi_{T})$ angular distribution of the $Q\bar{Q}$ pair is only due to the presence of linearly polarized gluons inside unpolarized nucleon, $C=C^{gg}$:
\begin{equation}\label{TMD20}
\bm{q}^4_{T}C^{gg}=C_1\left(z,\frac{m^2}{m^2_{\perp}}\right){\cal C}\left[w_4\, h_1^{\perp g}\, h_1^{\perp g}\right].
\end{equation}

The order $\alpha_s^2$ predictions for the coefficients $A_{0,1,2}$, $B_{0,1}$ and $C_1$ in Eqs.\eqref{TMD16}, \eqref{TMD19} and \eqref{TMD20} are presented in Ref.\cite{Pisano:2013cya}. Using these results, one can, in principle, extract the densities $h_{1}^{\perp q}(x,\ktsq)$ and $h_{1}^{\perp g}(x,\ktsq)$ from azimuthal distributions of the $D\bar{D}$ pairs produced in $pp$ collisions.

Before discussing the anticipated values of the azimuthal asymmetries, we need to resort to models for description of the TMD distributions. It is well known that the heavy flavor production is dominated by the $gg\rightarrow Q\bar{Q}$ subprocess. For this reason, it seems reasonable to begin from the contribution of the gluon fusion mechanism. Usually the TMD density of unpolarized gluons is taken to be a Gaussian:\footnote{For more details, see Ref.\cite{Collins:2015tmd} and references therein.}
\begin{equation}\label{TMD21}
f_1^g(x,\ktsq,\mu_F)=\frac{g(x,\mu_F)}{\pi \langle k_{T}^2\rangle}\exp\bigg(-\frac{\ktsq}{\langle k_{T}^2\rangle} \bigg),
\end{equation}
where $g(x,\mu_F)$ is the collinear gluon PDF and $\langle k_{T}^2\rangle$ depends implicitly on the scale $\mu_F$.

For description of the linearly polarized gluons in unpolarized proton, we use the parameterization motivated by previous TMD studies \cite{Boer:2012bt}:
\begin{equation}\label{TMD22}
h_1^{\perp g}(x,\ktsq,\mu_F)=\frac{2m_N^2}{\langle k_{T}^2\rangle}\frac{(1-\kappa)}{\kappa}\frac{g(x,\mu_F)}{\pi \langle k_{T}^2\rangle}\exp\bigg(1-\frac{\ktsq}{\kappa\langle k_{T}^2\rangle} \bigg).
\end{equation}
The model-independent positivity bound (\ref{TMDbound}) is satisfied by Eqs.(\ref{TMD21}) and (\ref{TMD22}) with $\kappa<1$. Usually the value $\kappa=2/3$ is used that maximizes the second $\ktv$ moment of $h_1^{\perp g}$. With the choice (\ref{TMD22}), all the TMD convolutions ${\cal C}\left[w_i\, f\, g\right]$ ($i=0$--4) with gluon PDFs are simple analytical functions.

Another widely used parameterization of the PDF for linearly polarized gluons is
\begin{equation}\label{TMD23}
h_1^{\perp g}(x,\ktsq,\mu_F)=\frac{2m_N^2}{\ktsq}f_1^{g}(x,\ktsq,\mu_F).
\end{equation}
With this choice, one can estimate the upper bounds of the expected values of the azimuthal asymmetries. Throughout this subsection, "Model 1" will refer to the Gaussian form (\ref{TMD22}) with $\kappa=2/3$ and "Model 2" to the form (\ref{TMD23}) saturating the positivity bound.
\begin{figure}[ht]
  \begin{minipage}[ht]{0.48\linewidth}
        \center{\epsfig{file=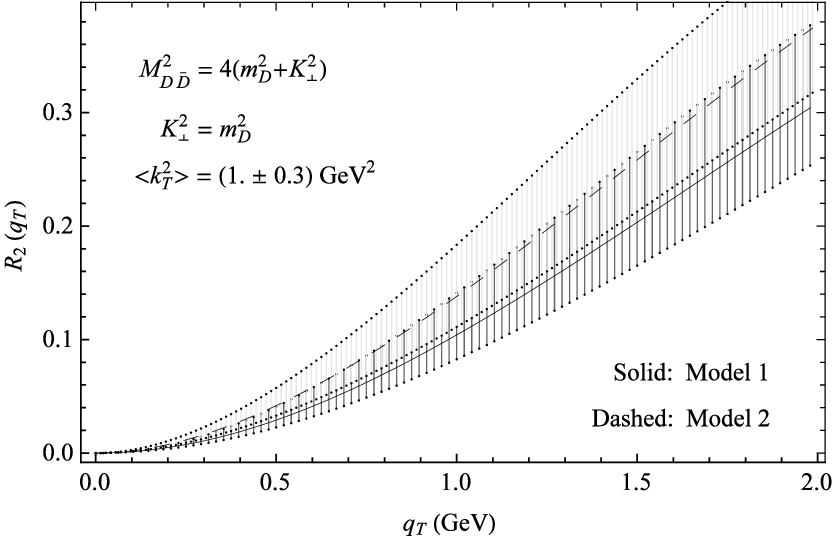,scale=0.53} \\ (a)}  
  \end{minipage}
    \begin{minipage}[ht]{0.48\linewidth}
        \center{\epsfig{file=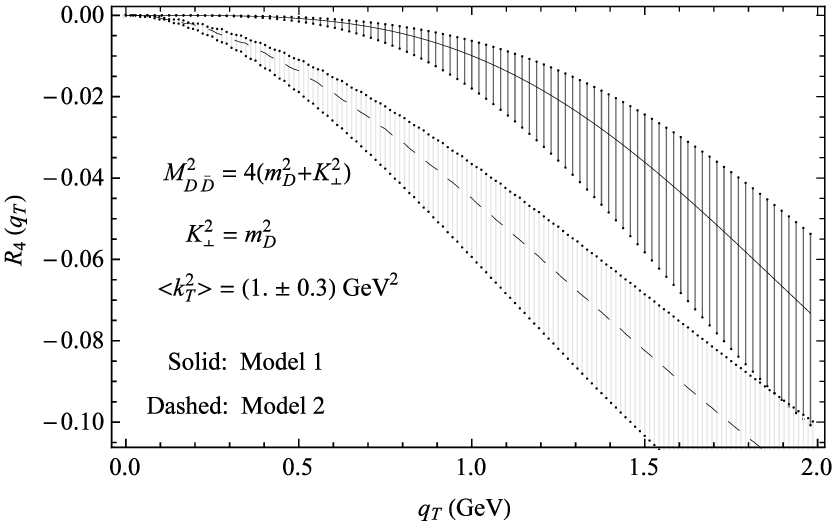,scale=0.55} \\ (b)} 
  \end{minipage}
\caption{(a) Gluon fusion contribution to the asymmetry $R_2(q_T)=\frac{B\bm{q}^2_T}{A}$ in $pp\rightarrow D\bar{D}X$ process at $M^2_{D\bar{D}}=4(K^2_{\perp}+m_D^2)$ and $K^2_{\perp}=m^2_D$. (b) Gluon fusion contribution to the asymmetry $R_4(q_T)=\frac{C\bm{q}^4_T}{A}$ in $pp\rightarrow D\bar{D}X$ process at $M^2_{D\bar{D}}=4(K^2_{\perp}+m_D^2)$ and $K^2_{\perp}=m^2_D$.}
\label{TMDFg.5}
\end{figure}

Let us discuss the gluon fusion contributions to the azimuthal asymmetries defined as
\begin{align} \label{TMD24}
R_2(q_T)&=\frac{B\bm{q}^2_T}{A}=2\langle \cos2(\phi_{\perp}-\phi_T)\rangle,& R_4(q_T)&=\frac{C\bm{q}^4_T}{A}=2\langle \cos4(\phi_{\perp}-\phi_T)\rangle.
\end{align}
Our analysis of the functions $A_{1,2}\left(z,m^2/m^2_{\perp}\right)$ and $B_{1}\left(z,m^2/m^2_{\perp}\right)$ shows that the asymmetry $R_2\left(q_T,z,m^2/m^2_{\perp}\right)$ has an extremum at $z=1/2$ and $m^2/m^2_{\perp}=1/2$ (i.e., at $M^2_{Q\bar{Q}}=4(m^2+K^2_{\perp})$ and $K^2_{\perp}=m^2$). This extremum value is $\frac{{\cal C}[w_1\, h_1^{\perp g}\, f_1^g]}{3{\cal C}[f_1^g\, f_1^g]-(1/8){\cal C}[w_0\, h_1^{\perp g}\, h_1^{\perp g}]}$. One can see that the function $R_2$ has a maximum at $z=m^2/m^2_{\perp}=1/2$ for all positive values of the convolution ${\cal C}\big[w_1\, h_1^{\perp g}\, f_1^g\big]$ independently of the form of $h_1^{\perp g}$ and $f_1^g$ densities we use.

The gluon fusion predictions for the asymmetry $R_2(q_T)=2\langle \cos2(\phi_{\perp}-\phi_T)\rangle$ in $pp\rightarrow D\bar{D}X$ reaction at $M^2_{D\bar{D}}=4(K^2_{\perp}+m_D^2)$ and $K^2_{\perp}=m^2_D$ are presented in Fig.\ref{TMDFg.5}(a). The solid line corresponds to the Model 1 described by the gluon density (\ref{TMD22}) with $\kappa=2/3$. The dashed line shows the predictions of the Model 2 saturating the positivity bound (\ref{TMDbound}). Both solid and dashed lines are given for  $\langle k_{T}^2\rangle$=1.0 GeV$^2$. The uncertainty bands result from the variation of $\langle k_{T}^2\rangle$ by 0.3 GeV$^2$. Note that the dashed line (Model 2) describes the largest $\cos2(\phi_{\perp}-\phi_T)$ asymmetry allowed by the gluon fusion mechanism.

Our analysis of the quantity $C_{1}\left(z,m^2/m^2_{\perp}\right)$ indicates that the distribution $R_4\left(q_T,z,m^2/m^2_{\perp}\right)$ takes its  maximum value at $M^2_{Q\bar{Q}}\rightarrow s$, i.e. for $x_1 x_2\rightarrow 1$. However, the heavy flavor production rates vanish at the threshold. For this reason, the gluon fusion predictions for the asymmetry $R_4(q_T)=2\langle \cos4(\phi_{\perp}-\phi_T)\rangle$ are also given in Fig.\ref{TMDFg.5}(b) at $M^2_{D\bar{D}}=4(K^2_{\perp}+m_D^2)$ and $K^2_{\perp}=m^2_D$. One can see from Figs.\ref{TMDFg.5}(a) and \ref{TMDFg.5}(b) that both the $R_2$ and $R_4$ distributions are predicted to be sizable (of the order of 10\%) within considered models for $q_T\gtrsim 1$ GeV.

Another processes proposed to probe the linearly polarized gluons in unpolarized proton are:  pseudoscalar $C$-even quarkonia (such as $\eta_c$ and $\chi_c$) \cite{Boer:2012bt,Echevarria:2019ynx}, di--gamma ($pp\rightarrow \gamma\gamma X$) \cite{Qiu:2011ai},  $J/\psi$-- pair ($pp\rightarrow J/\psi\, J/\psi\, X$) \cite{Lansberg:2017dzg,Scarpa:2019fol} and $J/\psi\, \gamma$ ($pp\rightarrow J/\psi\, \gamma\, X$) \cite{Dunnen:2014eta} production. 
%\JPL{In general, $J/\psi+\gamma$~\cite{Dunnen:2014eta} should be added, probably not at SPD though, but there is no mention of SPD in this paragraph.}  
These reactions are TMD factorizable but, unfortunately,  strongly suppressed in comparison with $pp\rightarrow D\bar{D}X$. 
%\JPL{They are maybe {\it suppressed} but they are in principle TMD factorisable, not this one !!!!!!! I am sorry that is not okay. In this context, why not to discuss single $J/\psi$ like in ~\cite{Mukherjee:2016cjw} or $J/\psi$+jet, which also non factorisable ??  All this part of the review should be reconsidered. \blue{NI: I agree- there is a lot of beautiful, TMD factorizable processes. However, whether they are measurable at SPD ?? Note once more: this is not a review on TMD factorization !!!!!!!! Here we discuss the things which can in  principle be measured at SPD...}}

\subsection{Non-nucleonic degrees of freedom in deuteron \label{un_non}}
The naive model describes the deuteron as a weakly-bound state of a proton and a neutron mainly in S-state with a small admixture of the D-state. However, such a simplified picture failed to describe the HERMES experimental results on the $b_1$ structure function of the deuteron~\cite{Airapetian:2005cb}. Modern models treat the deuteron as a six-quark state with the wave function 
\begin{equation}
|6q\rangle =  c_1 | NN\rangle + c_2 |\Delta \Delta \rangle + c_3 | CC \rangle,
\end{equation}
 that contains such terms as the nucleon $| NN\rangle$, $\Delta$-resonance $|\Delta\Delta\rangle$  and the so-called hidden color component $| CC \rangle$ in which two color-octet baryons combine to form a color singlet~\cite{Harvey:1988nk}. Such configurations can be generated, for example, if two nucleons exchange a single gluon. The relative contribution of the hidden-color term varies from about 0.1\% to 80\% in different models~\cite{Miller:2013hla}. The components other than $|NN\rangle$ should manifest themselves in the high-$Q^2$ limit. Possible contributions of the Fock states with a valent gluon like $|uuudddg\rangle$ could also be discussed~\cite{Hoyer:1997rh,Brodsky:2018zdh}.

 The unpolarized gluon PDF of the deuteron in the light-front quantization was calculated in the Ref.~\cite{Brodsky:2018zdh} under the approximation where the input nuclear wave function is obtained by solving the nonrelativistic Schrödinger equation with the phenomenological Argonne v18 nuclear potential as an input. Gluon PDFs calculated per nucleon are very similar for the proton ones in the range of small and intermediate $x$ values while for $x>0.6$ the difference becomes large due to the Fermi motion (see Fig.~\ref{fig:unpolarized}(b)). A similar work was performed in Ref.~\cite{Mantysaari:2019jhh} for determination of spatial gluon distribution in deuteron for low-$x$ that could be tested in the $J/\psi$ production at EIC. Today the gluon content of deuteron and light nuclei becomes the matter of interest for the lattice QCD studies~\cite{Winter:2017bfs}. Apart from the general understanding of the gluon EMC effect, the measurement of the gluon PDF at high-$x$ for deuteron could provide a useful input for high-energy astrophysical calculation~\cite{Brodsky:2018zdh}.
 
 SPD can perform an explicit comparison of the differential inclusive production cross-sections $d\sigma/dx_{F}$ for all three gluon probes: charmonia, open charm, and prompt photons using $p$-$p$ and $d$-$d$ collisions at $\sqrt{s_{NN}}=13.5$~GeV and possibly below. Such results could be treated in terms of the difference of unpolarized gluon PDFs in deuteron and nucleon.

\section{Gluon content of polarized proton and deuteron \label{pol}}
\subsection{Gluon helicity with longitudinally polarized beams \label{hel}}

The gluon helicity distribution function $\Delta g(x)$\footnote{Here we adopt historical notation for gluon helicity-PDF $\Delta g(x)$ which is also known as $g_1^g(x)$.} is a fundamental quantity characterizing the inner structure of the nucleon.
It describes the difference of probabilities to find a gluon with the same and opposite helicity orientation w.r.t. the spin of the longitudinally polarized nucleon.
 The integral $\Delta G=\int \Delta g(x) dx$ can be interpreted as the gluon spin contribution to the nucleon spin, as it was discussed in Sec.~\ref{lqcd}. After the EMC experiment discovered that only a small part of proton spin is carried by the quarks: $\Delta \Sigma \approx 0.25$, see Refs.~\cite{Aidala:2012mv,Ashman:1989ig}, the gluon spin was assumed to be another significant contributor, see Eq.~(\ref{JM}) in Sec.~\ref{lqcd}.
%\begin{equation}
%\frac{1}{2} = \frac{1}{2}\Delta \Sigma + \Delta G + L_{q}+L_{g},
%\end{equation}
%The quark helicity contribution was found to be ~\cite{}.

\begin{figure}[!t]
  \begin{minipage}[ht]{0.49\linewidth}
      \center{\epsfig{file=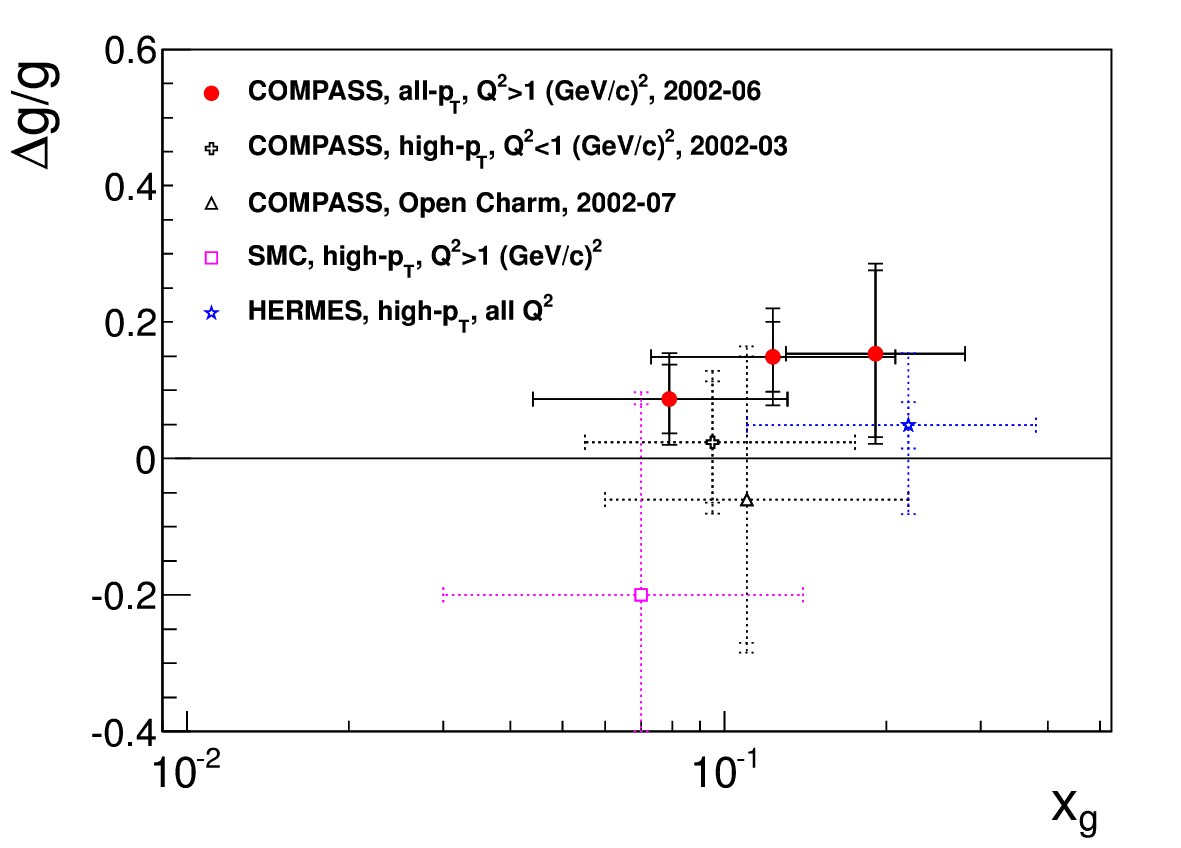,scale=0.4} \\ (a)} 
  \end{minipage}
  \hfill
  \begin{minipage}[ht]{0.49\linewidth}
        \center{\epsfig{file=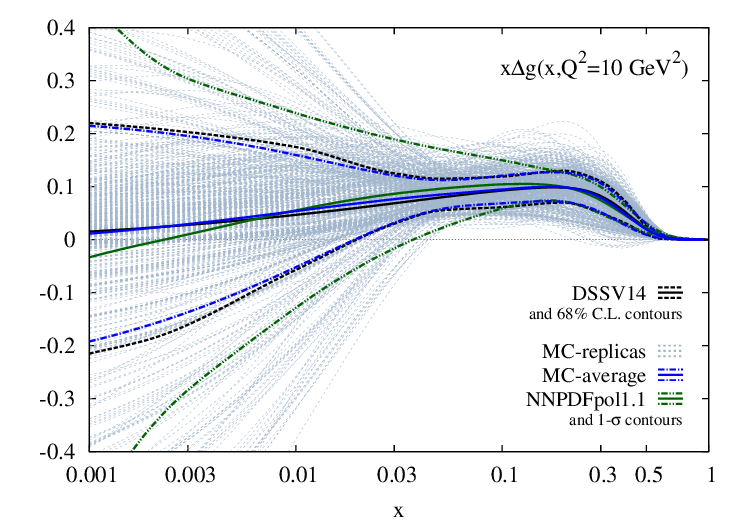,scale=0.64} \\ (b)} 
  \end{minipage}
  \caption{ (a) SIDIS data on $\Delta g(x)/g(x)$ extracted in LO~\cite{Adolph:2015cvj}. With kind permission of The European Physical Journal (EPJ). (b) Global fit results for the gluon helicity distribution $\Delta g(x)$. Reprinted figure from~\cite{deFlorian:2019zkl} \textcopyright~(2019) by the American Physical Society under Creative Commons Attribution 4.0 International License.}
  \label{fig:dgg_1}  
\end{figure}

The first attempt to measure the gluon polarization in the nucleon was made by the FNAL E581/704 Collaboration using a 200~GeV polarized proton beam and a polarized proton target~\cite{Adams:1994bg}. They measured the longitudinal double-spin asymmetries $A_{LL}$
     for inclusive multi-$\gamma$ and $\pi^0\pi^0$ production to be consistent with zero within their sensitivities. In the following years a set of SIDIS measurements was performed by the HERMES~\cite{Airapetian:2010ac}, SMC~\cite{Adeva:2004dh} and COMPASS~\cite{Ageev:2005pq, Alekseev:2009ad,Adolph:2012vj,Adolph:2012ca,Adolph:2015cvj} experiments. The production of hadron pairs with
high transverse momenta and the production of the open charm where the photon-gluon fusion mechanism dominates were studied.
 It was figured out that with a large uncertainty the value of $\Delta G$ is close to zero. Nevertheless, for gluons carrying a large fraction $x$ of the nucleon momentum, an evidence of a positive polarization has been observed, see Fig.~\ref{fig:dgg_1}(a). The summary of theoretical and experimental results for $\Delta G$ can be found in Ref. \cite{Deur:2018roz}.
 
New input for $\Delta G$ estimation was obtained from the measurement of the $A_{LL}$ asymmetries in the inclusive production of high-$p_T$ neutral ~\cite{Adam:2018cto,Adare:2014hsq, Adare:2008aa, Adare:2008qb} and charged pions \cite{Acharya:2020fgb},  $\eta$-mesons~\cite{Adare:2014hsq}, jets~\cite{Djawotho:2013pga},  di-jets \cite{Adamczyk:2016okk, Adam:2018pns}, heavy flavors~\cite{Adare:2012vv} and, recently, $J/\psi$-mesons~\cite{Adare:2016cqe} in polarized $p$-$p$ collisions at RHIC. These results in general are in agreement with the SIDIS measurements, which indicates the universality of the helicity-dependent parton densities and QCD factorization. 

\begin{figure}[!t]
  \begin{minipage}[ht]{0.58\linewidth}
        \center{\epsfig{file=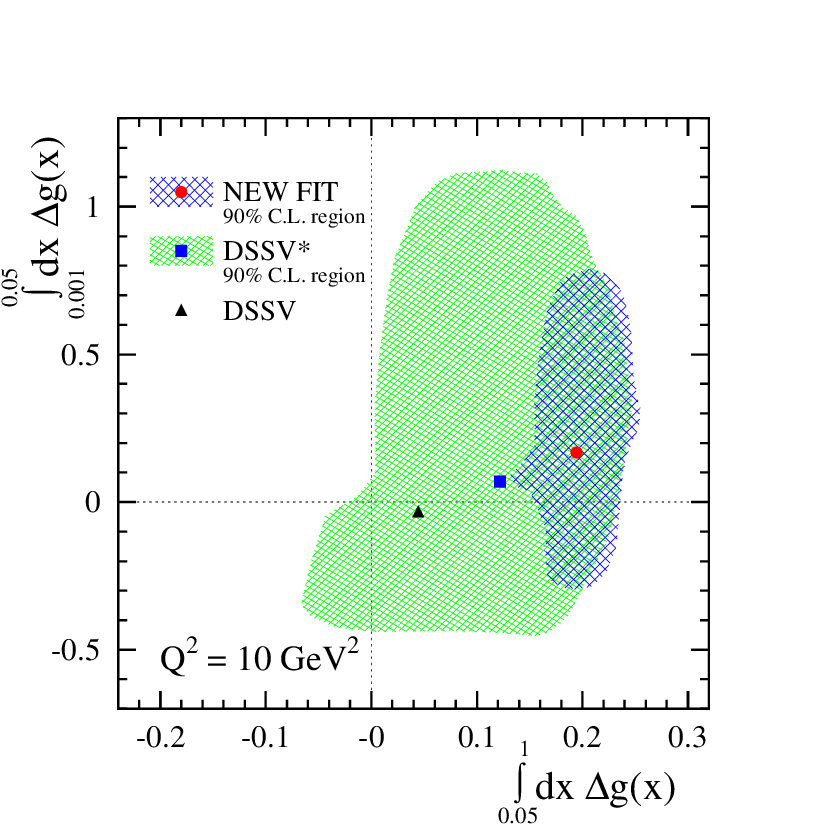,scale=0.7} \\ (a)} 
  \end{minipage}
  \hfill
  \begin{minipage}[ht]{0.41\linewidth}
          \center{\epsfig{file=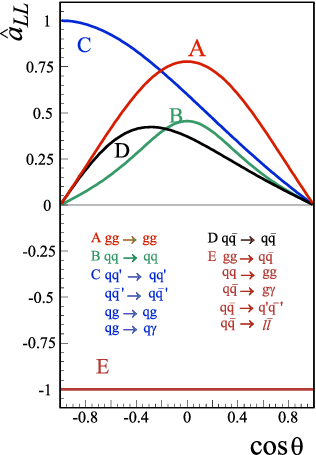,scale=1.} \\ (b)} 
  \end{minipage}
  \caption{ (a) Estimates of contributions of low-$x$ and high-$x$ kinematic ranges into $\Delta G$ for the DSSV series of the global fit.  The 90\% C.L. areas are shown. Reprinted figure with permission from Ref.~\cite{deFlorian:2014yva}  \textcopyright~(2014) by the American Physical Society. (b) Partonic longitudinal double-spin asymmetries $\hat{a}_{LL}$ for different hard processes as a function of center-of-mass scattering angle~\cite{Aidala2005RESEARCHPF}.}
  \label{fig:dgg_2}  
\end{figure}

%At the moment two main series of global fit of polarized PDF for nucleon do exist: DSSV and 
At the moment the most recent sets of polarized PDFs extracted in the NLO approximation are LSS15~\cite{Leader:2014uua}, DSSV14~\cite{deFlorian:2014yva,deFlorian:2019zkl}, NNPDF-pol1.1~\cite{Nocera:2014gqa}, and JAM17~\cite{Ethier:2017zbq}. To obtain them, different approaches, parameterizations, and sets of experimental data were used, see Ref.~\cite{Ethier:2020way} for more details. Fit results for $\Delta g(x)$ from  DSSV14 and NNPDF--pol1.1 are presented in Fig. \ref{fig:dgg_1}(b)~\cite{deFlorian:2019zkl}. The RHIC $p$-$p$ data put a strong constraint on the size of $\Delta g(x)$
in the range $0.05<x<0.2$ while a constraint on its sign is  weaker since in some of the processes only $\Delta g$ squared is probed (see discussion below). 
 The small $x$ region remains still largely unconstrained and could be covered in the future by measurements at EIC~\cite{Accardi:2012qut}. 
Region of high $x$ is covered at the moment only by SIDIS measurements which still lack a proper NLO description~\cite{deFlorian:2008mr}.
The uncertainty of the contribution to $\Delta G$ from the kinematic range $0.001<x<0.05$ vs. the corresponding contribution from the range $x>0.05$ for the DSSV global fits is shown in Fig.~\ref{fig:dgg_2}(a)~\cite{deFlorian:2014yva}.

\begin{figure}[!h]
  \begin{minipage}[ht]{0.498\linewidth}
      \center{\epsfig{file=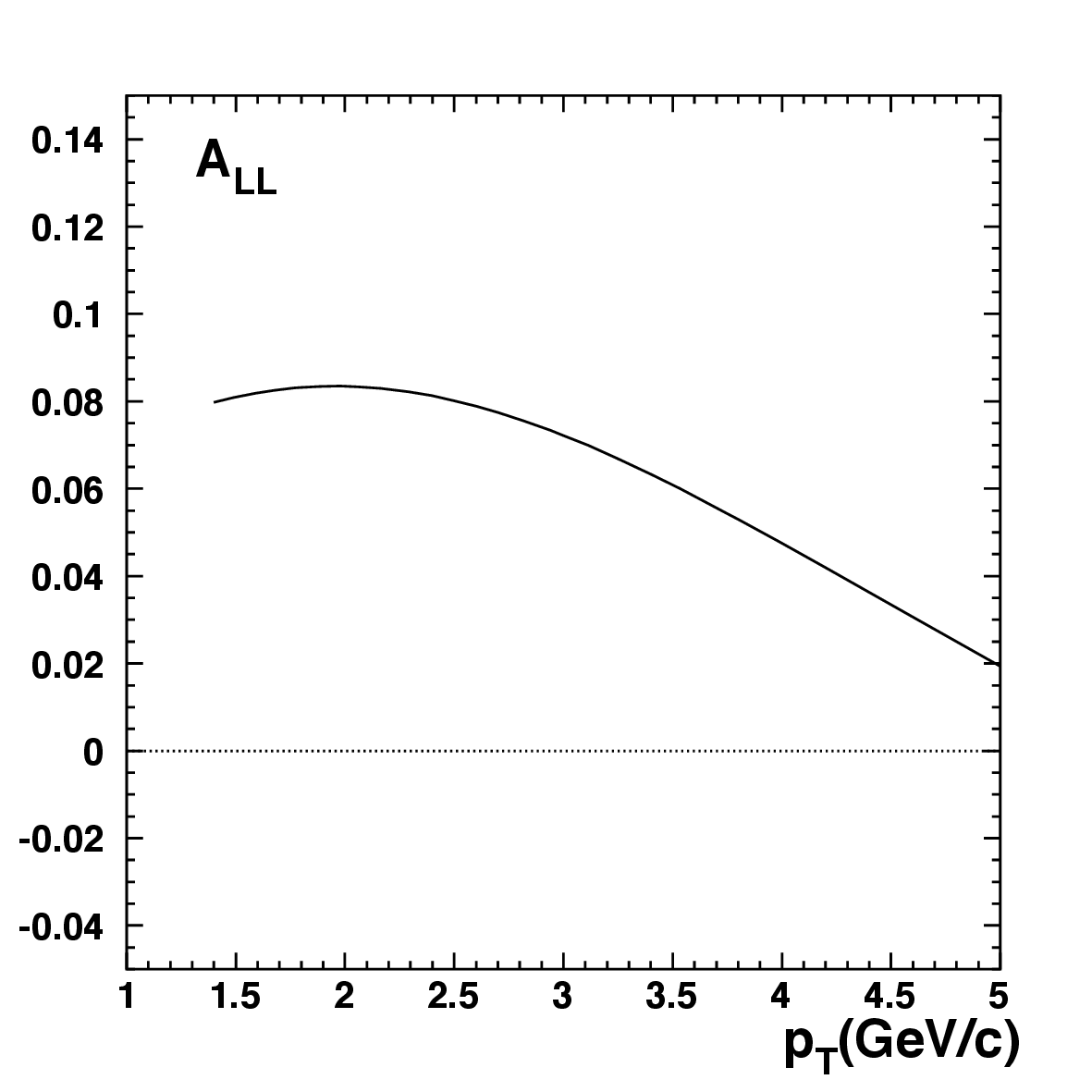,scale=0.4} \\ (a)}     
  \end{minipage}
  \hfill
  \begin{minipage}[ht]{0.498\linewidth}
        \center{\epsfig{file=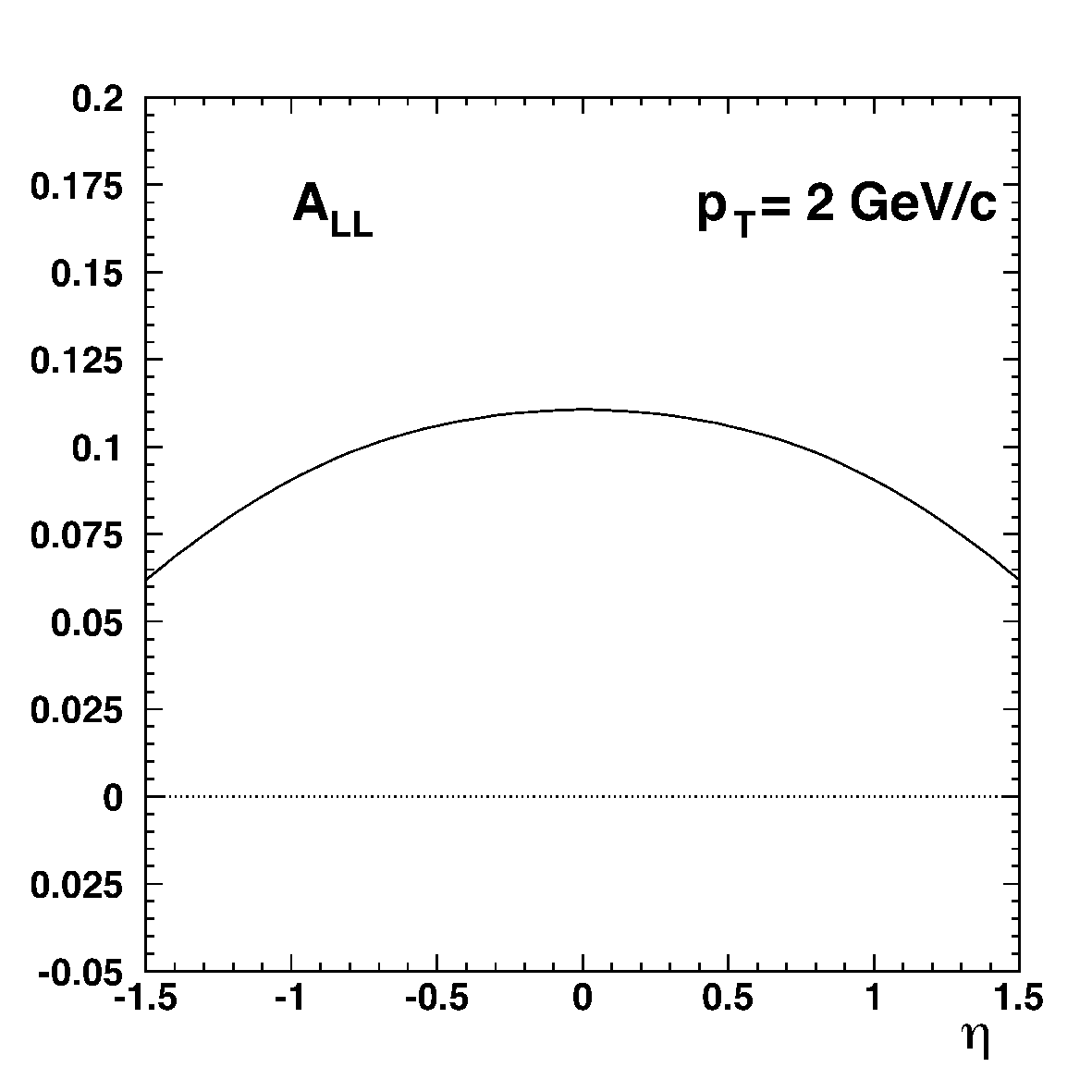,scale=0.4} \\ (b)}   
  \end{minipage}
  \caption{ Longitudinal double spin asymmetry $A_{LL}$  for inclusive $J/\psi$ production calculated using LO PDF set A form Ref. \cite{Gehrmann:1994rb} for $p$-$p$ collisions at $\sqrt{s}=39$~GeV in the LO approximation as a function of a) transverse momentum $p_T$ and b) pseudorapidity $\eta$~\cite{Anselmino:1996kg}.}
  \label{fig:ALL_JPSI}  
\end{figure}

\begin{figure}[!h]
  \begin{minipage}[ht]{0.498\linewidth}
      \center{\epsfig{file=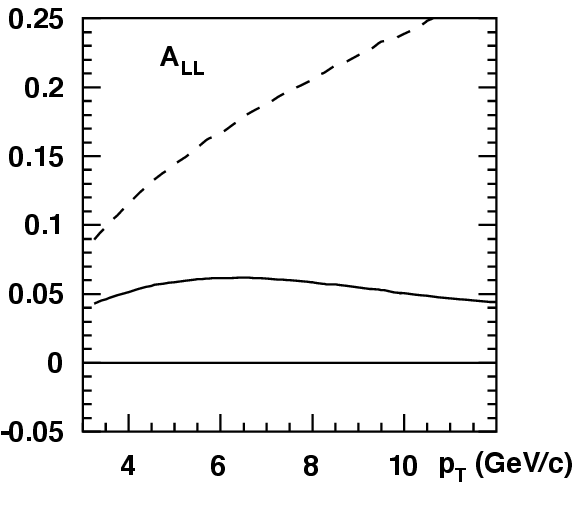,scale=0.86} \\ (a)}     
  \end{minipage}
  \hfill
  \begin{minipage}[ht]{0.498\linewidth}
        \center{\epsfig{file=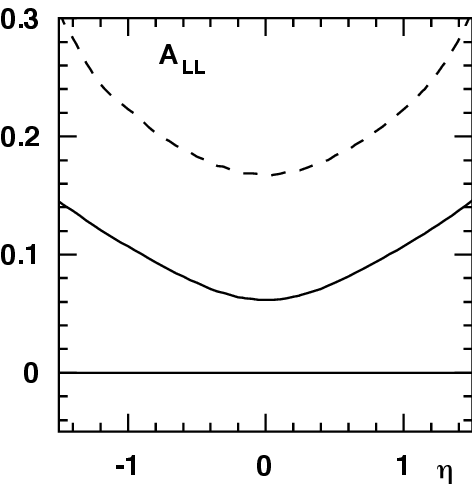,scale=0.9} \\ (b)}   
 \end{minipage}
  \caption{ Longitudinal double spin asymmetry $A_{LL}$ for inclusive prompt-photon production calculated for $p$-$p$ collisions at $\sqrt{s}=39$~GeV in the LO approximation as a function of a) transverse momentum $p_T$ and b) rapidity $\eta$ ($p_T=6$ GeV/$c$)~\cite{Anselmino:1996kg}. The full line corresponds to the NLO 'valence' PDF set from Ref. \cite{Gordon:1996ft}, while the dashed line -- to the set A form Ref. \cite{Gehrmann:1994rb}.}
  \label{fig:ALL_PP}  
\end{figure}

In case of the longitudinally polarized $p$-$p$ collisions the asymmetry $A_{LL}$ is defined as
\begin{equation}
A_{LL}=\frac{\sigma^{++}-\sigma^{+-}}{\sigma^{++}+\sigma^{+-}},
\end{equation}
where $\sigma^{++}$ and $\sigma^{+-}$ denote the cross-sections with the same and opposite proton helicity combinations, respectively. For the prompt
photons produced via the gluon Compton scattering 
\begin{equation}
A^{\gamma}_{LL}\approx \frac{\Delta g(x_1)}{g_(x_1)}\otimes A_{1p}(x_2) \otimes \hat{a}_{LL}^{g q(\bar{q}) \to \gamma q (\bar{q})} + (1 \leftrightarrow 2).
\end{equation}
Here $A_{1p}(x)$ is the asymmetry well measured in a wide range of $x$ and $\hat{a}_{LL}^{g q(\bar{q}) \to \gamma q (\bar{q})}$ is the asymmetry 
of the corresponding hard process. %The expressions for $\hat{a}_{LL}$ for different hard processes are presented in Tab. \ref{hard_asymmetries}.
The Fig.~\ref{fig:dgg_2}(b) shows the behavior of $\hat{a}_{LL}$ for different hard processes as a function of the center-of-mass scattering angle.
For charmonia and open charm production via the gluon-gluon fusion process the expression for the corresponding asymmetry reads
\begin{equation}
A^{c\bar{c}}_{LL}\approx \frac{\Delta g(x_1)}{g(x_1)}\otimes \frac{\Delta g(x_2)}{g_(x_2)} \otimes \hat{a}_{LL}^{gg\to c\bar{c}X}.
\end{equation}
This asymmetry on the one hand is more sensitive to the gluon polarization than the corresponding one for the prompt photons due to the quadratic dependence on $\Delta g$. % and larger absolute value of $\hat{a}_{LL}$.
 On the other hand the sign of the $\Delta g$ value can not be determined from it. So the measurements with prompt photons and heavy-quark states are complementary. The contribution of $q\bar{q}$ annihilation processes to the above-mentioned asymmetries is negligible despite $\hat{a}_{LL}=-1$ because of the smallness of the sea-quark polarization in the nucleon.
 
 It is important to emphasise that a sizable systematic uncertainty of $A_{LL}$ measurements in the inclusive $J/\psi$ production comes from our limited knowledge of charmonia production mechanisms including the feed-down contribution. Each of them has different partonic asymmetries $\hat{a}_{LL}$~\cite{Leader:2001gr}. 
 %The feed-down contribution from the decay of higher charmonia states like $\psi(2S)$, $\chi_{cJ}$ is about 20\% at the {\color{red} discussed energies~\cite{Abt:2008ed}}. 
 For the $\Delta g$ estimation in Ref.~\cite{Adare:2016cqe} the value of $\hat{a}^{J/\psi}_{LL}$ has been forced to $-1$. The SPD setup will have the possibility to reconstruct $\chi_{cJ}$ states via their radiative decays and resolve $J/\psi$ and $\psi(2S)$ signals in a wide kinematic range and disentangle contributions of direct and feed-down production mechanisms.
 %{The asymmetry $A_{LL}$ could be measured for each charmonium state separately that should improve quality of the $\Delta g$ estimation.}
 The quality of the $\Delta g$ estimation could be significantly improved by measuring $A_{LL}$ separately for each charmonium state.
 
Predictions for the longitudinal double-spin asymmetries $A_{LL}$ in $p$-$p$ collisions can be found in Ref.~\cite{Feng:2018cai} ($J/\psi$) and~\cite{ Vogelsang:2000bx} (prompt photons). They mostly cover the kinematic range of the RHIC experiments. Some estimates for $A_{LL}$ in charmonia~\cite{Anselmino:1996kg} and prompt-photon~\cite{Gordon:1996ft, Gordon:1996rk,Anselmino:1996kg} production at $\sqrt{s}=39$~GeV (see Figs.~\ref{fig:ALL_JPSI} and \ref{fig:ALL_PP}, respectively) have been done in preparation of the unrealized HERA-$\bm{N}$ project.

The authors of the Ref.~\cite{Xu:2004es} proposed to extract information about the gluon helicity $\Delta g$ via studying of the production of high-$p_T$ prompt photons accompanied by $\Sigma^+$ hyperons. To do that the single longitudinal spin asymmetry $A^{\gamma\Sigma}_L$ and the polarization of the produced $\Sigma^+$ hyperons should be measured. However, further elaboration of this method is needed.

\subsection{Gluon-related TMD and twist-3 effects with transversely polarized beams \label{tmd_t3}}

One of the promising ways to investigate the spin structure of the nucleon is the study of transverse single-spin asymmetries (SSAs) in the inclusive production of different final states in high-energy interactions. The  SSA $A_N$ is defined as
\begin{equation}
A_{N}=\frac{\sigma^{\uparrow}-\sigma^{\downarrow}}{\sigma^{\uparrow}+\sigma^{\downarrow}},
\end{equation}
where $\sigma^{\uparrow}$ and $\sigma^{\downarrow}$ denote the inclusive production cross-sections with opposite transverse polarization 
%\AK{
with respect to the production plane, defined by momenta of polarized projectile and produced hadron (or jet).
%}
%\UD{with respect to the production plane} \AK{(Looks strange, for example in CMS we have only one line for colliding particles momenta. I believe that it have to be written 'with respect to  momentum'. \UD{The plane is defined by the colliding beam direction and the final hadron momentum. If you consider transverse w.r.t. the colliding particle momenta you should have a dependence on the azimuthal angle of the transverse spin. This is also possible, but then you have to take moments of $A_N$ and you cannot refer to it as left-right asymmetry} )} of one of the colliding particles.
At the moment, more than forty years after the transverse spin phenomena were discovered, a wealth of experimental data indicating non-zero $A_N$ in the lepton-nucleon and nucleon-nucleon interactions were
%was \BP{(were)} 
collected, see e.g.~\cite{Anselmino:2020vlp,Avakian:2019drf,Perdekamp:2015vwa,Boglione:2015zyc,Aidala:2012mv,LIANG_2000,Aschenauer_2016,Aidala_2018} and references therein. 
%\BP{(Refs)}.
However, our understanding of the SSA phenomenon is not yet conclusive.

Theoretically two dual approaches are used to explain transverse single-spin azimuthal asymmetries: the collinear twist-3 formalism and the TMD factorization approach. 
In the first one at large transverse momenta $p_T \gg \Lambda_{QCD}$ of a produced particle, the collinear factorization involving twist-3 contributions for three-parton (Efremov-Teryaev-Qiu-Sterman) correlations Ref.~\cite{Efremov:1981sh, Efremov:1984ip, Qiu:1991pp,Efremov:1994dg} are used, for the review see Ref.~\cite{ Pitonyak:2016hqh}.\footnote{Here $\Lambda_{QCD}\approx 200$ MeV is the QCD scale.}An alternative approach, 
suitable also for less inclusive processes, is based on the TMD factorization, valid for $p_T\ll Q$, in terms of transverse-momentum-dependent parton distributions.
In this case, the SSAs originate from the   initial-state quark and Gluon Sivers Functions (GSF), which may be related to the mentioned twist-3 correlators, or the final-state Collins fragmentation functions.
%{A Sivers function $\Delta^{q(g)}_N (x,k_{T})$ is a TMD PDF that describes the left-right asymmetry of parton distributions in the transversely polarized nucleon as a function of the transverse momentum $k_T$ of a parton. It is related with the orbital angular momentum of partons and is an important aspect of the three-dimensional picture of the nucleon.}

The Sivers function $f_{1T}^{\perp,q(g)}(x,\ktsq)$ is a TMD PDF that describes the left-right asymmetry in the distribution of  partons w.r.t.  the plane defined by the nucleon spin and momentum vectors. Originating from the correlation between the spin of the nucleon and the orbital motion of partons, it is an important detail of the three-dimensional picture of the nucleon. 
This function is responsible for the so-called Sivers effect (for both quarks and gluons) that was first suggested in Ref.~\cite{Sivers:1989cc} as an explanation for the large transverse single-spin asymmetries $A_N$ measured in the inclusive pion production off transversely polarized nucleons.
More details on the theoretical and experimental status of the transverse spin structure of the nucleon can be found in Refs.~\cite{DAlesio:2007bjf, Perdekamp:2015vwa, Boer:2015vso}.
%
%{The first attempt to access the gluon Sivers function (GSF) using the azimuthal asymmetry in production of high-$p_T$ hadron pairs in SIDIS was performed by COMPASS}~\cite{Adolph:2017pgv}. \st{The resulting asymmetry is away from zero by more than two standard deviations that supports the possible existence of a non-zero gluon Sivers function.}
The first attempt to access the GSF studying azimuthal asymmetries in high-$p_T$ hadron pair production in SIDIS off transversely polarized deuterons and protons, was performed by COMPASS~\cite{Adolph:2017pgv}.
Using neural network techniques the contribution originating from the Photon–Gluon Fusion (PGF) subprocess has been separated from the leading-order virtual-photon absorption and QCD Compton scattering subprocesses.
The extracted combined proton-deuteron PGF-asymmetry was found to be negative and more than two standard deviations below zero, 
the possible existence of a non-zero GSF. In the meantime, COMPASS did not see any signal for the PGF Collins asymmetry, which can analogously be related to the gluon transversity distribution in nucleon. COMPASS studied the GSF also through Sivers asymmetry in the $J/\psi$-production channel~\cite{Szabelski:2016wym}, again obtaining an indication of a negative asymmetry.

Several inclusive processes were proposed to access the gluon-induced spin effects in transversely polarized $p$-$p$ collisions. Single spin asymmetries for production of charmonia~\cite{Godbole:2017syo,DAlesio:2017rzj} (RHIC, AFTER), open charm ~\cite{Anselmino:2004nk,Koike:2011mb, Kang:2008ih,Godbole:2016tvq,DAlesio:2017rzj} (RHIC)~\cite{Godbole:2016tvq} (AFTER),  and prompt photons ~\cite{Qiu:1991pp, Hammon:1998gb} (E704), ~\cite{Kanazawa:2012kt} (RHIC). The possible gluon induced effects were estimated using both approaches TMD and the collinear twist-three approaches for the experimental conditions of the past, present, and future experiments.
%were estimated using both approaches, \UD{the TMD and the twist-three ones}, for the experimental conditions of the past, present, and future experiments.
%{An indirect estimate of the GSF was obtained, within the GPM framework in} Ref.~\cite{DAlesio:2015fwo}, by fitting the data on the SSA in $\pi^0$ production at RHIC~\cite{Adare:2013ekj}. First $k_T$-moments of the GSF for SIDIS1~\cite{Anselmino:2005ea} and SIDIS2~\cite{Anselmino:2008sga} extractions the quark Sivers functions are shown in Fig. \ref{fig:GSF} (a) and (b), respectively.

In Ref.~\cite{DAlesio:2015fwo} a first estimate of the GSF was obtained using the midrapidity data on $A_N$, as measured in $\pi^0$ production in $pp$ collisions at RHIC~\cite{Adare:2013ekj}.
The extraction was performed within the GPM framework, a phenomenological extension of the TMD factorization scheme, where TMDs are assumed conditionally universal, using the GRV98-LO set for the unpolarized PDF and available parameterizations for the quark Sivers functions (SIDIS1 from Ref.~\cite{Anselmino:2005ea} and SIDIS2 from Ref.~\cite{Anselmino:2008sga}). 

The GSF is parameterized assuming a factorized Gaussian-like form, as follows:
\begin{equation}
\Delta^N\! f_{g/p^\uparrow}(x,\ktsq) =   \left (-2\frac{|{\bf k}_T|}{M_p}  \right )f_{1T}^{\perp\,g} (x,\ktsq)  = 2 \, {\cal N}_g(x)\,f_{g/p}(x)\,
h(\ktsq)\,\frac{e^{-\ktsq/\langle \ktsq \rangle}}
{\pi \langle \ktsq \rangle}\,,
\label{eq:siv-par-1}
\end{equation}
where $f_{g/p}(x)$ is the standard unpolarized collinear gluon distribution, and
\begin{equation}
{\cal N}_g(x) = N_g x^{\alpha}(1-x)^{\beta}\,
\frac{(\alpha+\beta)^{(\alpha+\beta)}}
{\alpha^{\alpha}\beta^{\beta}}\,, \hspace*{1cm}
%\label{eq:nq-coll}
%\end{equation}
%with $|N_g|\leq 1$, and
%\begin{equation}
h(\ktsq) = \sqrt{2e}\,\frac{|{\bf k}_T|}{M'}\,e^{-\ktsq/M'^2}\,.
\label{eq:siv-par}
\end{equation}
We can also define a suitable parameter $\rho = M'^2/(\langle \ktsq \rangle +M'^2)$, with $0<\rho<1$. By imposing $|N_g|\leq 1$ the positivity bound for the GSF is automatically fulfilled for any value of $x$ and $\ktv$. 

The first $\ktv$-moment of the GSF is also of relevance
\begin{equation}
\Delta^N \! f_{g/p^\uparrow}^{(1)}(x) = \int d^2 \ktv \frac{|\ktv|}{4 M_p} \Delta^N \! f_{g/p^\uparrow}(x,\ktsq) \equiv - f_{1T}^{\perp (1) g}(x) \,.
\label{siversm1}
\end{equation}

\begin{figure}[!h]
  \begin{minipage}[ht]{0.5\linewidth}
  \center{\epsfig{file=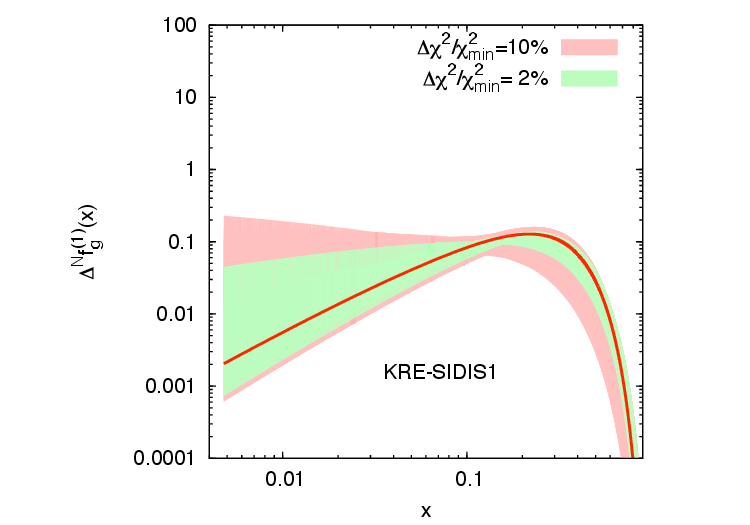,scale=0.71} \\ (a)}  
  \end{minipage}
  \hfill
  \begin{minipage}[ht]{0.5\linewidth}
    \center{\epsfig{file=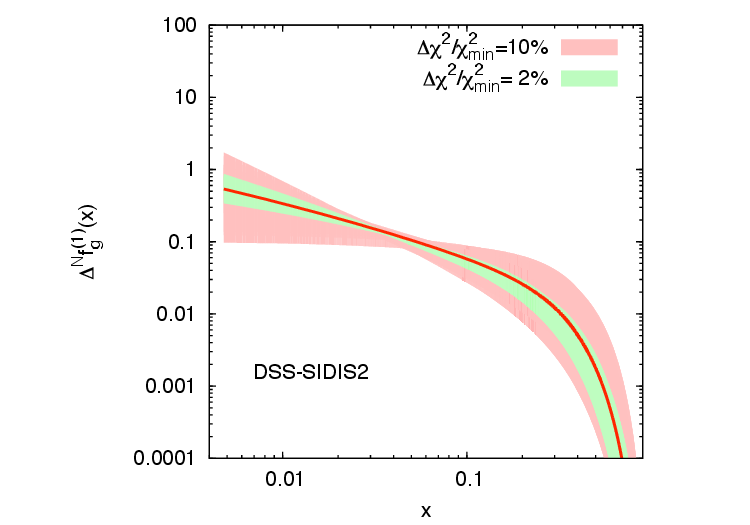,scale=0.71} \\ (b)}  
  \end{minipage}
  \caption{ The first $\ktv$-moment of the gluon Sivers function~\cite{DAlesio:2015fwo} obtained using the SIDIS1~\cite{Anselmino:2005ea} -- panel (a) and SIDIS2~\cite{Anselmino:2008sga} -- panel (b) extractions of the quark Sivers functions. Reprinted figures from Ref.~\cite{DAlesio:2015fwo} \href{https://doi.org/10.1007/JHEP09(2015)119}{\textcopyright~(2015)} Springer under the Creative Commons Attribution 4.0 International License.}
  \label{fig:GSF} 
\end{figure}

Two different  parameterization of the GSF  were obtained using different sets for the fragmentation functions,
%\UD{(consistently with the two quark Sivers extractions)}
  namely the Kretzer~\cite{Kretzer:2000yf} and DSS07~\cite{deFlorian:2007aj} sets, which give significantly different results for gluons.  The latter point has a strong impact on the extracted GSF, especially in the low-$x$ region, see Fig.~\ref{fig:GSF}, where the first moments of the GSF are shown in Fig.~\ref{fig:GSF}(a) and  ~\ref{fig:GSF}(b), respectively for the SIDIS1 and SIDIS2 sets. 

%{However, there are no hard constraints on the size of the gluon Sivers function apart from the positivity bound}~\cite{Boer:2015vso}.
We recall that the gluon Sivers function is expected to satisfy the positivity bound defined as twice the unpolarized TMD gluon distribution.
Although, some theoretical expectations are that the gluon Sivers function at relatively high $x$ should be about 1/3 of the quark one~\cite{Boer:2015vso}.

An extended version of the GPM, with inclusion of initial- and final-state interactions (ISIs and FSIs respectively) under a one-gluon exchange approximation, was developed, limiting to the quark sector, in Refs.~\cite{Gamberg:2010tj,DAlesio:2011kkm}: the so-called color-gauge-invariant GPM (CGI-GPM). 
%In this approach, the process dependence of the quark Sivers function, induced by the ISI and FSI effects, is shifted to the partonic cross-sections. Therefore the Sivers function can still be considered universal, but it has to be convoluted with modified partonic cross-sections, which turn out to have the same form, in terms of Mandelstam variables, of the hard functions of the twist-three collinear approach~\cite{Gamberg:2010tj}. In particular, this model is able to reproduce the expected opposite relative sign of the Sivers functions for SIDIS and DY, due to the effects of FSIs and ISIs, respectively~\cite{Collins:2002kn,Brodsky:2002rv}.
In this approach, the effects induced by ISIs and FSIs, leading to the process dependence of the Sivers function, can be collected into modified color factors and moved to the hard partonic parts. In this way, one can still consider a universal Sivers function, as extracted from SIDIS, but this time convoluted with properly modified partonic cross-sections. Interestingly, these coincide with the corresponding hard partonic parts appearing in the collinear twist-three formalism~\cite{Gamberg:2010tj}. Moreover, this modified GPM allows to get the expected opposite sign of the Sivers function when moving from SIDIS to DY processes~\cite{Collins:2002kn,Brodsky:2002rv}.

In Ref.~\cite{DAlesio:2017rzj} this approach was extended to the gluon Sivers contribution to $A_N$ for $D$-meson and $J/\psi$ production and, subsequently, in Ref.~\cite{DAlesio:2018rnv} to the case of inclusive pion and photon production. The main difference w.r.t.~the quark case is that for the gluon sector one needs to introduce two different classes of modified partonic cross-sections, corresponding to the two different ways in which a color-singlet state can be formed out of three gluons, {\it i.e.}\ either through an anti-symmetric or a symmetric color combination. Each one of them has to be convoluted with a different gluon Sivers distribution. These two universal and independent distributions are named, respectively, the $f$-type and $d$-type gluon Sivers functions~\cite{Bomhof:2006ra}: the former is even under charge conjugation, while the latter is odd.

\begin{figure}[!h]
\begin{center}
  \center{\epsfig{file=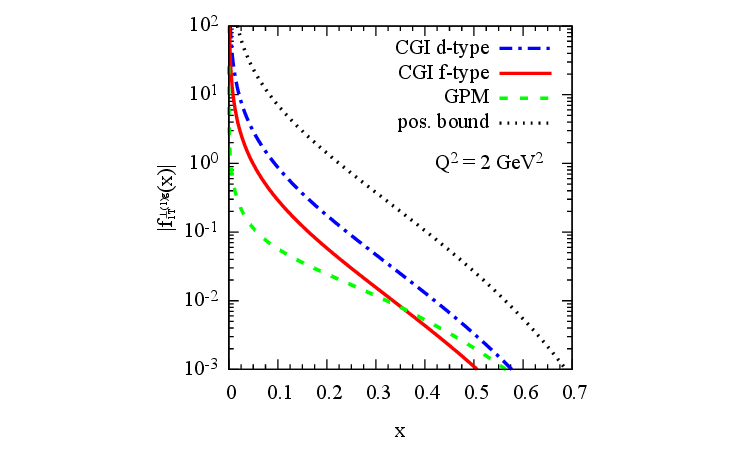,scale=1.}}
\caption{Allowed upper values for the first $\ktv$-moments of the gluon Sivers function in different approaches and scenarios at $Q^2 = 2$ GeV$^2$ \cite{DAlesio:2018rnv}: GPM approach (dashed line), CGI-GPM $d$-type ($N_g^{(d)}=0.15$, dot-dashed line) and $f$-type ($N_g^{(f)}=0.05$, solid line). The positivity bound (dotted line) is also shown.}
\label{fig:1stmom}
\end{center}
\end{figure}

In Ref.~\cite{DAlesio:2018rnv} a detailed phenomenological study was carried out, leading to the first preliminary determination of the allowed upper bounds for the two GSFs entering the CGI-GPM approach. This was obtained by analysing the mid-rapidity pion $A_N$ data together with those for $D$-meson production, from the PHENIX Collaboration \cite{Adare:2013ekj,Aidala:2017pum}. 
The main outcome of this study is that while pion SSAs are very sensitive to the $f$-type GSF, being the $d$-type one  dynamically suppressed, the opposite is true in the SSAs for $D$-meson production, at least at forward rapidities. A combined extraction (in terms of allowed upper bounds) of these two GSFs was therefore possible. The corresponding results for the first moments of the GSF are shown in Fig.~\ref{fig:1stmom}. 

Notice that, while in Ref.~\cite{DAlesio:2015fwo} a single value $\langle \ktsq\rangle = 0.25$ GeV$^2$~\cite{Anselmino:2005nn} was adopted for the unpolarized quark and gluon TMDs, in Ref.~\cite{DAlesio:2018rnv} for the unpolarized gluon TMD the authors used a different value, $\langle \ktsq\rangle = 1$ GeV$^2$. This, indeed, gives a better account of the unpolarized cross-sections for $J/\psi$ production at not so large $p_T$ values, without spoiling the description of the inclusive pion production data. Moreover, a different set for the unpolarized PDFs (the CTEQ6L1  set~\cite{Pumplin:2002vw}) was employed. 
For these reasons, an updated extraction of the GSF within the GPM was performed (still adopting the DSS07 set for the unpolarized collinear pion FFs), finding  results very similar to those reported in Ref.~\cite{DAlesio:2015fwo} for the first $\ktv$-moments of the GSF (SIDIS2 case), although with slightly different parameters.

\begin{figure}[!h]
  \begin{minipage}[ht]{0.5\linewidth}
   \center{\epsfig{file=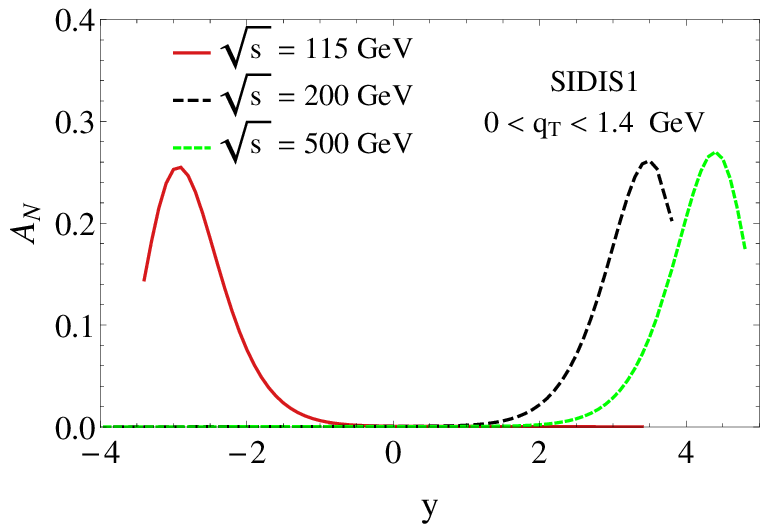,scale=0.55} \\ (a)}  
  \end{minipage}
  \hfill
  \begin{minipage}[ht]{0.5\linewidth}
     \center{\epsfig{file=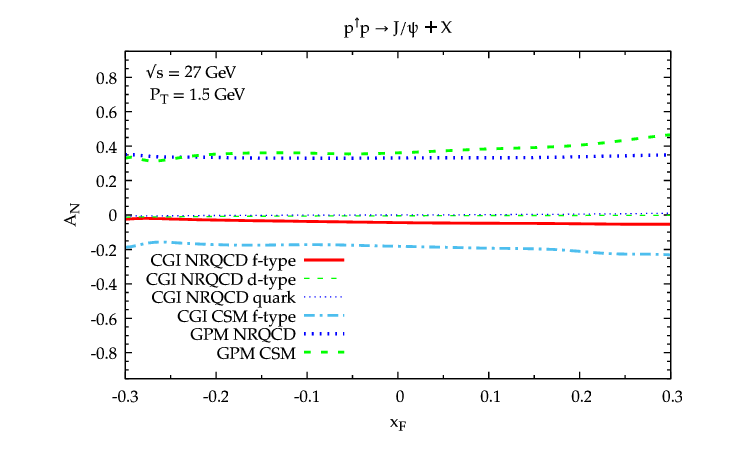,scale=0.66} \\ (b)} 
  \end{minipage}
  \caption{(a) Predictions for $A^{J/\psi}_{N}$ for $\sqrt{s}=115$ GeV (AFTER), 200~GeV and 500~GeV (RHIC) as a function of rapidity $y$. Reprinted figure with permission from \cite{Godbole:2017syo}  \textcopyright~(2017) by the American Physical Society. 
  %(b) $A_N^D$ predicted basing on the twist-3 approach for $D^0$ (red solid curve) and $\bar{D^0}$ (green dotted curve)  as a function of $x_F$ from Ref.~\cite{Koike:2011mb}. Experimental points from PHENIX~\cite{Aidala:2017pum} are shown in blue.
  (b)  Maximized contributions to $A_N$ estimates as a function of $x_F$ for the process $p^\uparrow p\to J/\psi + X$ at $\sqrt s=27$ GeV and $p_T=1.5$ GeV adopting the CGI-GPM and GPM approaches, within the CS model and NRQCD. \label{ANjpsi-NICA}  }
\end{figure}

%Several inclusive processes were proposed to access the gluon-induced spin effects in transversely polarized $p$-$p$ collisions. Single spin asymmetries for production of charmonia~\cite{Godbole:2017syo,DAlesio:2017rzj} (RHIC, AFTER), open charm ~\cite{Anselmino:2004nk,Koike:2011mb, Kang:2008ih,Godbole:2016tvq,DAlesio:2017rzj} (RHIC)~\cite{Godbole:2016tvq} (AFTER),  and prompt photons ~\cite{Qiu:1991pp, Hammon:1998gb} (E704), ~\cite{Kanazawa:2012kt} (RHIC) were estimated using both approaches, \UD{the TMD and the twist-three ones}, for the experimental conditions of the past, present, and future experiments. 
%
The SSA $A^{J/\psi}_{N}$ in the $J/\psi$ production was measured by the PHENIX Collaboration in $p$-$p$ and $p$-$A$ collisions at $\sqrt{s_{NN}}=200$ GeV/$c$~\cite{Adare:2010bd, Aidala:2018gmp}. The obtained asymmetries
%\UD{experimental data} \old{values} for 
$A^{J/\psi}_{N}$ are consistent with zero for negative and positive $x_F$ values.
%An indication of negative SSA $A^{J/\psi}_{N}=-0.086\pm0.026 \pm 0.003$ was found for the positive $x_{F}$ that could be a glimpse of the nonzero gluon Sivers function. 
Theoretical predictions~\cite{Godbole:2017syo} based on the Color Evaporation Model within a TMD approach and employing the GSF
%gluon Sivers function 
from Ref.~\cite{Boer:2003tx} for different center-of-mass energies are shown in Fig.~\ref{ANjpsi-NICA}(a) as functions of the rapidity $y$.
Since the $J/\psi$ production mechanism is not well understood, the measurement of the $A^{J/\psi}_{N}$ may bring a valuable input to that matter as well.

\begin{figure}[h]
  \begin{minipage}[ht]{0.5\linewidth}
  \center{\epsfig{file=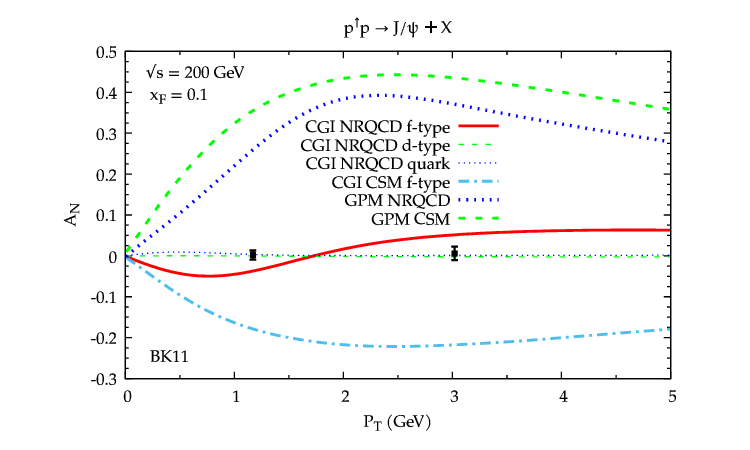,scale=0.7} \\ (a)}  
  \end{minipage}
  \hfill
  \begin{minipage}[ht]{0.5\linewidth}
    \center{\epsfig{file=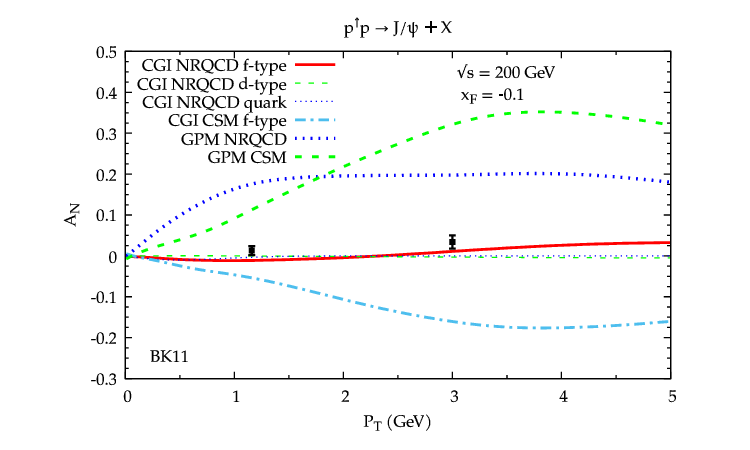,scale=0.7} \\ (b)}  
  \end{minipage}
\caption{Maximized contributions to $A_N$ estimates as a function of $p_T$ for the process $p^\uparrow p\to J/\psi + X$ at $\sqrt s=200$ GeV and $x_F=0.1$ (a) and $x_F=-0.1$ (b) adopting the CGI-GPM and GPM approaches, within the CS model and NRQCD. Data are taken from \cite{Aidala:2018gmp}. Reprinted figures from  Ref.~\cite{DAlesio:2020eqo}, \href{https://doi.org/10.1103/PhysRevD.102.094011}{\textcopyright (2020)} by the American
Physical Society under the Creative Commons Attribution 4.0 International License.}
\label{fig:allxF}
\end{figure}

In this context, a comprehensive analysis adopting different mechanisms for quarkonium production, namely the Color-Singlet Model and its extension to NRQCD, and two TMD schemes, the GPM and the CGI-GPM, has been recently performed in Refs.~\cite{DAlesio:2019gnu,DAlesio:2020eqo}. In Fig.~\ref{fig:allxF} we show a comparison among these options, adopting maximized gluon Sivers functions (that is by using ${\cal N}_g(x)=1$ and $\rho=2/3$ for all GSFs), for $A_N$ at RHIC kinematics, $\sqrt s = 200$ GeV and $x_F=0.1$ (a) and $x_F=-0.1$ (b), vs.~$p_T$. This indeed could be a valuable tool to quantify the potential role of the GSFs and the feasibility of their corresponding extraction. As one can see the $f$-type contribution in NRQCD (solid lines) suffers from the most effective cancellations, coming from the relative sign of the modified partonic cross-sections. The $d$-type one (thin dashed lines) is suppressed by the absence of the $gg$ channel, that dominates the unpolarized cross-section (this reflects also into the negligible contribution of the quark Sivers function). On the opposite side the GPM approach, both within the CS model (thick dashed lines) and in NRQCD (dotted lines), gives the potentially largest contributions. It is worth to remark that within the GPM the differences coming from the production mechanism are small. Moreover, even if to a much lesser extent as compared to the GPM, also within the CGI-GPM one can potentially put some constraints on the size of the $f$-type GSF even with the few data points available.

These estimates can be certainly affected by uncertainties, intrinsic in the model and/or induced by the nonperturbative parameters used. On the other hand, the relative size of these maximized contributions to the asymmetry, coming from different terms and models, is certainly under better control. One has to keep in mind that the ultimate goal of this comparison is to estimate the impact of ISIs and FSIs in the computation of such SSAs.

In Fig.~\ref{ANjpsi-NICA}(b) we show the analogous estimates for $pp$ collisions at NICA energy, $\sqrt s = 27$ GeV, this time at fixed $p_T$ as a function of $x_F$. A word of caution is mandatory: at these energies the quark-initiated contribution to the unpolarized cross-section, when adopting the NRQCD framework, is not negligible and, starting around $p_T\simeq 2$ GeV, becomes comparable with the gluon-initiated ones. On the other hand, when employing the parametrizations of the quark Sivers functions (as extracted from SIDIS data) their contribution to the SSA turns out to be completely negligible.  
Once again, in Fig.~\ref{ANjpsi-NICA}(b), we see that the largest maximized estimates are obtained within the GPM, both in the CSM and in NRQCD. In the CGI-GPM, the only sizeable contribution comes from the $f$-type GSF in the CS model. In this respect, a measurement of $A_N$ at NICA could represent an important tool to disentangle among the different approaches and to constrain the GSF. Moreover, a detailed comparison with the analogous results obtained at RHIC kinematics could help in better understanding the role of quark and gluon initiated subprocesses both in the unpolarized cross-sections and in SSAs. 

\begin{figure}[!h]
  \begin{minipage}[ht]{0.50\linewidth}
    \center{\epsfig{file=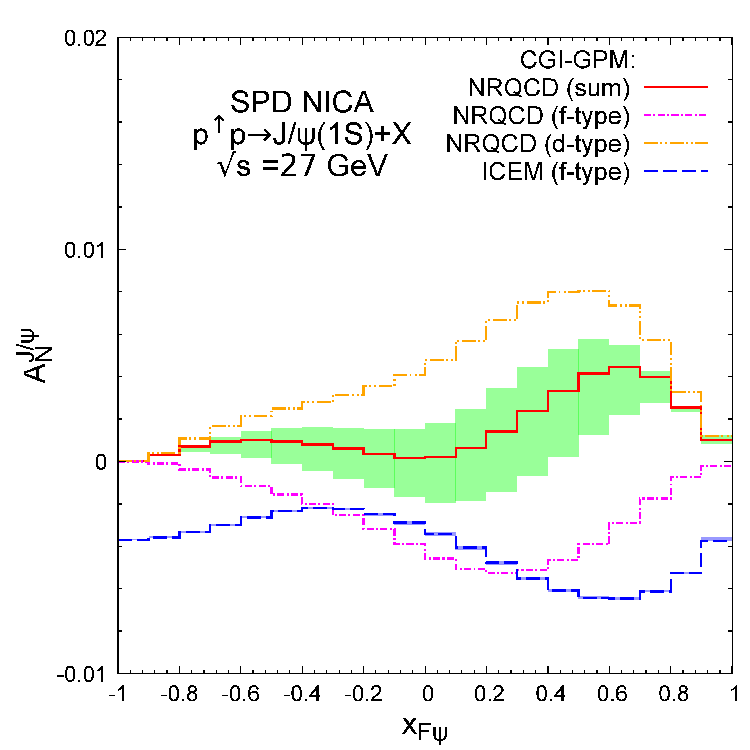,scale=0.55} \\ (a)} 
  \end{minipage}
  \hfill
  \begin{minipage}[ht]{0.50\linewidth}
     \center{\epsfig{file=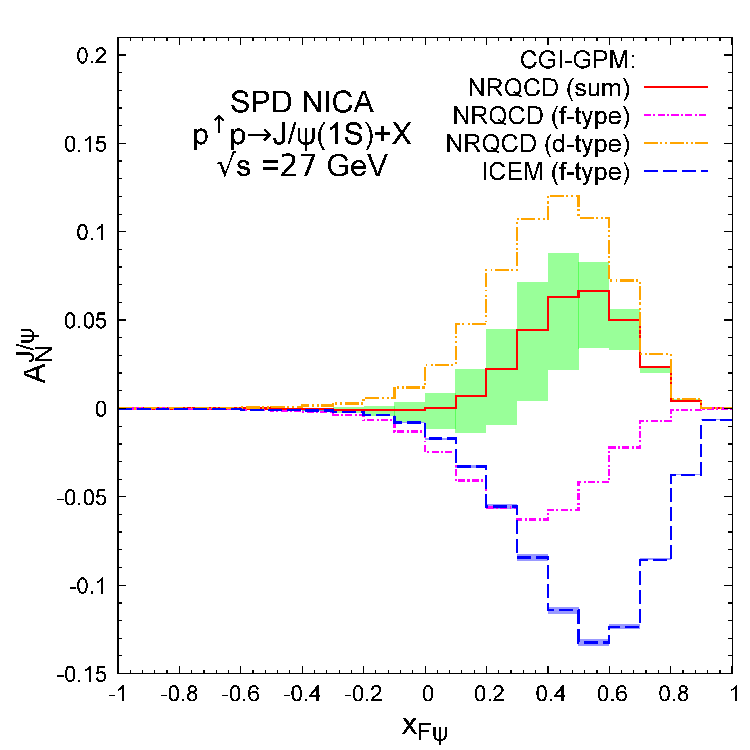,scale=0.55} \\ (b)} 
  \end{minipage}
  \caption{Comparisons of NRQCD and ICEM predictions for $A_N^{J/\psi}$ as function of $x_F$ in $p$-$p$ collisions at the energy $\sqrt{s}=27$~GeV obtained in CGI-GPM with D'Alesio et al. (a) and SIDIS1 (b) parameterizations of Sivers function. 
  }
  \label{fig:AN_Jpsi_VS} 
\end{figure}

Predictions for $A^{J/\psi}_{N}$ in proton-proton collisions at NICA energy $\sqrt{s}=27$ GeV, obtained in GPM $+$ NRQCD approach, as
function of $x_F$ and $p_T$ are shown in the Figure (\ref{fig:AN_Jpsi_VS}). For
comparison, results are presented for SIDIS1 \cite{Anselmino:2005ea} and D'Alesio et al.
\cite{DAlesio:2017rzj,DAlesio:2019gnu} parameterizations of proton Sivers
function.
%{SSA for charmonia may also shed light on the $J/\psi$ production mechanism}.
     
\begin{figure}[!h]
\begin{center}
 \begin{minipage}[ht]{0.49\linewidth}
     \center{\epsfig{file=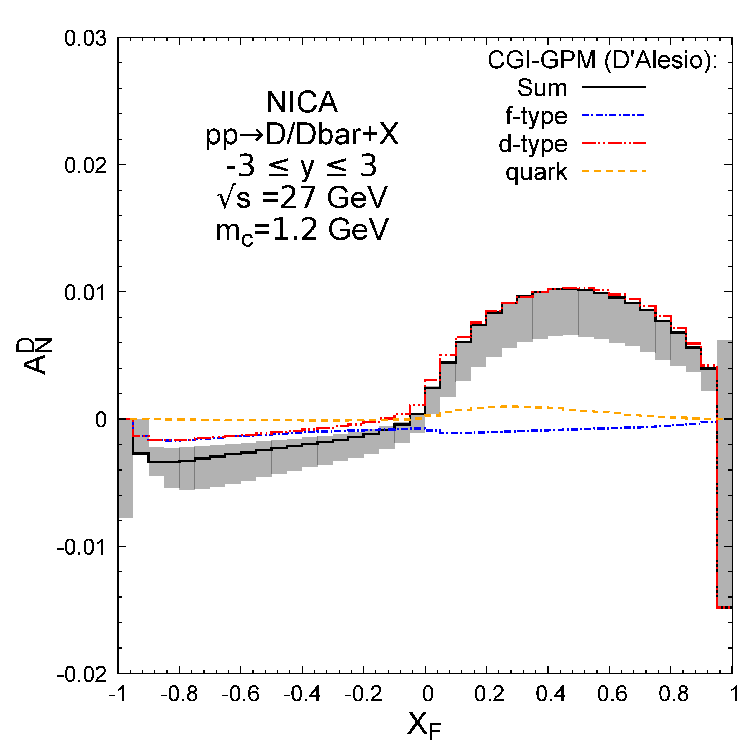,scale=0.55} \\ (a)} 
  \end{minipage}
  \hfill
  \begin{minipage}[ht]{0.49\linewidth}
       \center{\epsfig{file=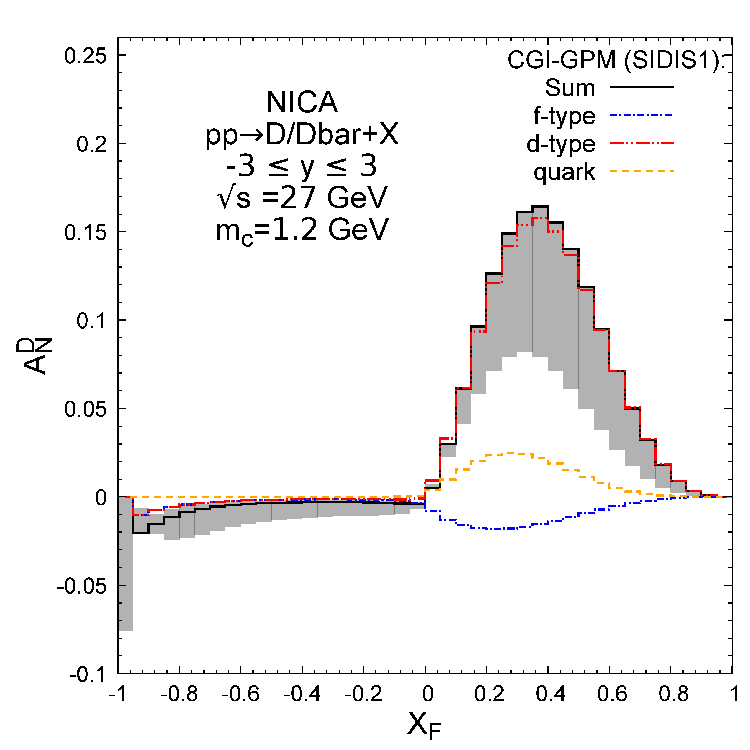,scale=0.55} \\ (b)} 
  \end{minipage}
\caption{CGI-GPM predictions for single-spin asymmetry in inclisive $D$-meson production with gluon Sivers function extracted in the D'Alesio {\it et.al.}~\cite{DAlesio:2017rzj,DAlesio:2019gnu} fit (a) and SIDIS1~\cite{Anselmino:2005ea} fit (b)}
\label{fig:AN_D}
\end{center}
\end{figure}

 A measurement with open-heavy hadrons (both $D$- and $B$-mesons) was performed at RHIC (PHENIX, $\sqrt{s}=200$ GeV)~\cite{Aidala:2017pum} using high-$p_T$ muons from their semileptonic decays. 
%{No clear indication of a nonzero SSA  was obtained in the results which have relatively large statistical uncertainties.} 
Obtained results are affected by relatively large statistical uncertainties and do not exhibit any significant non-zero asymmetry. Nevertheless, the results do not contradict the predictions of the twist-3 approach from Ref.~\cite{Koike:2011mb}. The Sivers effect contribution to the $A_N^D$ asymmetry calculated within the Generalized Parton Model for $\sqrt{s}=27$ GeV is presented in Fig.~\ref{fig:AN_D}.
 
 %An example of such prediction for $D^0$ and $\bar{D^0}$  as a function of $x_F$ together with the experimental points are presented in Fig. \ref{fig:AN_D} (b).

\begin{figure}[!h]
  \begin{minipage}[ht]{0.50\linewidth}
       \center{\epsfig{file=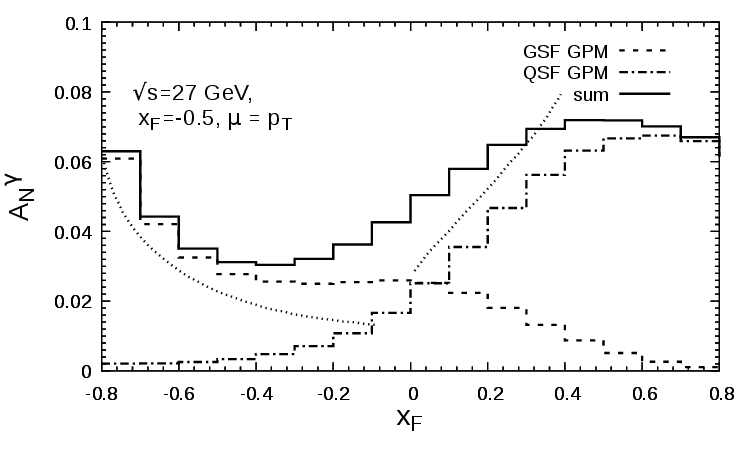,scale=0.6} \\ (a)} 
  \end{minipage}
  \hfill
  \begin{minipage}[ht]{0.50\linewidth}
         \center{\epsfig{file=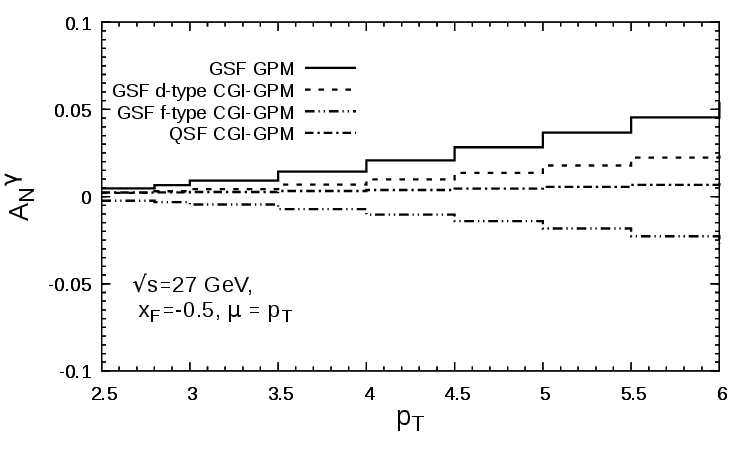,scale=0.6} \\ (b)} 
  \end{minipage}
  \caption{(a) $x_F$ dependence of the asymmetry $A_N^{\gamma}$ calculated adopting the SIDIS1 Sivers function for $\sqrt{s}=27$ GeV and $4<p_T<6$ GeV. Gluon and quark contributions are shown separately, see the legend in the plots. Dotted lines illustrate the twist-3 predictions for  $\sqrt{s} = 30$ GeV and $p_T = 4$ GeV for negative~\cite{Hammon:1998gb} and positive~\cite{Qiu:1991pp} values of $x_F$. (b) $p_T$ dependence of the $A_N^{\gamma}$ asymmetry for different values of $\sqrt{s}$ at $x_F=-0.5$.}
  \label{fig:AN_PP} 
\end{figure}

Measurement of the $A^{\gamma}_N$ for prompt photons provides a unique opportunity to study the Sivers PDF and twist-3 correlation functions
%{as soon as the corresponding hard process is free from the Collins effect since it does not involve fragmentation in the final state.} 
, since the corresponding hard process does not involve fragmentation in the final state and thus is exempt from the Collins effect.
The first attempt to measure $A^{\gamma}_N$ at $\sqrt{s}=19.4$ GeV was performed at the fixed target experiment E704 at Fermilab in the kinematic range $-0.15<x_F<0.15$ and 2.5 GeV/$c$ $<p_T<$ 3.1 GeV/$c$. The results were consistent with zero within large statistical and systematic uncertainties~\cite{Adams:1995gg}. Figure~\ref{fig:AN_PP}(a) shows the 
%expected 
results of our calculations of $A_N^{\gamma}$ asymmetry as a function of $x_F$ for $\sqrt{s}=27$~GeV based on the SIDIS1 parameterization
%extraction 
of the GSF. Quark and gluon contributions from the gluon Compton scattering, dominating at positive and negative values of $x_F$, respectively, are shown separately. The $q\bar{q}$ annihilation contribution is also presented. Dotted lines illustrate the twist-3 predictions for  $\sqrt{s} = 30$ GeV and $p_T = 4$ GeV/$c$ for negative~\cite{Hammon:1998gb} and positive~\cite{Qiu:1991pp} values of $x_F$.  The $p_T$ dependence of the $A_N^{\gamma}$ asymmetry at $x_F=-0.5$ is shown for different values of $\sqrt{s}$ in Fig.~\ref{fig:AN_PP}(b).
 
%As it has been described above, at low $p_T \ll Q$ the TMD Sivers function is responsible for the main contribution to the SSAs, while twist-3 contribution  dominates for  $p_T\sim Q$.  Nearly $4\pi$ acceptance of the SPD detector allows to operate in the wide range of $p_T$ and scan the transition between these two approaches. Running at the very challenging energy range where the condition $\Lambda_{QCD}\ll p_T \ll Q < \sqrt{s}$ does not work well. SPD has a unique chance to investigate the applicability of the TMD factorization approach at low $\sqrt{s}$ and test other approaches such as [].

 %    as far as experimentally demonstrating and measuring a gluon Sivers effect in transversely polarized protons, several complementary future possibilities exist, in which AFTER@LHC can play a very important role.
     
 %    As it has been described above, at low pT, the non-perturbative TMD Sivers function will be respon- sible for its SSA, while twist-3 dominates the contributions to the SSA when pT ∼ Q. At intermediate pT , one can see the transition between these two frameworks and a relation between Sivers function and Qiu-Sterman function has been shown in Ref. [31].

\subsection{Gluon transversity in deuteron \label{trans}}

%Nucleon spin structure has been investigated for many years   using mainly the longitudinal polarization of hadrons and  partons for solving the issue of the nucleon spin puzzle.Presently, however,
The transversely polarized parton densities   (called transversity distributions) within transversely polarized proton and deuteron are extensively investigated both theoretically and experimentally. A remarkable difference of the parton transverse distributions from the longitudinal ones is that the gluon transversity distribution does not exist in the spin-1/2 nucleons. For this reason, the $Q^2$- evolution of the quark transversity in a nucleon is independent of the gluon contributions \cite{Kumano:1997qp, Kumano:1997ng, Hirai:1997mm}, whereas the quark and gluon helicities evolve with $Q^2$ in a close relation to each other according to the DGLAP equations \cite{Hirai:1997gb}.

The quark transversity PDFs are less constrained experimentally compared to the helicity functions~\cite{Anselmino:2020vlp}, while the gluon transversity is presently unknown~\cite{Kumano:2019igu}. The latter one can be potentially studied in experiments with linearly polarized  deuterons.\footnote{In the case of the transversely polarized
deuteron, one also needs to account for the contributions of the transverse and tensor spin components, see Table \ref{table:polarizations} and discussion below.} %Main issue of such measurements can be formulated as follows. 
Traditionally, the deuteron is considered as a weakly coupled spin-1 bound state of a proton and neutron. Since proton and neutron themselves do not contain the gluon transversity, one could expect vanishing gluon transversity in the deuteron. However, if sizable values for the gluon transversity are measured, this might indicate the presence of new (different from  $|pn\rangle$ state) degrees of freedom in the deuteron.

%%%%%%%%%%%%%%%%%%%%%%%%%%%% figure %%%%%%%%%%%%%%%%%%%%%%%%%%%%
\begin{wrapfigure}[9]{r}{0.40\textwidth}
 %\vspace{-0.80cm}
\begin{center}
        \center{\epsfig{file=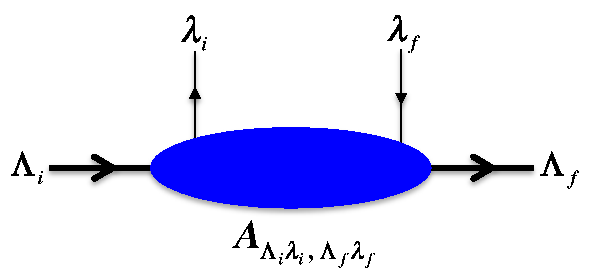,scale=0.6}}
\end{center}
%\vspace{-0.9cm}
\caption{Parton-hadron forward scattering amplitude 
$A_{\Lambda_i \lambda_i,\, \Lambda_f \lambda_f}$.}
\label{fig:helicity-amp}
%\vspace{-0.30cm}
\end{wrapfigure}
%%%%%%%%%%%%%%%%%%%%%%%%%%%% figure %%%%%%%%%%%%%%%%%%%%%%%%%%%%

The longitudinally and transversely polarized quark  distributions in a nucleon are defined as matrix elements on the light-cone  \cite{Kumano:2019igu}\footnote{ Functions $g_1^{q}$ and $h_1^{q}$ are also referred to as $\Delta q(x)$ and $\Delta_T q(x)$ correspondingly. }:
\begin{align}
g_{1}^{q} (x) & = \! \int \! \frac{d \xi^-}{4\pi} \, e^{i x p^+ \xi^-}
\left\langle \, p \, s_L \! \right | \, \bar\psi (0) 
\gamma^+ \gamma_5 \psi (\xi)  \, \left | p \, s_L \, \right\rangle 
_{\xi^+=\bm\xi_\perp=0}  ,
\nonumber \\[-0.00cm]
h_{1}^{q} (x) & = \! \int  \frac{d \xi^-}{4\pi} \, e^{i x p^+ \xi^-}
\left\langle \, p \, s_{Tj} \right | \, \bar\psi (0)  
\, i \, \gamma_5 \, \sigma^{j +} 
 \psi (\xi) \left | p \, s_{Tj} \, \right\rangle 
_{\xi^+=\bm\xi_\perp=0} ,
\label{eqn:delta-deltaT-qx}
\end{align}
where gauge links which ensure gauge invariance are omitted; $s_{L}$ and $s_{T}$ indicate the longitudinal and transverse polarizations of the nucleon, respectively. In Eqs.(\ref{eqn:delta-deltaT-qx}), the antisymmetric tensor $\sigma^{\mu\nu}$ is defined by
$\sigma^{\mu\nu} = \frac{i}{2} (\gamma^\mu \gamma^\nu -\gamma^\nu \gamma^\mu )$ while $a^\pm = (a^0 \pm a^3)/\sqrt{2}$.

\begin{figure}[h!]
  \begin{minipage}[ht]{0.48\linewidth}
  \vspace{0.60cm}
       \center{\epsfig{file=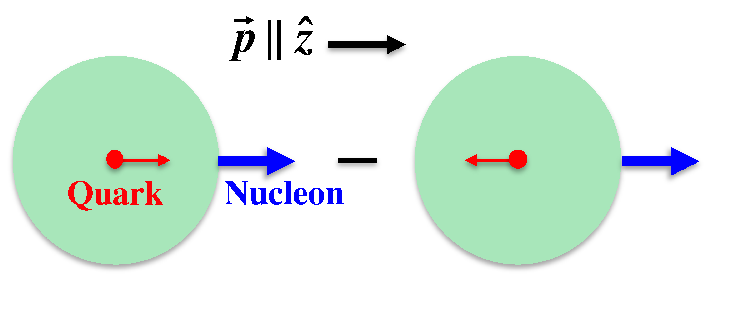,scale=0.6} \\ (a)} 
  \end{minipage}
    \begin{minipage}[ht]{0.48\linewidth}
       \center{\epsfig{file=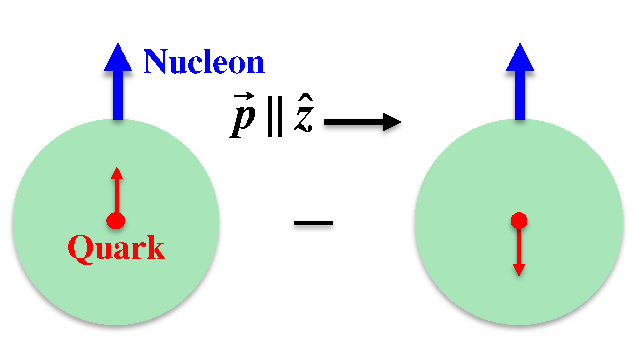,scale=0.6} \\ (b)}       
  \end{minipage}
\caption{Schematic illustration for quark helicty $(a)$ and transversity $(b)$ distribution functions.}
\label{fig:long-trans-polarizations}
\end{figure}

According to the optical theorem, hadron structure functions are determined by the imaginary part of corresponding  forward scattering amplitudes.
Such an amplitude, $A_{\Lambda_i \lambda_i ,\, \Lambda_f \lambda_f}$, is illustrated in Fig.\,\ref{fig:helicity-amp}, where the initial and final hadron (parton) helicities are denoted as $\Lambda_i$ and $\Lambda_f$ ($\lambda_i$ and $\lambda_f$). The helicity conservation yields $\Lambda_i - \lambda_i = \Lambda_f - \lambda_f$.
The quark densities \eqref{eqn:delta-deltaT-qx} are expressed in terms of helicity amplitudes as follows:
\begin{align}
g_{1}^{q} (x) &   = q_+ (x) - q_- (x) 
                 \sim \text{Im} \, (A_{++,\, ++} - A_{+-,\, +-}) ,
\nonumber \\
h_{1}^{q} (x) & = q_\uparrow (x) - q_\downarrow (x) 
                 \sim \text{Im} \, A_{++,\, --} \ .
\label{eqn:delta-deltaT-qx-amplitudes}
\end{align}
%One should be careful with the amplitude notation 
%because the amplitudes are often expressed
%by spin components along the quantization axis.
%%%%%
The density $g_{1}^{q} (x)$ is determined by the distributions $q_+ (x)$ and $q_- (x)$ which describe the quarks with spin oriented parallel and anti-parallel to the longitudinal nucleon spin, as shown in Fig.\,\ref{fig:long-trans-polarizations}(a). 
The quark transversity distribution is given by
$h_{1}^{q} (x) = q_\uparrow (x) - q_\downarrow (x)$,
where subscripts ``$\uparrow$" and ``$\downarrow$" indicate the quark polarizations which are parallel and anti-parallel 
to the transverse nucleon spin, as illustrated in Fig.\,\ref{fig:long-trans-polarizations}(b).
The relation $h_{1}^{q} (x)  \sim \text{Im} \, A_{++,\, --}$ indicates that the quark transversity distribution in a nucleon is described by the helicity-flip ($\lambda_i =+$, $\lambda_f =-$)  amplitude. Note that the density $g_{1}^{q} (x)$ is a chiral-even function while the distribution $h_{1}^{q} (x)$ is a chiral-odd one.

%%%%%%%%%%%%%%%%%%%%%%%%%%%% figure %%%%%%%%%%%%%%%%%%%%%%%%%%%%

%%%%%%%%%%%%%%%%%%%%%%%%%%%% figure %%%%%%%%%%%%%%%%%%%%%%%%%%%%
\begin{wrapfigure}[19.5]{r}{0.50\textwidth}
\vspace{-1.0cm}
\begin{center}
        \center{\epsfig{file=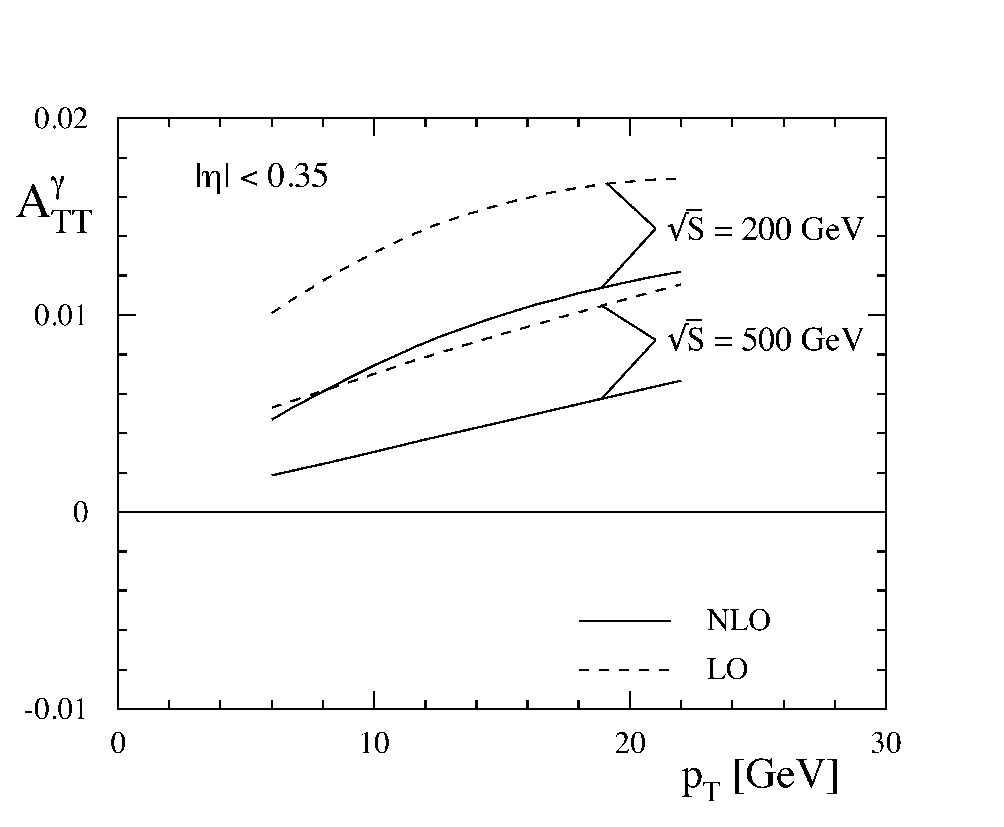,scale=0.5}}
\end{center}
\vspace{-0.3cm}
\caption{$A^{\gamma}_{TT}$ asymmetry for the prompt-photon production 
  at 200 and 500~GeV coming from the $q\bar{q}$ annihilation process 
  calculated in LO~\cite{Soffer:2002tf}  and NLO~\cite{Mukherjee:2003pf}. Adapted figure from~\cite{Mukherjee:2003pf} \textcopyright (2003) by the American Physical Society.}
  \label{fig:Deuteron-Agamma-TT} 
\vspace{-0.0cm}  
\end{wrapfigure}
%%%%%%%%%%%%%%%%%%%%%%%%%%%% figure %%%%%%%%%%%%%%%%%%%%%%%%%%%%

We see that the helicity and transversity functions, $g_{1}^{q} (x)$ and $h_{1}^{q} (x)$, have completely different nature and  properties. For this reason, their contributions to the spin asymmetries differ essentially from each other. Possible way to access the transversity are  measurements of single transverse spin asymmetries in lepton-proton and proton-proton collisions with interference dihadron fragmentation function used as a chiral-odd probe \cite{Radici:2018iag}.

The transversity PDF may also be studied through the measurement of the double transverse spin asymmetry, $A_{TT}$, defined analogously to the double longitudinal spin asymmetry, $A_{LL}$. This is a unique opportunity for SPD NICA, where transverse polarization of both colliding beams will be available. Due to the absence of transversity gluon contributions, the transverse asymmetries in $pp$ collisions are expected to be much smaller than the longitudinal ones, $A_{TT} \ll A_{LL}$. As an example, the asymmetry $A^{\gamma}_{TT}$ in the prompt-photon 
production at 200 and 500 GeV coming from the $q\bar{q}$ annihilation process calculated in LO \cite{Soffer:2002tf} 
and NLO \cite{Mukherjee:2003pf} is shown in 
Fig.~\ref{fig:Deuteron-Agamma-TT}. Predicted values for the quantity $A^{\gamma}_{TT}$ are of the order of one per cent. On the other hand, in case of $pd$ or $dd$ collisions a measurable value of $A_{TT}$ could be expected in case of presence of sizable contributions of non-nucleon degrees of freedom in deuteron.

Note that the quark transversity distributions play important role in correct interpretation of data on the neutron electric dipole moment (EDM), which is used in search for physics beyond the standard model (BSM).
The current experimental lower bound of the neutron EDM is  close to the typical BSM predictions, $10^{-28}$--$10^{-26}$\,e$\cdot$cm.
For more accurate comparison with data, one needs to convert the quark-level BSM results for EDMs to the neutron one by using the quark transversity distributions \cite{Kumano:2019igu}:
$d_n   =  \sum_q d_q \, h_{1}^{q}$, where
$h_{1}^{q}  \equiv \int_0^1 dx \, 
        \left [ h_{1}^{q} (x) - h_{1}^{\bar{q}} (x) \right ]$
and $d_q$ is the quark EDM.
%The studies of the quark transversity distributions are also 
%valuable for choosing a BSM signature.

Let us discuss the polarized gluon distributions in the deuteron. The spin-1 states are described by the polarization vectors $\bm E$ defined as
\begin{align}
\bm E_\pm =&\frac{1}{\sqrt{2}} \left ( \, \mp 1,\, -i,\, 0 \, \right ) ,&  
\ \ 
\bm E_0  =& \left ( \, 0,\, 0,\, 1 \, \right ) ,&
\ \ 
\bm E_x  =& \left ( \, 1,\, 0,\, 0 \, \right ) ,&
\ \ 
\bm E_y  = &\left ( \, 0,\, 1,\, 0 \, \right ) .
\label{eqn:dct}
\end{align}
The vectors $\bm{E}_+$, $\bm{E}_0$, and $\bm{E}_-$
correspond to the spin states with $z$ component equal to 
$s_z =+1$, $0$, and $-1$; $\bm{E}_x$ and $\bm{E}_y$ describe the linear polarization.
The vector, $\bm{S}$, and tensor, $T^{ij}$, polarizations of a spin-1 particle can be expressed in terms of vectors $\bm E$ as
\begin{align}
\bm{S} =&\, {\rm Im}\,(\,\bm{E}^{*} \times \bm{E}\,),& 
T^{ij}  =&\, \frac{1}{3}\, \delta^{ij} 
       - {\rm Re}\,(\,{E^i}^{*} E^j\,).
\label{eqn:spin-1-vector-tensor-2}
\end{align}
We parameterize these polarizations in the form
\begin{align}
\bm{S}  =& (S_{T}^x,\, S_{T}^y,\, S_L),&
T^{ij}   =& \frac{1}{2} 
\left(
    \begin{array}{ccc}
     - \frac{2}{3} S_{LL} + S_{TT}^{xx}    & S_{TT}^{xy}  & S_{LT}^x  \\[+0.20cm]
     S_{TT}^{xy}  & - \frac{2}{3} S_{LL} - S_{TT}^{xx}    & S_{LT}^y  \\[+0.20cm]
     S_{LT}^x     &  S_{LT}^y              & \frac{4}{3} S_{LL}
    \end{array}
\right) ,
\label{eqn:st-1}
\end{align}
where $\Big\{S_{T}^x, S_{T}^y, S_L, S_{LL}, S_{TT}^{xx}, S_{TT}^{xy}, S_{LT}^x, S_{LT}^y\Big\}$ is the full set of parameters describing the vector and tensor polarizations.

For example, in the case of the polarization vector $\bm E_+$, Eqs.~(\ref{eqn:spin-1-vector-tensor-2}) lead to the following set of parameters:
\begin{align}
S_L  =& 1,& S_{LL} =& 1/2,& 
S_{T}^x =& S_{T}^y = S_{TT}^{xx} 
= S_{TT}^{xy} = S_{LT}^x =S_{LT}^y =0 .
\label{eqn:spin-parameters+1}
\end{align}

The parameters of longitudinal, transverse, 
and linear polarizations of a spin-1 particle are listed in Table~\ref{table:polarizations}.
The polarization $\bm E_+$ means that the spin is directed  along $z$ axis, $S_L=1$. Note however that this polarization can also be described by the tensor parameter $S_{LL}=1/2$.

%%%%%%%%%%%%%%%%%%%%%%%%%%%%%%%%%%%%%%%%%%%%%%%%%%%%%%%%%%%%%%%%%%
\begin{table}[t!]
\renewcommand{\arraystretch}{1.5}
\begin{center}
\begin{tabular}{|l|c|c|c|c|c|c|} \hline
Polarizations & $\bm E$ & 
       $S_T^x$ & $S_T^y$ & $S_L$ & $S_{LL}$ & $S_{TT}^{xx}$ 
\bigstrut \\ \hline
%%%%%
Longitudinal $+z$ & $\frac{1}{\sqrt{2}} (-1,\, -i,\, 0)$ &
          0    &   0     &  $+$1 & $+\frac{1}{2}$ &   0  \bigstrut \\ \hline
Longitudinal $-z$ & $\frac{1}{\sqrt{2}} (+1,\, -i,\, 0)$ &
          0    &   0     &  $-$1 & $+\frac{1}{2}$ &   0  \bigstrut \\ \hline
%%%%%
Transverse $+x$ & $\frac{1}{\sqrt{2}} (0,\, -1,\, -i)$ &
         $+$1    &   0     &  0  & $-\frac{1}{4}$ & $+\frac{1}{2}$ \bigstrut \\ \hline
Transverse $-x$ & $\frac{1}{\sqrt{2}} (0,\, +1,\, -i)$ &
         $-1$    &   0     &  0  & $-\frac{1}{4}$ & $+\frac{1}{2}$ \bigstrut \\ \hline
%%%%%
Transverse $+y$ & $\frac{1}{\sqrt{2}} (-i,\, 0,\, -1)$ &
         0   &   $+$1      &  0  & $-\frac{1}{4}$ & $-\frac{1}{2}$ \bigstrut \\ \hline
Transverse $-y$ & $\frac{1}{\sqrt{2}} (-i,\, 0,\, +1)$ &
         0   &   $-1$     &  0  & $-\frac{1}{4}$ & $-\frac{1}{2}$ \bigstrut \\ \hline
%%%%%
Linear  $x$  &  $(1,\, 0,\, 0)$ &
         0   &   0      &  0  & $+\frac{1}{2}$ & $-1$ \bigstrut \\ \hline
Linear  $y$  &  $(0,\, 1,\, 0)$ &
         0   &   0     &  0  & $+\frac{1}{2}$ & $+1$ \bigstrut \\ \hline
%%%%%
\end{tabular}
\caption{Longitudinal, transverse, and linear polarizations
of the deuteron: vectors $\bm{E}$ and parameters 
of the vector, $\bm{S}$, and tensor, $T^{ij}$, polarizations.
All other parameters vanish for the considered cases,
$S_{TT}^{xy} = S_{LT}^{x} = S_{LT}^{y}=0$.
}
\label{table:polarizations}
\end{center}
\end{table}
%%%%%%%%%%%%%%%%%%%%%%%%%%%%%%%%%%%%%%%%%%%%%%%%%%%%%%%%%%%%%%%%%%
%%%%%%%%%%%%%%%%%%%%%%%%%%%% figure %%%%%%%%%%%%%%%%%%%%%%%%%%%%
\begin{wrapfigure}[12]{r}{0.40\textwidth}
 %\vspace{-0.80cm}
\begin{center}
        \center{\epsfig{file=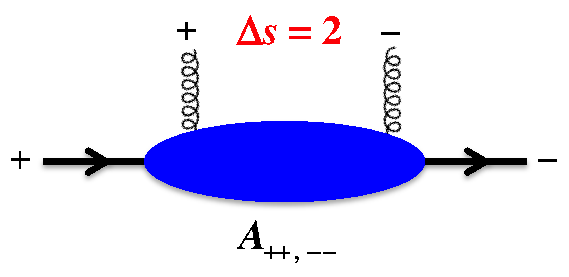,scale=0.6}}
\end{center}
%\vspace{-0.9cm}
\caption{Gluon-hadron forward scattering amplitude 
with the spin flip of 2.}
\label{fig:helicity-amp-transversity}
%\vspace{-0.30cm}
\end{wrapfigure}
%%%%%%%%%%%%%%%%%%%%%%%%%%%% figure %%%%%%%%%%%%%%%%%%%%%%%%%%%%

As mentioned above, the gluon transversity distribution is forbidden in the spin-1/2 nucleons; it however may contribute to a spin-1 hadron \cite{Jaffe:1989xy}. Similarly to the quark distribution \eqref{eqn:delta-deltaT-qx-amplitudes}, the gluon transversity\footnote{Denoted as $h_{1}^{g}$ or $\Delta_T g$.} is given by the helicity-flip amplitude:
\begin{align}
h_{1}^{g} (x) & \sim \text{Im} \, A_{++,\, --} \ .
\label{eqn:delta-deltaT-gx-amplitudes}
\end{align}
We see that the function $h_{1}^{g} (x)$ is associated with the spin flip $\Delta s = 2$ amplitude as shown in Fig.~\ref{fig:helicity-amp-transversity}. Therefore, in search for the gluon transversity, one should use hadrons with spin $ \ge 1$. The most appropriate candidate is the deuteron: it is stable and can be polarized. 

The gluon transversity distribution in the deuteron is defined 
in the matrix-element form as
\begin{align}
h_{1}^{g} (x) = \varepsilon_{TT}^{\alpha\beta}
\int  \frac{d \xi^-}{2\pi} \, x p^+ \, e^{i x p^+ \xi^-} 
\langle \, p \, E_{x} \left |  A_{\alpha} (0) \, A_{\beta} (\xi)  
\right | p \, E_{x} \, \rangle 
_{\xi^+=\bm\xi_\perp=0} ,
\label{eqn:delta-deltaT-gx}
\end{align}
where $E_x$ descries the linear polarization of the deuteron
along the positive $x$-axis. The tensor $\varepsilon_{TT}^{ \alpha \beta}$ in Eq.\,(\ref{eqn:delta-deltaT-gx})
is defined by
$ \varepsilon_{TT}^{ \alpha \beta} \equiv 
      \varepsilon_x^\alpha \varepsilon^{*\beta}_x
       -\varepsilon_y^\alpha \varepsilon^{*\beta}_y $, 
where $\varepsilon_{x,y}^{\alpha}=(0, \bm{\varepsilon}_{x,y} \,)$ describe the linearly polarized gluons in the deuteron. So, one needs the linearly polarized deuteron target (or beam) for measuring the gluon transversity $h_{1}^{g} (x)$. As shown in Refs.~\cite{Kumano:2019igu,Kumano:2020gfk}, measurements of  the difference of cross-sections with the deuteron linearly polarized in $"x"$ and $"y"$ directions, $d\sigma (E_x)-d\sigma (E_y)$, will allow to determine the function $h_{1}^{g} (x)$.

Presently, experimental data on the gluon transversity are not available. There is only the JLab Letter of Intent \cite{jlab-gluon-trans} which proposes to measure $h_{1}^{g} (x)$ in the electron scattering off polarized deuteron \cite{Keller:2020wan}. The gluon-induced 
Drell-Yan process $qg\to q\gamma^*\to q\mu^+\mu^-$ was also 
proposed in Refs.~\cite{Kumano:2019igu,Kumano:2020gfk}
as a way to access $h_{1}^{g} (x)$ in polarized $pd$ collisions in the SpinQuest experiment at Fermilab~\cite{Fermilab-dy}. 

\begin{wrapfigure}[17]{r}{0.5\textwidth}
\vspace{-0.9cm}
\begin{center}
        \center{\epsfig{file=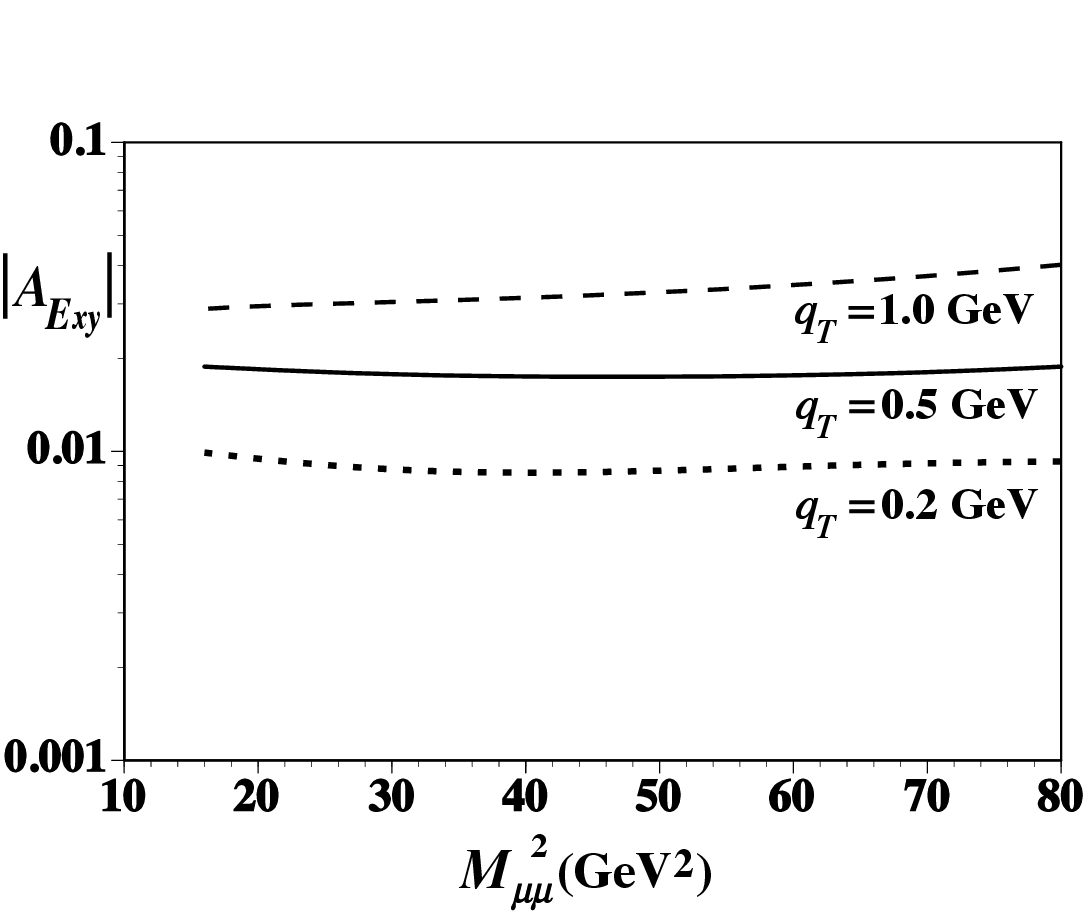,scale=0.43}}
\end{center}
\vspace{-0.0cm}
\caption{Spin asymmetry $|A_{E_{xy}}|$ 
for the proton-deuteron Drell-Yan process.}
\label{fig:AExy}
\vspace{-0.30cm}
\end{wrapfigure}

In Fig.~\ref{fig:AExy}, we present predictions for the asymmetry $A_{E_{xy}}=[d\sigma (E_x)-d\sigma (E_y)]
/[d\sigma (E_x)+d\sigma (E_y)]$ in the $pd$ Drell-Yan process with unpolarized proton as a function of $M_{\mu\mu}^2$. The results are given for the di-muon transverse momenta $q_T=0.2$, 0.5, and 1.0 GeV at the azimuthal angle $\phi=0$ and rapidity $y=0$. Note that, in these predictions, the gluon transversity is assumed to be equal to the longitudinally polarized gluon  distribution. For this reason, the asymmetry $|A_{E_{xy}}|$ presented in Fig. \ref{fig:AExy} may be overestimated. At NICA, the $J/\psi$ production in $pd$ and $dd$ collisions with a linearly polarized deuteron can be used to access the gluon transversity. The asymmetry $A_{E_{xy}}$ in these processes is  expected to be of the same order of magnitude, i.e. of the order of one per cent.

%%%%%%%%%%%%%%%%%%%%%%%%%%%% figure %%%%%%%%%%%%%%%%%%%%%%%%%%%%
%%%%%%%%%%%%%%%%%%%%%%%%%%%% figure %%%%%%%%%%%%%%%%%%%%%%%%%%%%
In the above examples, the single spin asymmetry in the reactions with linearly polarized deuterons was discussed. In principle, non-vanishing contributions of the gluon transversity $h_{1}^{g}(x)$ to the double transverse spin asymmetry, $A_{TT}$, are also possible. However, in the case of $A_{TT}$,  one should take into account two more types of contributions: 
%(1) 
quark and anti-quark transversities and 
%(2) 
tensor-polarized PDFs. For this reason, the single spin asymmetry in the processes with linearly polarized deuterons,  $A_{E_{xy}}$, seems to be most direct way to probe the function $h_{1}^{g}(x)$ \cite{Kumano:2019igu,Kumano:2020gfk}. 

%\vfill\eject
%%%%%%%%%%%%%%%%%%%%%%%%%%%%%%%%%%%%%%%%%%%%%%%%%%%%%%%%%%%%%%%%%%%%%%%%%%%%%%%%
\subsection{Tensor-polarized gluon distribution in deuteron \label{tensor-g}}

The polarized $ep$ DIS is described in terms of four structure functions ($F_{1,2}$, $g_{1,2}$), whereas the $ed$ DIS contains eight functions. The additional four structure functions, $b_{1-4}$, can be measured using an unpolarized lepton beam \cite{Frankfurt:1983qs,Hoodbhoy:1988am}. Among them, $b_1$ and $b_2$ are twist-2 structure functions which are
related to each other by the Callan-Gross type relation $b_2 = 2 x b_1$, while $b_3$ and $b_4$ are higher-twist ones. In the LO in $\alpha_s$, the $b_1$ (and $b_2$) structure function can be expressed in terms of tensor-polarized quark and anti-quark distribution functions $\delta_{_T} q$ and $\delta_{_T} \bar q$ as
\begin{align}
b_1 (x,Q^2)_{\text{LO}} 
    = \frac{1}{2} \sum_i e_i^2 \left [ \delta_{_T} q_i (x,Q^2) 
      + \delta_{_T} \bar q_i (x,Q^2) \right ].
\label{eqn:b1-parton}
\end{align}
The tensor-polarized PDFs are defined by
\begin{align}
\delta_{_T} f (x,Q^2) \equiv f^0 (x,Q^2) 
          - \frac{f^{+1} (x,Q^2) +f^{-1} (x,Q^2)}{2},\qquad
(f=q, \bar{q}, \text{or}\,\, g),
\label{eqn:tensor-pdf}
\end{align}
where $f^\lambda$ indicates an unpolarized parton distribution
in the hadron spin state $"\lambda"$.
In practice, the deuteron is best suited for measuring $b_1$ because it is stable and simplest spin-one hadron. In the framework of the parton model, the following sum rule was derived \cite{Close:1990zw,Kumano:2014pra}:
\begin{align}
 \int dx \, b_1 (x)_{\text{LO}}
    = - \frac{5}{24}\lim_{t \to 0} \, t \, F_Q (t) 
     + \sum_i e_i^2 
     \int dx \, \delta_{_T} \bar q_i (x) ,
\label{eqn:b1-sum}
\end{align}
where $F_Q (t)$ is the electric quadrupole form factor of the considered hadron.
Since the first term vanishes, non-zero values of $b_1$ could  indicate nonvanishing tensor-polarized anti-quark distributions.
Eq. \eqref{eqn:b1-sum} is an analogue of the Gottfried sum rule which was derived for the unpolarized $ep$ DIS \cite{Gottfried:1967kk,Kumano:1997cy}:
\begin{align}
\int \frac{dx}{x}
 \, [F_2^p (x) - F_2^n (x) ] _{\text{LO}}
   =  \frac{1}{3} 
   +\frac{2}{3} \int dx \, [ \bar u(x) - \bar d(x) ] .
\label{eqn:gottfried}
\end{align}

This sum rule indicates that the light anti-quark distributions
$\bar u$ and $\bar d$ are not identical if Eq. \eqref{eqn:gottfried} deviates from 1/3. The first terms in  r.h.s. of Eqs.\eqref{eqn:b1-sum} and \eqref{eqn:gottfried} (0 and 1/3, respectively) differ essentially from each other because the number of valence quarks depends on the flavor but is independent of the tensor polarization. The deviation of the sum rule \eqref{eqn:gottfried} from 1/3 was experimentally confirmed by the New Muon Collaboration (NMC) \cite{Amaudruz:1991at,Arneodo:1994sh} and later $pp$ and $pd$ Drell-Yan measurements~\cite{Baldit:1994jk,Hawker:1998ty,Towell:2001nh}.
Since large deviation from 1/3 cannot be explained within the  perturbative QCD, the NMC result created a new field of hadron physics related to the flavor-asymmetric anti-quark distributions \cite{Kumano:1997cy,Garvey:2001yq,Peng:2014hta}.
%%%%%
Similarly, a new field of tensor-polarized anti-quark 
distributions could be created if the $b_1$ sum \eqref{eqn:b1-sum} will be non-zero.
Such an indication was obtained in the HERMES experiment, 
$\int dx b_1 (x) 
 =  \, 0.35 \pm 0.10 \, (\text{stat}) \pm 0.18 \, (\text{sys})$
\cite{Airapetian:2005cb}.
Furthermore, the conventional convolution calculations of $b_1$ based on the standard deuteron model are very different from the HERMES measurements  \cite{Airapetian:2005cb} in magnitude and $x$ dependence \cite{Cosyn:2017fbo}.

%%%%%%%%%%%%%%%%%%%%%%%%%%%% figure %%%%%%%%%%%%%%%%%%%%%%%%%%%%
\begin{wrapfigure}[19]{r}{0.50\textwidth}
\vspace{-0.30cm}
\begin{center}
        \center{\epsfig{file=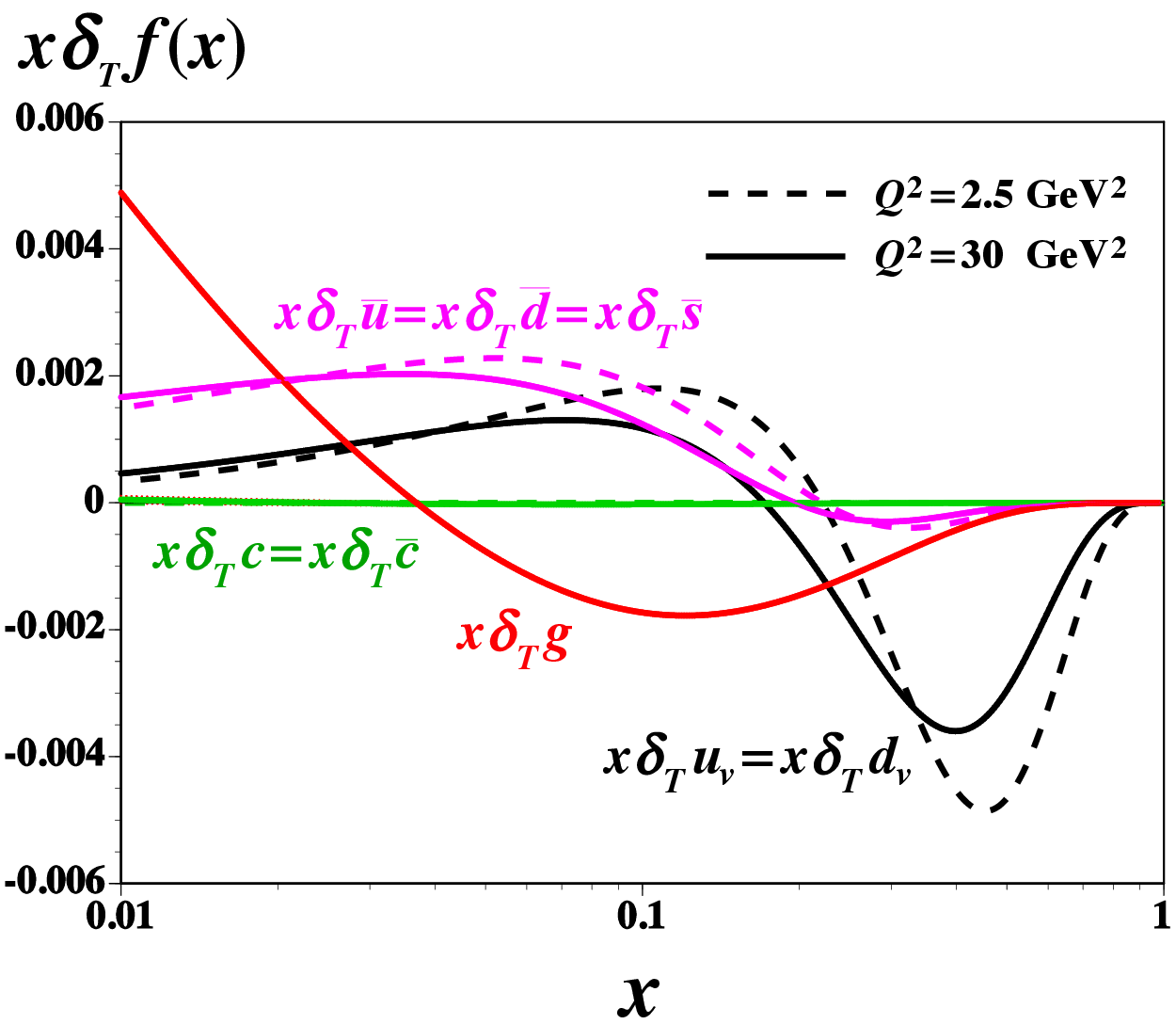,scale=0.37}}
\end{center}
\vspace{-0.cm}
\caption{$Q^2$ evolution of the tensor-polarized PDFs.}
\label{fig:tensor-gluon}
\vspace{-0.30cm}
\end{wrapfigure}
%%%%%%%%%%%%%%%%%%%%%%%%%%%% figure %%%%%%%%%%%%%%%%%%%%%%%%%%%%

The tensor-polarized quark and anti-quark distributions were fitted to the HERMES data in Ref.~\cite{Kumano:2010vz}. The obtained results for the quark and anti-quark densities at $Q^2=2.5$ GeV$^2$ (the averaged value of the HERMES measurements) are shown in Fig.~\ref{fig:tensor-gluon} by dashed curves. Since experimental information on the gluon tensor polarization is presently unavailable, $\delta_{_T} g=0$ for $Q^2=2.5$ GeV$^2$ has been assumed. However the tensor-polarized PDFs satisfy the same standard DGLAP evolution equations as the unpolarized ones \cite{Hoodbhoy:1988am,Kumano:2016ude}. Due to this evolution, a non-zero tensor-polarized gluon distribution will appear at  large values of $Q^2$ even if it does not exist at small $Q^2$. The DGLAP predictions for $\delta_{_T} f$ ($f=q,\bar{q},g$) at $Q^2=30$ GeV$^2$ are given in Fig.~\ref{fig:tensor-gluon} by solid curves.

Further experimental studies of $b_1$ will be performed at JLab 
\cite{Jlab-b1}. The tensor-polarized deuteron target for this experiment is under development~\cite{Keller:2020wan}.
Tensor-polarized quark and anti-quark distributions could be probed at Fermilab \cite{Fermilab-dy} using the proton-deuteron Drell-Yan process \cite{Hino:1998ww,Hino:1999qi,Kumano:2016ude}.
%%%%%%
The tensor-polarized gluon distribution could be studied at NICA using the tensor-polarized deuteron in prompt-photon and J/$\psi$ production processes. 
%\old{It should be a unique project to measure at NICA the gluon part complementary to the JLab and Fermilab projects for measuring the tensor-polarized quark and anti-quark distribution functions.}

\subsection{Deuteron tensor polarization and shear forces \label{TensorD}}

The availability of tensor polarized deuteron beam opens a possibility to study shear forces 
generated by quarks and gluons~\cite{Teryaev:2019ale}.  
%The structure of spin -1 particle energy momentum tensor matrix %elements is quite rich~\cite{Cosyn:2019aio}. 
The natural way to get the traceless part of the energy-momentum tensor related to a shear 
is provided just by the tensor polarization, as the relevant tensor $S^{\mu \nu}$ is a traceless 
one by construction. The contribution of the "tensor-polarized" parton distribution 
$C^T$~\cite{Hoodbhoy:1988am,Close:1990zw} (introduced as an "aligned" one~\cite{Efremov:1981vs}) 
is constrained by the zero sum rule~\cite{Efremov:1981vs} for its second moment 
(complementing the Close-Kumano sum rule~\cite{Close:1990zw}) which may be decomposed into 
quark and gluon components~\cite{Efremov:1994xf}:
\begin{eqnarray}
\label{sr}
\sum_{i=q, \bar q} \int_0^1 \delta_{T i} (x.Q^2) x dx = \delta_T (Q^2), \\
\int_0^1 \delta_{T G} (x,Q^2) x dx = - \delta_T (Q^2).
\end{eqnarray}
As a result, the matrix elements of the energy momentum tensors of quarks and gluons look as
\begin{eqnarray}
\sum_{i} \langle P, S|T_{i}^{\mu \nu}|P, S \rangle _{Q^2}=
2 P^{\mu} P^{\nu} (1-\delta(Q^2)) +2 m_N^2 S^{\mu \nu} \delta_T(Q^2)\\
\langle P, S|T_g^{\mu \nu}|P, S \rangle _{\mu ^2}=
2 P^{\mu} P^{\nu} \delta(Q^2) - 2 m_N^2 S^{\mu \nu} \delta_T(Q^2),
\end{eqnarray}
where the second terms describe the average (integrated over transverse distance) shear force. 
Here $m_N$ is the nucleon mass. 

The zero sum rules (\ref{sr}) were later interpreted~\cite{Teryaev:2009zz} as yet another manifestation 
of Equivalence Principle (EP), as it was done earlier~\cite{Teryaev:1999su} for the Ji sum rules. 
In turn, the smallness of $\delta_T$, compatible with the existing HERMES data, 
was suggested~\cite{Teryaev:2009zz} to be the new manifestation of the Extended Equivalence Principle 
(ExEP)~\cite{Teryaev:2003ch,Teryaev:2006fk, Teryaev:2016edw} valid separately for quarks and gluons 
in non-perturbative QCD due to the confinement and chiral symmetry violation. 
It was originally suggested for anomalous gravitomagnetic moments~\cite{Teryaev:2003ch,Teryaev:2016edw}.
In particular, it provides the rotation of spin in a terrestrial experiment with the angular velocity 
of Earth's rotation. Let us stress, that it may seem trivial if spin is considered just as a vector. 
However, it became highly non-trivial if one takes into account that the device which measures the spin
is rotating together with Earth. This is a particular example of the practical importance of 
the quantum theory of measurement. Another example may be represented by the Unruh radiation 
in heavy-ion collisions~\cite{Prokhorov:2019cik}, which implies that the particles production 
may be also considered as a quantum-mechanical measurement in a non-inertial hadronic medium. 

Recently, ExEP was also discovered for the pressure~\cite{Polyakov:2018exb}.

To check ExEP for the shear force, one may use future studies of DIS at JLab and of Drell-Yan processes 
with tensor polarized deuterons~\cite{Kumano:2017uel}~\footnote{Complementary probes are provided by 
vector mesons~\cite{Teryaev:2006fk}.}.

Note that tensor polarized parton distribution may be also measured in {\it any} hard process with 
the relevant combination of deuteron polarizations, in particular, for large $p_T$ pion production, 
providing much better statistics. The correspondent quantity can be the P-even single spin asymmetry
\begin{equation}
A_T=\frac{d \sigma(+)+ d \sigma(-)- 2 d \sigma(0)}{d \sigma(+)+d \sigma(-) + d \sigma(0)} \sim \frac{\sum_{i=q,\bar{q},g} \int d\hat \sigma_i \delta_{Ti}(x)}{\sum_{i=q,\bar{q},g} \int d\hat \sigma_i f_i(x)} ,
\end{equation}
where the differential cross-section with a definite deuteron polarization appears. 

Note that since the polarization tensor is traceless, the sum  rule is valid for the three mutually orthogonal orientations 
of coordinate frame~\cite{Efremov:1981vs}:
\begin{equation}
\sum_{i} \rho^i_{00}=1; \,\ \sum_{i} S_{ii}=0.
\end{equation}
As a result, the leading twist kinematically dominant "longitudinal" tensor polarization can be obtained by 
accelerating {\it transversely} polarized deuterons which will be accessible at NICA.

\section{Summary \label{sum}}
%{Possibilities of the Spin Physics Detector at NICA to contribute to ourcunderstanding of the gluon content in the proton and deuteron are discussed. High-luminosity $p$-$p$ and $d$-$d$ collisions with longitudinal or transverse beam polarization at the center-of-mass energy up to 27~GeV are planned for that. A unique possibility to use in parallel three gluon-induced processes as probes: inclusive production of charmonia ($J/\psi$ and higher states), open charm, and prompt photons, will be realized at SPD.}
In this Review we have discussed theoretical aspects of past and possible future measurements related to the study of the (un)polarized gluon content of the proton and deuteron, particularly focusing on the opportunities opening for the future Spin Physics Detector project at the NICA collider in Dubna. Proposed measurements at SPD are foreseen to be carried out performing a high-luminosity $p$-$p$, $p$-$d$ and $d$-$d$ collisions at the center-of-mass energy up to 27~GeV using longitudinally or transversely polarized proton and vector- or tensor-polarized deuteron beams. The SPD will have a unique possibility to probe gluon content employing simultaneously three gluon-induced processes: the inclusive production of charmonia ($J/\psi$ and higher states), open charm production, and production of the prompt photons.
The kinematic region to be covered by the SPD is unique and has never been accessed purposefully in polarized hadronic collisions. The data are expected to provide inputs for gluon physics, mostly in the region around $x\sim 0.1$ and above. % to give a glimpse on higher $x$ as well.
The expected event rates for all three aforementioned production channels are sizable in the discussed kinematic range and the experimental setup is being designed to increase the registration efficiency for the final states of interest. 
%{Mostly gluons with $x\sim0.1$ can be accessed although higher $x$ will also be touched. The kinematic region covering by SPD has never accessed before in polarized hadronic interactions.}

%\textcolor{red}{[M.N.] Charmonium production will become invaluable tool to study gluon component of polarized and un-polarized proton, once heavy quarkonium production theory solves it's phenomenological ``puzzles'' and reaches maturity. In a meanwhile, studies of heavy-quarkonium production mechanism and hadron structure studies should become complementary and SPD-NICA can provide many useful measurements to distinguish different models, simultaneously constraining our knowledge of gluon content of the proton.}

Despite certain phenomenological difficulties currently present in the heavy quarkonium production theory, charmonium production is an invaluable tool to study the gluon content of (un)polarized protons. Future measurements at the NICA SPD will provide complementary inputs for the study of heavy-quarkonium production mechanisms and explicit hadron structure aspects. In particular, SPD data should help to validate and constrain different theoretical models available on the market.

Precise data on the total cross-section of the open charm production at energies not so far from the production threshold should significantly reduce the present  uncertainties in the $c$-quark mass and $\alpha_s$ at a GeV scale. This will essentially improve description of the processes with charmed particles in the framework of  perturbative QCD. 
Measuring the $c\bar{c}$ pair production with large enough invariant mass, we will probe the gluon density at high values of $x$. Detailed information on the gluon distribution at large $x$ is very important for various phenomenological applications: from infrared renormalon ambiguities in cross-sections to intrinsic charm content of the proton.

%\BP{These data will also} shed light on the role of gluons in the charmonia production mechanisms.
One of the important measurements would be the extraction of the classic double longitudinal spin asymmetry $A_{LL}$, which provides access to the gluon helicity function $\Delta g(x)$.
Due to the quadratic dependence of the asymmetry on $\Delta g(x)$ in the charm production process and linear dependence in case of the prompt-photon production, the corresponding measurements at SPD will be highly complementary and will help to determine both the sign and the size of the gluon polarization.

A special attention should be drawn to the TMD observables.
In particular, both the unpolarized gluon distribution and the gluon Boer-Mulders functions can be probed in (un)polarized collisions at SPD.
The gluon Sivers function $\Delta^{g}_N (x,k_{T})$ and higher-twist effects are planned to be studied via measurements of the single transverse spin asymmetry $A_N$.
Due to the relatively small $\sqrt{s}$, the SPD results will provide inputs for the investigation of possible 
application of the TMD factorization approach in the region of relatively large $x$. This, in turn, is related to gluonic contribution in the valence region  to spin structure of particular hadrons and nuclei (protons and deuterons).       

Unpolarized gluon content of the deuteron can be studied from the explicit comparison of the differential cross-sections for the $p$-$p$ and $d$-$d$ collisions for each of the probes. The effects related to possible non-nucleonic degrees of freedom and the Fermi motion can be investigated at high $x$. Exploration of the deuteron, the simplest nuclear system, is a bridge to another physics program at NICA -- the study of hot and dense hadronic matter in heavy-ion collisions~\cite{Golovatyuk:2019rkb}.
The single spin asymmetry in $pd$ collisions with linearly polarized deuterons, and the double transverse spin asymmetry $A_{TT}$ in $dd$ reactions provide unique opportunities to access for the first time the gluon transversity function $h_{1}^{g}(x)$.
The J/$\psi$ and prompt-photon production measurements with tensor-polarized deuterons at the SPD would help to determine the tensor-polarized gluon distribution.

Main proposed measurements to be carried out at the SPD are summarized in Table~\ref{tab:summary}.

\begin{table}[htp]
\caption{Study of the gluon content in proton and deuteron at SPD.}
\begin{center}
\begin{tabular}{|l|c|c|}
\hline
Physics goal & Observable & Experimental conditions  \bigstrut \\
\hline
Gluon helicity $\Delta g(x)$ & $A_{LL}$ asymmetries & $p_L$-$p_L$, $\sqrt{s}=$27 GeV \bigstrut \\
\hline
Gluon Sivers PDF  $f_{1T}^{\perp g}(x,\ktsq)$,      & $A_N$ asymmetries,    & $p_T$-$p$, $\sqrt{s}=$27 GeV \bigstrut \\
Gluon Boer-Mulders PDF $h_1^{\perp g}(x,\ktsq)$ &   Azimuthal asymmetries        &  $p$-$p$, $\sqrt{s}=$27 GeV \bigstrut \\
TMD-factorization test         &  Diff. cross-sections,               &      $p_T$-$p$,        energy scan                                \bigstrut \\
                                            &      $A_N$ asymmetries          &         \bigstrut \\          
\hline
Unpolarized gluon  &       & $d$-$d$, $p$-$p$, $p$-$d$ \bigstrut \\   
density $g(x)$ in deuteron   &  Differential               &                       $\sqrt{s_{NN}}=$ 13.5 GeV                             \bigstrut \\         
Unpolarized gluon  & cross-sections  & $p$-$p$, \bigstrut \\   
density $g(x)$ in proton   &                 &                       $\sqrt{s}\leq$ 27 GeV                             \bigstrut \\      
\hline  
Gluon transversity $h_1^g(x)$ & Double vector/tensor asymmetries& $d_{tensor}$-$d_{tensor}$, $\sqrt{s_{NN}}=$ 13.5 GeV \bigstrut \\
''Tensor porlarized'' PDF $C_G^T(x)$ & Single vector/tensor asymmetries & $d_{tensor}$-$d$, $p$-$d_{tensor}$ \bigstrut \\
\hline
\end{tabular}
\end{center}
\label{tab:summary}
\end{table}%

The study of the gluon content in the proton and deuteron at NICA SPD will serve as an important contribution to our general understanding of the spin structure of hadrons and QCD fundamentals. The expected inputs from the SPD will be highly complementary to the ongoing and planned measurements at RHIC, and future facilities such as EIC at BNL and fixed-target LHC projects at CERN. 

The physics program of the SPD facility is open for exciting and challenging ideas from theorists and experimentalists worldwide.

\section*{Acknowledgements}
%We thank W. Vogelsang,  for helpful discussion
We thank D. Boer and all participants of the workshop {"Gluon content of proton and deuteron with the Spin Physics Detector at the NICA collider"} for useful and inspiring discussions. 
M.G.E. is supported by the Spanish MICINN grant PID2019-106080GB-C21. M.N. and V.S. are supported by the Ministry of education and science of Russia via the State assignment to educational and research institutions under project FSSS-2020-0014.

%\section*{References}
%[1] https://inspirehep.net/search?p=find+eprint+1309.4235

\bibliography{gluon_paper/case_PPa}
\end{document}